\documentclass[aps,prd,amssymb,groupedaddress]{revtex4}
\usepackage{epsfig}
\usepackage[english]{babel}

\usepackage{amsfonts}
\usepackage{amssymb}
\usepackage{amsmath}

\def\a{\alpha}
\def\b{\beta}
\def\D{\Delta}
\def\d{\delta}
\def\e{\epsilon}

\def\g{\gamma}

\def\l{\lambda}

\def\O{\Phi}

\def\s{\sigma}

\def\t{\tau}

\def\W{\Omega}
\def\w{\omega}

\def\z{\zeta}

\def\pr{\prime}

\def\={\nonumber &=}
\def\nn{\nonumber}
\def\&{{}&}

\def\({\left(}
\def\){\right)}
\def\[{\left[}
\def\]{\right]}
\def\<{\left\langle}
\def\>{\right\rangle}

\def\uk{{\bf \hat{k}}}
\def\un{{\bf \hat{n}}}

\def\ux{{\bf \hat{x}}}
\def\bk{{\bf k}}

\def\curl{\mathcal}

\def\eq{\begin{eqnarray}}
\def\qe{\end{eqnarray}}

\def\and{\quad \mbox{and} \quad}

\def\fnl{f_\textrm{NL}}
\def\barfnl{\bar f_\textrm{NL}}
\def\Fnl{ F_\textrm {NL}}
\def\Fnllocal{F_\textrm {NL}^\textrm{loc}}
\def\fnllocal{f_\textrm {NL}^\textrm{loc}}

\def\bfnl{\kern2pt\overline{\kern-2ptf}_\textrm{NL}}

\def\lmax{l_\textrm{max}}
\def\lall{l_1,l_2,l_3}
\def\lsum{l_1+l_2+l_3}

\def\Blll{B_{l_1l_2l_3}}
\def\blll{b_{l_1l_2l_3}}
\def\bllllocal{b_{l_1l_2l_3}^\textrm{local}}

\def\blllconst{b_{l_1l_2l_3}^\textrm{const}}
\def\blllequil{b_{l_1l_2l_3}^\textrm{equil}}
\def\blllstring{b_{l_1l_2l_3}^\textrm{string}}

\def\kmax{k_\textrm{max}}
\def\kall{k_1,k_2,k_3}
\def\ksum{k_1+k_2+k_3}

\def\Bkkk{B(\kall)}
\def\Skkk{S(\kall)}

\def\Cl{C_l}
\def\alm{a_{lm}}

\def\Vtetra{{{\cal V}_{\cal T}}}

\def\Q{\curl{Q}}
\def\Qn{\curl{Q}_n}

\def\Qp{\curl{Q}_p}
\def\Qnxyz{\curl{Q}_n(x,y,z)}

\def\R{\curl{R}}
\def\Rn{\curl{R}_n}
\def\Rm{\curl{R}_m}
\def\Rp{\curl{R}_p}

\def\barQ{\kern2pt\overline{\kern-2pt\curl{Q}}}
\def\barQn{\barQ_n}

\def\barQp{\barQ_p}

\def\barR{\kern2pt\overline{\kern-2pt\curl{R}}}
\def\barRn{\barR_n}

\def\nmax{n_\textrm{max}}

\def\aR{\alpha^{\scriptscriptstyle{\cal R}}}
\def\aRn{\aR_n}
\def\aRp{\aR_p}
\def\baR{\bar{\alpha}^{\scriptscriptstyle{\cal R}}}
\def\baRn{\baR_n}

\def\aQ{\alpha^{\scriptscriptstyle{\cal Q}}}
\def\aQn{\aQ_n}
\def\aQp{\aQ_p}
\def\baQ{\bar{\alpha}^{\scriptscriptstyle{\cal Q}}}
\def\baQn{\baQ_n}
\def\baQp{\baQ_p}

\def\bbR{\bar{\beta}^{\scriptscriptstyle{\cal R}}}
\def\bbRn{\bbR_n}

\def\bQ{\beta^{\scriptscriptstyle{\cal Q}}}
\def\bQn{\bQ_n}
\def\bQp{\bQ_p}
\def\bbQ{\bar{\beta}^{\scriptscriptstyle{\cal Q}}}
\def\bbQn{\bbQ_n}

\def\Qndefn{q_{\{p}\,q_r\,q_{s\}}}

\def\balpha{\mbox{\boldmath$\alpha$}_l}
\def\bbeta{\mbox{\boldmath$\beta$}_l}
\def\bgamma{\mbox{\boldmath$\gamma$}_l}
\def\bdelta{\mbox{\boldmath$\delta$}_l}
\def\Malpha{M_\alpha(x,\hat{\bf n})}
\def\Mbeta{M_\beta(x,\hat{\bf n})}
\def\Mgamma{M_\gamma(x,\hat{\bf n})}
\def\Mdelta{M_\delta(x,\hat{\bf n})}

\def\setsize{\csname @setfontsize\endcsname \setsize}

\begin{document}


\title{General CMB and Primordial Bispectrum Estimation I: \\
Mode expansion, map-making 
and measures of $\Fnl$}

\author{J.R.~Fergusson}

\author{M.~Liguori}

\author{E.P.S.~Shellard}

\affiliation
{Centre for Theoretical Cosmology,\\
Department of Applied Mathematics and Theoretical Physics,\\
University of Cambridge,
Wilberforce Road, Cambridge CB3 0WA, United Kingdom}

\date{\today}

\begin{abstract}

{\setsize{9}{12}
\noindent
We present a detailed implementation of two parallel bispectrum estimation methods which
can be applied to general non-separable primordial and CMB bispectra.  The method 
exploits bispectrum mode decompositions on the tetrahedral domain of allowed wavenumber
or multipole values, using both separable basis functions and related orthonormal 
modes.   We provide concrete examples of such modes constructed from 
symmetrised tetrahedral polynomials, demonstrating the rapid convergence of 
expansions describing nonseparable bispectra.   We use these modes to create 
rapid and robust pipelines for generating simulated CMB maps of high resolution
($l > 2000$) given an arbitrary primordial power spectrum  and 
bispectrum or  an arbitrary late-time CMB angular power spectrum and bispectrum.  
By extracting coefficients for the same separable basis functions from an observational 
map, we are able to present an efficient $\fnl$ estimator for a given theoretical model with 
a nonseparable bispectrum.   The estimator has two manifestations, comparing the 
theoretical and observed coefficients at either primordial or late times, thus 
encompassing a wider range of models, including secondary anisotropies and lensing 
as well as active models, such as cosmic strings.  We provide examples and 
validation of both $\fnl$ estimation methods by direct comparison with simulations 
in a WMAP-realistic context.    In addition, we demonstrate how the full primordial 
and CMB bispectrum can be extracted from observational maps using these 
mode expansions, irrespective of the theoretical model under study. 
We also propose a universal definition of the bispectrum parameter $\Fnl$, 
so that the integrated bispectrum on the observational domain can be more 
consistently compared between theoretical models.  We obtain WMAP5 estimates 
of $\fnl$ for the equilateral model from both our primordial and late-time estimators 
which are consistent with each other, as well as results already published in the literature.
These general bispectrum estimation methods should prove useful for 
nonGaussianity analysis with the Planck satellite data, as well as in other contexts. 
}
\end{abstract}


\maketitle


\setsize{11}{13}

\section{Introduction}

\noindent Standard  inflationary scenarios predict the Universe to be close to flat with primordial curvature perturbations which are nearly scale-invariant and Gaussian. All these predictions are in very good accord with cosmic microwave background (CMB) and large-scale structure measurements, such as those provided by WMAP and SDSS.  Despite this remarkable agreement, present observations are not able to completely rule out alternatives to inflation, nor to effectively discriminate among the vast number of different inflationary models that have been proposed.  However, almost all such quantitative comparisons derive from inferred measurements of the primordial two-point correlator or power spectrum $P(k)$ from $\langle \zeta\zeta\rangle$, where $\zeta$ is the curvature perturbation. If we wish to subject inflation to more stringent tests and to distinguish between competing models then perhaps the best prospects are offered by studying nonGaussianity, that is, the higher order correlators beyond the power spectrum.   The three-point correlator of the CMB or bispectrum $B_{l_1l_2l_3}$ is a projection
on the sky of the evolved primordial bispectrum $B(\kall)$ arising from $\langle \zeta\zeta\zeta \rangle$, consisting of contributions from triangle configurations with sidelengths given by the wavenumbers $\kall$. The bispectrum has attracted most attention in the literature to date and its study 
is usually simplified to the characterization of a single nonlinearity parameter $\fnl$, which schematically is given by the ratio $\fnl \approx  B(k,k,k)  / P(k)^2$.

Standard inflation, that is, single field slow-roll inflation, predicts a very small bispectrum with $\fnl \sim 0.01$ 
\cite{Maldacena:2002vr,Acquaviva:2002ud}, possessing a characteristic scale-invariant local shape.  (This local shape is dominated by squeezed triangle configurations, that is, those for which one side is much smaller
than the others, e.g.\ $ k_1 \ll k_2, k_3$.)   In fact, such a low signal would be undetectable even by an ideal noiseless CMB experiment, because it is below the level of NG contamination expected from secondary anisotropies $\fnl \approx {\cal O}(1)$.   However,  measurement of a significantly larger primordial $\fnl \gtrsim 1$ would have profound consequences because it would signal the need for new physics during inflation or even a paradigm shift away from it.  Present measurements of this local $\fnl$ are equivocal with the WMAP team reporting \cite{Komatsu:2008hk}
\eq\label{eq:cmblocal}
\fnl = 51\pm 60 ~(95\%)
\qe
and with other teams obtaining higher \cite{07121148} (WMAP3) or equivalent values \cite{Curto:2008ym,Rudjord:2009mh}, while with improved WMAP5 noise analysis a  lower value was found $\fnl = 38 \pm 42$, but at a similar 2$\sigma$ significance \cite{Smith:2009jr}.    The Planck satellite experiment is expected to markedly improve precision measurements with $\Delta \fnl = 5 $ or better
\cite{PlanckBlueBook}.   

Further motivation for the study of the bispectrum comes from the prospect of distinguishing alternative more complex models of inflation which can produce nonGaussianity  with potentially observable amplitudes $\fnl \gtrsim 1$, but also in a  variety of different bispectrum shapes, that is, with the nonGaussian signal peaked for different triangle configurations of wavevectors. To date only special separable bispectrum shapes have been constrained by CMB data, that is, those that can be expressed (schematically) in the form $B(\kall) = X(k_1)Y(k_2) Z(k_3)$, or else can be accurately approximated in this manner.   All CMB analysis, such as those quoted above for the local shape (\ref{eq:cmblocal}), exploits this separability to reduce the dimensionality of the required integrations and summations to bring them to a tractable form. The separable approach reduces the problem from one of $\curl{O}(\lmax^5)$ operations to a  manageable $\curl{O}(\lmax^3)$ \cite{Komatsu:2003iq}. Other examples of meaningful constraints on separable bispectrum shapes using WMAP5 data include those for the equilateral shape \cite{Komatsu:2008hk} and another shape `orthogonal' to both equilateral and local \cite{Senatore:2009gt}.   Despite these three shapes being a good approximation to non-Gaussianity from a number of classes of inflation models,  they are not exhaustive in their coverage of known primordial models \cite{Fergusson:2008ra}, nor other types of late-time non-Gaussianity, such as that from cosmic strings \cite{Hindmarsh:2009qk,Regan:2009hv}; they cannot be expected to be, given the functional degrees of freedom available.  Bringing observations to bear on this much broader class of cosmological models, therefore, is the primary motivation for this paper.

\begin{figure}[t]
\centering
\includegraphics[width=0.99\linewidth]{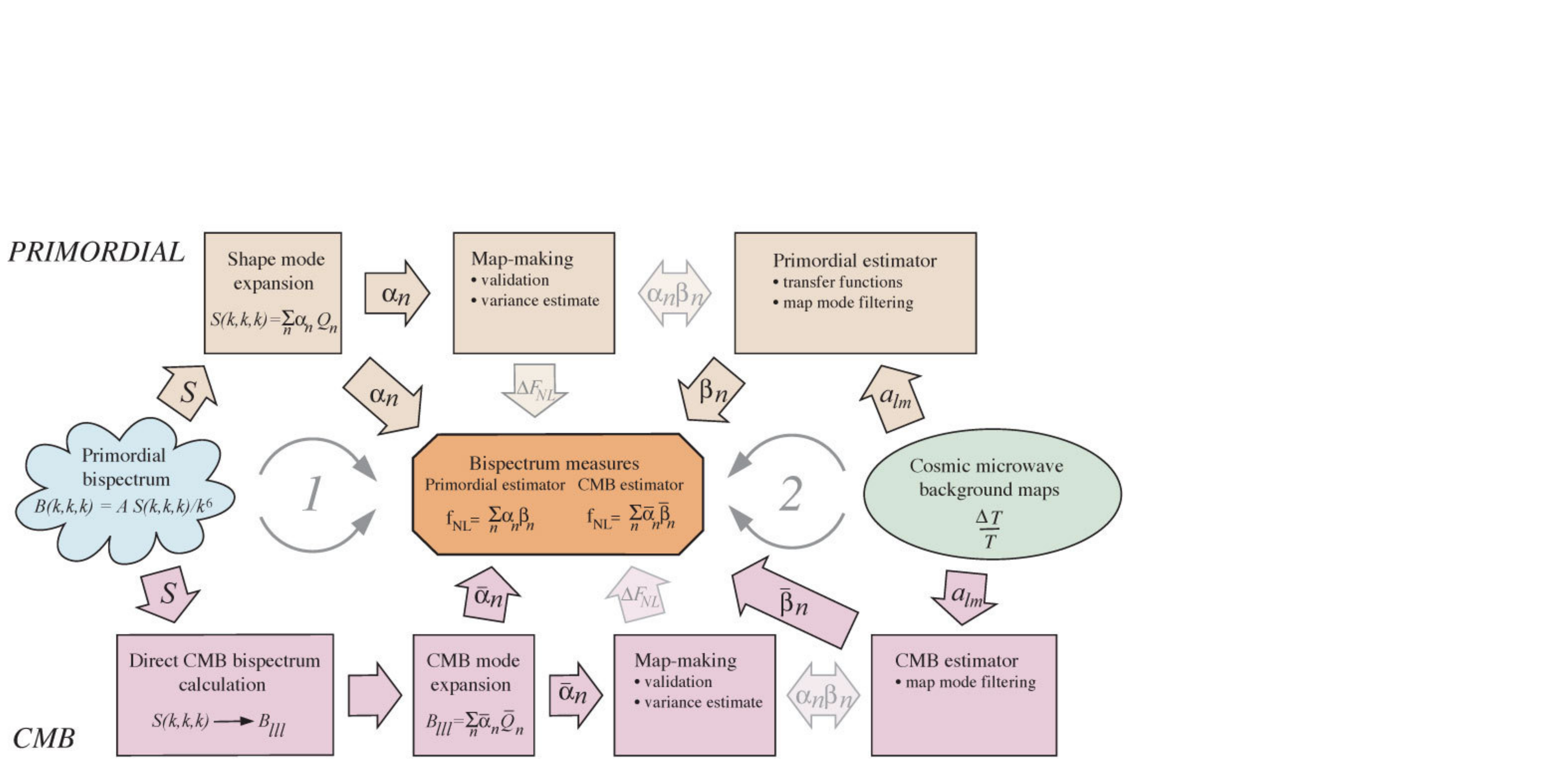}
\caption[Estimator flow chart]{\small Flow chart for the two general estimator methodologies described and implemented
in this article using complete separable mode expansions.   Note the overall redundancy which assists estimator validation 
and the independence of the extraction of expansion coefficients from theory $\alpha_n$ (cycle 1) and data $\beta_n$ (cycle 2).  Explanations for the schematic equations can be found in the main text. }
\label{fig:estimators}
\end{figure}

In a previous paper \cite{Fergusson:2006pr}, we described a general approach to 
the estimation of non-separable CMB bispectra.  The method has developed out of the first direct 
calculations of the reduced CMB bispectrum $\blll$ which surveyed a wide variety of non-separable 
primordial models, revealing smooth coherent patterns of acoustic peaks on the tetrahedral domain of 
allowed multipole values.    Since the $\blll$ could be well represented using a  limited 
number of bins, we could similarly decompose them into orthogonal mode functions which converged
in relatively short mode expansions \cite{Fergusson:2008ra}.  Here, we describe the detailed 
implementation of these 
methods in  a comprehensive dual approach to estimating bispectrum parameters which is illustrated
in fig.~\ref{fig:estimators}.   We present concrete examples of separable basis functions $\Qn$ (symmetrised
tetrahedral polynomials) and corresponding orthonormal modes $\Rn$ on the domain of allowed wavenumbers $\kall$;  these are then deployed within a more general mode 
expansion methodology.   In the first primordial implementation, we decompose an arbitrary 
non-separable shape $S$ using separable basis functions with coefficients $\alpha_n$.   This expansion
can be used for a  fast calculation of the full CMB bispectrum $ \Blll$ (section III), as well as leading 
to a robust method for generating simulated maps from a given power spectrum $P(k)$
and bispectrum $B(\kall)$ (section IV).  Our main emphasis here, however, is on a primordial 
estimator for $\fnl$ which is achieved by a confrontation between theory, represented by the 
$\alpha_n$ coefficients, and a set of observational coefficients $\beta_n$ found by extracting 
the same modes from the observed CMB map (section III).   Examples of simulated maps and 
recovery of the input $\fnl$ are given in section V in a WMAP-realistic context. 

In the second and parallel late-time implementation (see fig.~\ref{fig:estimators}), we assume the 
theoretical CMB bispectrum $\Blll$ is calculated already from the primordial shape \cite{Fergusson:2008ra} 
or because it is a late-time effect ranging from secondary 
anisotropies through to fluctuations induced by cosmic strings.   A separable mode 
expansion of $\Blll$ allows for a simpler and more direct approach to $\fnl$ estimation, as well 
as simulated map generation, in a wider variety of scenarios.   Here, as well as primordial 
models we consider the antithetical example of cosmic strings. 
These two estimator methods are complementary with each having distinct advantages 
depending on the properties and generation mechanism of the non-Gaussianity under 
investigation.  They provide independent validation in situations where both are applicable.    

It remains to point out recent and related developments, especially those by colleagues in 
Planck Working Group 4 (NonGaussianity).   To date most primordial shapes have been assumed 
to be scale-invariant, but in ref.~\cite{Sefusatti:2009xu} some deviations from the local shape were 
considered in developing a more general approach.  Spherical Mexican wavelets, using a limited 
number of scales, were employed in ref.~\cite{Curto:2008ym, Curto:2009pv} to estimate $\fnl$ for the local shape
with WMAP3 data, providing a constraint consistent with (\ref{eq:cmblocal}).  Similar work has been 
achieved for needlets with corresponding constraints \cite{Rudjord:2009mh}), again essentially 
tailoring the method to the local template using local shape map simulations.   
Another approach to a late-time estimator has also 
exploited the smoothness of the reduced CMB bispectrum by using a limited number of 
multipole bins \cite{Bucher:2009nm}.  The method was tested for the local shape using 
map simulations, and emphasised Planck forecasts investigating the pattern of acoustic peaks 
in the local model.   We shall discuss here how these late time approaches -- whether 
wavelets, bins or other alternatives -- fall within the general mode expansion methodology 
outlined previously \cite{Fergusson:2006pr} and can be applied, in principle, to explore 
nonseparable primordial models  beyond local nonGaussianity.
We point out in the implementation presented here, however, that direct estimation of the CMB bispectrum 
can be achieved without reference to the calculated bispectrum for a particular model and without
relying on corresponding CMB map simulations.

\section{The CMB bispectrum and $\fnl$ estimation}

\subsection{Relation between primordial and CMB bispectra}

In this section we will review some basic definitions and mathematical formulae that will be used throughout the rest
of the paper. Our work will be concerned with the analysis of the three-point function induced by a NG primordial
gravitational potential $\Phi({\bf k})$ in the CMB temperature fluctuation field.
Temperature anisotropies are represented using the $a_{lm}$ coefficients of a spherical harmonic decomposition of 
the cosmic microwave sky, 
$$ 
\frac{\Delta T}{T}(\hat {\bf n}) = \sum_{lm} a_{lm} Y_{lm}(\hat {\bf n})\,.
$$
The primordial potential $\Phi$ is imprinted on the CMB mutipoles $a_{lm}$ by a convolution with transfer
functions $\D_l(k)$ representing the linear perturbation evolution, through the integral 
\begin{align} \label{eq:alm}
a_{lm} = 4\pi (-i)^l \int \frac{d^3 k}{(2\pi)^3}\, \D_l(k) \,\O(\bk)\, Y_{lm}(\uk)\,.
\end{align}
The CMB bispectrum is the three point correlator of the $a_{lm}$, so substituting we obtain 
\eq\label{eq:bispectrum1}
B^{l_1 l_2 l_3}_{m_1 m_2 m_3} &=& \<a_{l_1 m_1} a_{l_2 m_2} a_{l_3 m_3}\>\\
&=& (4\pi)^3 (-i)^{l_1+l_2+l_3} \int \frac{d^3 k_1}{(2\pi)^3} \frac{d^3 k_2}{(2\pi)^3} \frac{d^3 k_3}{(2\pi)^3} \D_{l_1}(k_1) \D_{l_2}(k_2) \D_{l_3}(k_3) \times\\
&&\qquad\qquad\qquad\qquad \<\O(\bk_1)\O(\bk_2)\O(\bk_3)\> Y_{l_1 m_1}(\uk_1) Y_{l_2 m_2}(\uk_2) Y_{l_3 m_3}(\uk_3)\,,
\qe
where $k_1 = |{\bf k}_1|$,  $k_2 = |{\bf k}_2|$ and  $k_3 = |{\bf k}_3|$. 
Here, we define the primordial bispectrum  as
\begin{align}\label{eq:primbispect}
\<\O(\bk_1)\O(\bk_2)\O(\bk_3)\> = (2\pi)^3 B_\O(k_1,k_2,k_3)\, \d(\bk_1+\bk_2+\bk_3)\,,
\end{align}
where the delta function enforces the triangle condition, that is, the constraint imposed by translational invariance
that wavevectors in Fourier space must close to form a 
triangle, ${\bf k}_1+{\bf k}_2+{\bf k}_3=0$. We replace the delta function in (\ref{eq:primbispect}) with its exponential integral form, substitute this into equation (\ref{eq:bispectrum1}) and integrate out the angular parts of the three ${\bf k}_i$ integrals in the usual manner to yield
\begin{align}\label{eq:bispectrum3}
\nn B^{l_1 l_2 l_3}_{m_1 m_2 m_3} = \(\frac{2}{\pi}\)^3 \int x^2d x \int  \& d k_1 d k_2 d k_3 (k_1 k_2 k_3)^2 B_\O(k_1,k_2,k_3) \,\D_{l_1}(k_1) \D_{l_2}(k_2) \D_{l_3}(k_3)\\
& \times j_{l_1}(k_1 x) j_{l_2}(k_2 x) j_{l_3}(k_3 x) \int d\W_x \,Y_{l_1 m_1}(\ux) Y_{l_2 m_2}(\ux) Y_{l_3 m_3}(\ux) \,.
\end{align}
The last integral over the angular part of $x$ is known as the Gaunt integral which can be expressed in terms of Wigner-$3j$ symbols as
\begin{align}\label{eq:Gaunt}
\nn \curl{G}^{l_1 l_2 l_3}_{m_1 m_2 m_3} &\equiv \int d\W_x Y_{l_1 m_1}(\ux) Y_{l_2 m_2}(\ux) Y_{l_3 m_3}(\ux) \\
&= \sqrt{\frac{(2l_1+1)(2l_2+1)(2l_3+1)}{4\pi}} \( \begin{array}{ccc} l_1 & l_2 & l_3 \\ 0 & 0 & 0 \end{array} \) \( \begin{array}{ccc} l_1 & l_2 & l_3 \\ m_1 & m_2 & m_3 \end{array} \)\,. 
\end{align}
Given that most theories we shall consider are assumed to be isotropic, it is usual to work with the angle-averaged bispectrum,
\begin{align}
B_{l_1 l_2 l_3} = \sum_{m_i} \( \begin{array}{ccc} l_1 & l_2 & l_3 \\ m_1 & m_2 & m_3 \end{array} \)\<a_{l_1 m_1} a_{l_2 m_2} a_{l_3 m_3}\>\,.
\end{align}
or the even more convenient reduced bispectrum which removes the geometric factors associated with the Gaunt integral,
\begin{align}\label{eq:Gauntredbispect}
B^{l_1 l_2 l_3}_{m_1 m_2 m_3} = \curl{G}^{l_1 l_2 l_3}_{m_1 m_2 m_3} b_{l_1 l_2 l_3}\,.
\end{align}
The reduced bispectrum from (\ref{eq:bispectrum1}) then takes the much simpler form
\begin{align} \label{eq:redbispect}
\nn b_{l_1 l_2 l_3}= \(\frac{2}{\pi}\)^3 \int x^2dx \int & d k_1 d k_2 d k_3\, \( k_1 k_2 k_3\)^2\, B_\O(k_1,k_2,k_3)\\
&\times  \D_{l_1}(k_1) \,\D_{l_2}(k_2)\, \D_{l_3}(k_3)\, j_{l_1}(k_1 x)\, j_{l_2}(k_2 x)\, j_{l_3}(k_3 x)\,.
\end{align}
Here, it is important to note that the Gaunt integral in (\ref{eq:Gauntredbispect}) encodes several constraints on the angle 
averaged bispectrum $\Blll$ which are no longer transparent in the 
reduced bispectrum $\blll$.  These are, first, that the sum of the three multipoles $l_i$ must be even and, secondly, that the $l_i$'s satisfy the 
triangle condition, analogously to the wavenumbers $k_i$.  
For wavenumbers, the triangle condition is enforced through the $x$-integral over the three spherical Bessel functions $j_l(k_ix)$ which 
evaluates to zero if the $k_i$'s cannot form a triangle, whereas in multipole space it is enforced by the angular integration $d\Omega_x$ over the 
spherical harmonics $Y_{l_im_i}$ in (\ref{eq:Gaunt}).    Appreciating the origin of these constraints is important 
when we later consider the separability of the reduced bispectrum expression
(\ref{eq:bispectrum3}).  

\subsection{Separable primordial shapes and CMB bispectrum solutions}

Given that the primordial power spectrum is very nearly scale-invariant, it is expected that the 
bispectrum will behave similarly.   In order to bring the bispectrum to a scale-invariant 
form we have to appropriately eliminate a $k^6$ scaling which naturally arises in (\ref{eq:primbispect}). 
This is usually achieved by multiplying through by the factor $(k_1 k_2 k_3)^2$ appearing in (\ref{eq:redbispect}) 
and defining a primordial shape function as
\begin{align} \label{eq:shapefn}
S(k_1,k_2,k_3) \equiv \frac{1}{N} (k_1 k_2 k_3)^2 B_\O(k_1,k_2,k_3)\,,
\end{align}
where $N$ is a normalisation factor which is often taken such that for equal $k_i$ the shape
function has unit value $S(k,k,k)=1$.  (This normalisation is also used for $\fnl$, but it only strictly 
applies for scale-invariance and, in any case, leads to inconsistent comparisons between different models, as 
we shall discuss in section IV.)    We thus characterise scale-invariant models in terms of an overall
amplitude, parametrised by $\fnl$, and their transverse shape, described by $S(\kall)$ on a triangular slice with 
$\ksum = \hbox{const.} $ \cite{0405356}.    This leaves a two-dimensional space
on which it is most elegant to use the two independent variables $\tilde \alpha, \,\tilde\beta$ \cite{0506704, Fergusson:2006pr}
\eq
\tilde \alpha = (k_2-k_3)/\tilde k \,,\qquad \tilde \beta = (\tilde k - k_1)/\tilde k\,,\quad
\quad \hbox{where}\quad \tilde k = {\textstyle \frac{1}{2}}(\ksum)= \hbox{const.}\,, 
\qe
with the following domains  $0 \le\tilde \b \le 1$ and $ -(1-\tilde\b)\le \tilde\a \le 1 - \tilde\b$.
For scale-dependent models with a non-trivial variation in $\tilde k$, the full three-dimensional 
 dependence on the $k_i$ must be retained.    In terms of the shape function (\ref{eq:shapefn}), the reduced bispectrum (\ref{eq:redbispect}) 
 can be rewritten as
 \begin{eqnarray}\label{eq:redbispect2}
b_{l_1 l_2 l_3} &=& \frac{1}{N}\(\frac{2}{\pi}\)^3 \int  x^2 dx\int d k_1 d k_2 d k_3\,  S(k_1,k_2,k_3)\,
 \D_{l_1}(k_1) \D_{l_2}(k_2) \D_{l_3}(k_3)\, j_{l_1}(k_1 x) j_{l_2}(k_2 x) j_{l_3}(k_3 x).
\end{eqnarray}

The simplest possible shape function is the constant model 
\eq\label{eq:constantS}
S(\kall) = 1\,,
\qe
for which a large-angle analytic solution for the reduced bispectrum was presented in ref.~\cite{Fergusson:2008ra},
\begin{align}\label{eq:constbispect}
\blllconst ~=~ \frac{\D^2_\O}{27 N} \frac{1}{(2l_1+1)(2l_2+1)(2l_3+1)}\[\frac{1}{l_1+l_2+l_3+3} + \frac{1}{l_1+l_2+l_3}\]\,, \qquad (l\ll 200)\,. 
\end{align}
Here, we take the Sachs-Wolfe approximation that $\D_{l}\(k\) = \frac{1}{3} j_{l}\((\t_o - \t_{dec}) \, k\)$ for $l\ll 200$ and exploit the manifest separability of 
the expression (\ref{eq:redbispect2}) to perform the one-dimensional  $k_i$ integrations individually.  The more 
general constant solution does not have an analytic solution for $l\gtrsim 200$, for the reason that 
the transfer functions cannot be expressed in a simple form, but  it can be evaluated 
numerically from the expression
\eq\label{eq:constbispectsep}\label{eq:Iterm}
\blllconst ~=~ \frac{\D^2_\O}{N} \int  x^2 dx \, {\cal I}_{l_1}(x)\, {\cal I}_{l_2}(x)\, {\cal I}_{l_3}(x)\,,  \qquad \hbox{where} \quad   {\cal I}_l(x) = \frac{2}{\pi} \int d k\, \D_{l}(k) \, j_{l}(k x)\,.
\qe
The large-angle solution (\ref{eq:constbispect}) is an important benchmark 
with which to compare the shape of late-time CMB bispectra from other models  $\blll$ (note the $l^{-4}$ scaling) and, 
additionally, it has some further recent physical motivation \cite{Chen:2009we}.

The most studied scale-invariant shape function is the local model,
\eq\label{eq:localS}
S(\kall) &=& \frac{1}{3} \(\frac{k_1^2}{k_2k_3}+\frac{k_2^2}{k_1k_3}+\frac{k_3^2}{k_1k_2}\)\\
\nn &\approx&~ \frac{(k_1k_2k_3)^2}{3\D^2_\O}\left[ P(k_1)P(k_2)+P(k_2)P(k_3)+P(k_3)P(k_1)\right]\,,
\qe
where, in the second line, we also allow for power spectra which are nearly scale invariant,  defined by 
$\langle \Phi({\bf k})\Phi^*({\bf k}') \rangle = (2\pi)^3 P(k) \delta ({\bf k} -{\bf k}')$ with $P(k) \sim k^{-3}$.  
Using the Sachs-Wolfe approximation again, this has  the corresponding large-angle analytic solutions
\eq\label{eq:localbispect}
\bllllocal ~&=&~  \frac{2\D^2_\O}{27\pi^2}\(\frac{1}{l_1(l_1+1)l_2(l_2+1)} + \frac{1}{l_2(l_2+1)l_3(l_3+1)} + \frac{1}{l_3(l_3+1)l_1(l_1+1)}\)
\qe
Here, we see that the divergences for the squeezed triangles ($k_1\ll k_2,k_3 ...$) in the primordial 
shape (\ref{eq:localS}) are also reflected in $\bllllocal$, making it a much less useful for relative comparison than the 
constant model (\ref{eq:constbispect}).   It is straightforward, in principle, to calculate the full bispectrum 
from the separable expressions arising from (\ref{eq:localS}),
\eq\label{eq:localbispectsep}
\bllllocal  =  \int x^2 dx \left [  \mbox{\boldmath$\alpha$}_{l_1} (x) \mbox{\boldmath$\beta$}_{l_2}(x)   \mbox{\boldmath$\beta$}_{l_3}(x) + (\hbox{2 perms})\right]\, ,
\qe 
where the separated integrals analogous to (\ref{eq:constbispectsep}) become 
\eq\label{eq:ABterm}
\balpha(x) = \frac {2}{\pi} \int dk \, k^2\, \D_{l}(k) \, j_{l}(k x)\,, \qquad \bbeta (x) = \frac {2}{\pi} \int dk \, k^2 P(k) \, \D_{l}(k) \, j_{l}(k x)\,. 
\qe
We note that these highly oscillatory integrals must be evaluated numerically with considerable care. 

The separable equilateral shape has also received a great deal of attention with  \cite{0405356}
\eq\label{eq:equilS}
S(\kall) &=& \frac{(k_1+k_2-k_3)(k_2+k_3-k_1)(k_3+k_1-k_2)}{k_1k_2k_3} \\
\nn &=& -2 - \left[\frac{k_1^2}{k_2k_3} + (\hbox{2 perms})\right] +\left[ \frac{k_1}{k_2} + (\hbox{5 perms})\right]\,.
\qe
This is a much more regular shape than local (\ref{eq:localS}) with the signal dominated by equilateral triangle
configurations $k_1\approx k_2\approx k_3$ (the apparent divergence of the local shape in the second term 
cancels against the third).  There is no simple large-angle analytic solution known for the equilateral 
model, unlike (\ref{eq:constbispectsep}) and  (\ref{eq:localbispectsep}).  In order to calculate the full equilateral 
bispectrum we evaluate the simplified expression 
\eq\label{eq:equilbispectsep}
\blllequil =  \int x^2 dx  \big\{ 2  \mbox{\boldmath$\delta$}_{l_1}  \mbox{\boldmath$\delta$}_{l_2}   \mbox{\boldmath$\delta$}_{l_3} + \left[ \mbox{\boldmath$\alpha$}_{l_1}  \mbox{\boldmath$\beta$}_{l_2}  \mbox{\boldmath$\beta$}_{l_3} + (\hbox{2 perms})\right]
+ \left[ \mbox{\boldmath$\beta$}_{l_1}  \mbox{\boldmath$\gamma$}_{l_2}  \mbox{\boldmath$\delta$}_{l_3} + (\hbox{5 perms})\right] \big\}\,,
\qe 
where $\balpha, \,\bbeta$ are given in (\ref{eq:ABterm}) and $\bgamma,\,\bdelta$ are defined by 
\eq\label{eq:CDterm}
\bgamma(x) = \frac {2}{\pi} \int dk \, k^2\, P(k) ^{1/3} \D_{l}(k) \, j_{l}(k x)\,, \qquad\bdelta(x) = 
\frac {2}{\pi} \int dk \, k^2\, P(k) ^{2/3} \D_{l}(k) \, j_{l}(k x)\,.
\qe

The equilateral shape is not derived 
directly from a physical model, but was
chosen phenomenologically as a good separable approximation to specific models including the non-local part
of Maldacena's original shape \cite{Maldacena:2002vr}, as well as non-canonical cases such as  higher derivative 
models \cite{0306122} and DBI 
inflation \cite{0404084} (for a review of single-field inflation shapes, see e.g.\ ref.~\cite{0605045}).  These shapes are, in general, non-separable from the perspective of the integral (\ref{eq:redbispect2}).
Here, we give a specific shape example for a model with  higher derivative operators (which is also identical to DBI inflation):
\eq\label{eq:dbiS}
S(\kall) = \frac{1}{k_1 k_2 k_3 (k_1+k_2+k_3)^2} \(\sum_i k_i^5 + \sum_{i \neq j}(2 k_i^4 k_j - 3 k_i^3 k_j^2) 
+ \sum_{i \neq j \neq l}(k_i^3 k_j k_l - 4 k_i^2 k_j^2 k_l)\).
\qe
Not only is the equilateral shape (\ref{eq:equilS}) an excellent approximation to (\ref{eq:dbiS}), a full Fisher matrix analysis of the respective CMB 
bispectra has shown they are 99\% correlated out to $\lmax \le 2000$ \cite{Fergusson:2008ra}.   However, a simple separable approximation is 
not necessarily available for arbitary primordial shapes, nor is a particular separable representation necessarily convenient from a calculational perspective (as we shall discuss in section V for the equilateral case above).  In ref.~\cite{Fergusson:2008ra}, we reviewed models currently proposed in the literature showing that families of CMB bispectra arising from non-separable shapes, such as feature and flattened models, are largely independent of the separable models currently constrained observationally (see also discussion of a `cosine' shape 
correlator in ref.~\cite{0405356}). The independence of two shapes $S$ and $S'$ can be calculated from the integral \cite{Fergusson:2008ra}
\begin{align}
F_\e(S,S^\pr) = \int_{\curl{V}_k}\, S(k_1,k_2,k_3)\, S^\pr(k_1,k_2,k_3)\, \w_\e(k_1,k_2,k_3) d\curl{V}_k\,,
\end{align}
where we choose the weight to be
\begin{align}
w(k_1,k_2,k_3) = \frac{1}{k_1+k_2+k_3}\,,
\end{align}
reflecting the scaling we see in the CMB correlator we meet in the next section. The shape correlator is then defined by 
\begin{align}\label{eq:shapecorrelator}
\bar{\curl{C}}(S,S^\pr) = \frac{F(S,S^\pr)}{\sqrt{F(S,S)F(S^\pr,S^\pr)}}\,.
\end{align}

By way of further illustration of the need to move beyond simple separable primordial shape functions, we present the 
late-time CMB bispectrum predicted analytically for cosmic strings \cite{Regan:2009hv} 
\begin{align}\label{eq:stringbispect}
\blllstring  = \frac{A}{(\z l_1 l_2 l_3)^2} \[(l_3^2 - l_1^2 - l_2^2)\(\frac{L}{2l_3} + \frac{ l_3}{50L}\)\sqrt{\frac{l_*}{500}}\,\mbox{erf}(0.3 \z l_3) ~+~ \mbox{2 perms}\] , \qquad (l \le 2000)\,,
\end{align}
where $l_{min} = \min(l_1,l_2,l_3)$,  $l_* = \min(500,l_{min})$, $\z = \min(1/500,1/l_{min})$ and 
\begin{align}
L = \z \sqrt{{\textstyle\frac{1}{2}}({l_1^2 l_2^2 + l_2^2 l_3^2 + l_3^2 l_1^2} )- {\textstyle\frac{1}{4}}{(l_1^4+l_2^4+l_3^4)}}\,.
\end{align}
Here, $A\sim (8\pi G\mu)^3$ is a model dependent amplitude with $G\mu = \mu/m_{\rm Pl}^2$ measuring the string tension $\mu$ relative
to the Planck scale. The cutoffs around $l\approx 500$ in (\ref{eq:stringbispect}) are associated with the 
string correlation length at decoupling (perturbations with $l\gtrsim500$ can only be causally seeded after 
last scattering).  (For the original small angle solution valid for $l\gg 2000$, see ref.~\cite{Hindmarsh:2009qk, Regan:2009hv}.)  Here, the non-separable nature and very different scaling  of the string CMB bispectrum are clear from a comparison 
with (\ref{eq:localbispect}).  Moreover, given the late-time origin of this signal from string metric perturbations,  the modulating effect of acoustic peaks from the transfer functions is absent.

\subsection{Estimators for $\fnl$ and related correlators}

The main purpose of this non-Gaussian CMB analysis is to measure the CMB bispectrum induced by non-Gaussianities in 
the primordial gravitational potential, the link being given by equation (\ref{eq:redbispect}).
Unfortunately, the bispectrum signal is too weak to measure individual multipoles directly, so to compare theory with observation we must use an estimator which sums over the available multipoles. An estimator can be thought of as performing a least squares fit of the bispectrum 
predicted by theory $\<a_{l_1 m_1} a_{l_2 m_2} a_{l_3 m_3}\>$ to the bispectrum actually obtained from observations $a^{obs}_{l_1 m_1} a^{obs}_{l_2 m_2} a^{obs}_{l_3 m_3}$. Ignoring sky cuts and inhomogeneous noise,  the estimator is weighted with the expected 
signal variance from $C_l$ and written in the simple form
\eq \label{eq:estimator}
\nn \curl{E} &=& \frac{1}{N} \sum_{l_i m_i} \frac{\<a_{l_1 m_1} a_{l_2 m_2} a_{l_3 m_3}\>\,\, a^{obs}_{l_1 m_1} a^{obs}_{l_2 m_2} a^{obs}_{l_3 m_3}}{C_{l_1}C_{l_2}C_{l_3}}\\
&=& \frac{1}{N} \sum_{l_i m_i}\curl{G}^{l_1 l_2 l_3}_{m_1 m_2 m_3}\,{b_{l_1 l_2 l_3} }\,\frac {a^{obs}_{l_1 m_1} 
a^{obs}_{l_2 m_2} a^{obs}_{l_3 m_3}}{C_{l_1} C_{l_2} C_{l_3}}\,.
\qe
where we have used (\ref{eq:Gauntredbispect}) and  the Gaunt integral is given in (\ref{eq:Gaunt}) and $N$ is the usual normalisation factor,
\begin{align}
N = \sum_{l_i} \frac{B_{l_1 l_2 l_3}B_{l_1 l_2 l_3}}{C_{l_1}C_{l_2}C_{l_3}}\,.
\end{align}
We note from the second line of (\ref{eq:estimator}) that, for a given theoretical model, we need only  calculate the reduced bispectrum
$\blll$ rather than the much more challenging full bispectrum, $\<a_{l_1 m_1} a_{l_2 m_2} a_{l_3 m_3}\>$.

The above estimator has been shown to be optimal \cite{0503375} for general bispectra in the limit where the non-Gaussianity is small and the observed map is free of instrument noise and foreground contamination. Of course, this is an idealised case and we need to consider taking into account the effect of sky cuts and inhomogeneous noise, which was considered in some detail in refs~\cite{0509029,07114933}. In the more general case the optimal estimator takes the form:
\eq\label{eq:optimalestimator}
\curl{E} &=& 
\frac{1}{N} 
\sum_{l_i m_i} 
\( \begin{array}{ccc} l_1 & l_2 & l_3 \\ m_1 & m_2 & m_3 \end{array} \) 
B_{l_1 l_2 l_3} \times\\
&&\nn\quad\left[ \left( C^{-1} a^{obs} \right)_{l_1 m_1} \left( C^{-1} a^{obs} \right)_{l_2 m_2} 
\left( C^{-1} a^{obs} \right)_{l_3 m_3} + 
C^{-1}_{l_1 m_1,l_2 m_2} \left( C^{-1} a^{obs} \right)_{l_3 m_3} \right] \, ,
\qe
where the covariance matrix $C$ is now non-diagonal due to mode-mode coupling introduced by the mask and anisotropic noise.
Moreover, due to the breaking of isotropy, an additional term linear in the $a_{lm}$ has now to be added 
in order to maintain the optimality of the estimator \cite{0503375}. In the ideal case one can easily see that the linear term 
is proportional to a monopole, while the covariance matrix is diagonal and equal to ${1/C_l}$, thus reproducing the initial 
formula (\ref{eq:estimator}).

In this paper we will follow the approach of \cite{08030547} and approximate the estimator (\ref{eq:optimalestimator}) as
\begin{align}\label{approxestimator}
\curl{E} = \frac{1}{\tilde{N}} \sum_{l_i m_i} \frac{\curl{G}^{l_1 l_2 l_3}_{m_1 m_2 m_3} \, \tilde{b}_{l_1 l_2 l_3} }{ \tilde{C}_{l_1}\tilde{C}_{l_2}\tilde{C}_{l_3} } \(a^{obs}_{l_1 m_1} a^{obs}_{l_2 m_2}  - 6 C^{sim}_{l_1 m_1 , l_2 m_2} \) a^{obs}_{l_3 m_3}\,,
\end{align}
where the tilde denotes modification to include experimental effects.  The normalisation becomes
\begin{align}
\tilde{N} = f_{sky} \sum_{l_i} \frac{\tilde{B}^2_{l_1 l_2 l_3}}{\tilde{C}_{l_1}\tilde{C}_{l_2}\tilde{C}_{l_3}}\,,
\end{align}
with the $C_l$'s and $\blll$ now incorporating beam and noise effects through  
\begin{align}
\tilde{C}_l = b_l^2 C_l + N_l   \qquad \mbox{and}\qquad \tilde{b}_{l_1 l_2 l_3} = b_{l_1}b_{l_2}b_{l_3}\, b_{l_1 l_2 l_3}\,.
\end{align}
Here, $b_l$ is the beam transfer function, $N_l$ the noise power spectrum, $f_{sky}$ the fraction of the sky remaining after application of the mask and  $C^{sim}_{l_1 m_1 , l_2 m_2}$ is the covariance matrix calculated from Gaussian simulations.
In what follows, it will be clear from the context whether beams, noise and masks are being incorporated in the analysis,
so for simplicity we shall continue with the original estimator notation (\ref{eq:estimator}). 

The estimator (\ref{eq:estimator}) also naturally defines a correlator for testing whether two competing bispectra could be differentiated by an ideal experiment. Replacing the observed bispectrum with one calculated from a competing theory we have,
\begin{align}\label{eq:cmbcor}
\curl{C}(B,B^\pr) = \frac{1}{N}\sum_{l_i} \frac{B_{l_1 l_2 l_3}B^\pr_{l_1 l_2 l_3}}{C_{l_1} C_{l_2} C_{l_3}}\,,
\end{align}
where now the normalisation $N$ is defined as follows,
\begin{align}
N = \sqrt{ \sum_{l_i}\frac{B^2_{l_1 l_2 l_3}}{C_{l_1} C_{l_2} C_{l_3}}} \sqrt{\sum_{l_i} \frac{{B^\pr}^2_{l_1 l_2 l_3}}{C_{l_1} C_{l_2} C_{l_3}}}\,.
\end{align}
While this late time correlator is the best measure of  whether two CMB bispectra are truly independent, it requires a full calculation of the CMB bispectrum which is time consuming in general. In \cite{Fergusson:2008ra} we determined that for the majority of models the shape correlator (\ref{eq:shapecorrelator}) introduced earlier is sufficent to determine independence.

An inspection of equations (\ref{eq:estimator},\ref{eq:optimalestimator}) shows that a brute force numerical implementation of the optimal estimator above would take $\curl{O}( l_{max}^5)$ operations. This means an implementation is not feasible for 
the angular resolutions achieved by present and forthcoming datasets (e.g.\ in the signal dominated regime we have $l_{max} \lesssim 500$ 
for WMAP and $l_{max} \lesssim 2000$ for Planck). However, as initially shown 
in ref.~\cite{Komatsu:2003iq}, if a specific theoretical bispectrum can be written in separable form 
as $B(k_1,k_2,k_3) = X(k_1)Y(k_2)Z(k_3)$  then the computational cost of the algorithm can be reduced to 
$\curl{O} (l_{max}^3)$ operations, making the estimation tractable even at very high angular resolutions. 
This establishes the fact that separability is a crucial property for realistic data analysis, even though it is 
not generic for well-motivated inflationary and other models.   As we have seen, the usual solution adopted has been to approximate 
the primordial non-separable shape under study using a separable form that is highly correlated with the original.
This kind of approach requires a case-by-case analysis of all non-separable 
bispectra arising from different 
models and an educated  ``guess'' of a good separable approximation, the close correlation of which must be verified numerically before moving on to the real analysis. Besides being impractical, this can also prove to be extremely 
difficult in specific cases. The aim
of this work is then to find a completely general mathematical framework to ``separate'' shapes, both primordial and late-time, 
and thus build a general
pipeline for $f_{\rm NL}$ estimation and simulation of non-Gaussian CMB maps, that can be applied to any shape 
of interest.

\section{Bispectrum mode decomposition}

Our goal is to represent arbitrary non-separable primordial bispectra $\Bkkk$ or CMB bispectra $\blll$ 
on their respective wavenumber or multipole domains using a rapidly convergent 
mode expansion \cite{Fergusson:2006pr}.   
Moreover, we need to achieve this in a separable manner, making tractable 
the three-dimensional integrals required for bispectrum estimation (\ref{eq:redbispect2}) by breaking 
them down into products of one-dimensional integrals.   In particular, this means that we wish to 
expand an arbitrary non-separable primordial shape function as
\eq \label{eq:separablebasis}
\Skkk = \sum_p\sum_r\sum_s \alpha _{prs} \,q_p(k_1) \, q_r(k_2) \, q_s(k_3) \, ,
\qe 
where the $q_p$ are appropriate basis mode functions which are convergent and complete, that is,  they 
span the space of all functions on the bispectrum wavenumber (or multipole) domain.  
In what follows below, we present one pathway for efficiently achieving this objective in stages.   
First, we create examples of one-dimensional mode 
functions $q_p(k_1)$ in the $k_1$-direction which are orthogonal and well-behaved over the full  
wavenumber (or multipole) domain.  We then construct three-dimensional products of these mode 
functions $q_p(k_1)q_r(k_2)q_p(k_3)\rightarrow \Qn$ creating a complete basis for all possible bispectra 
on the given domain.  Finally, by orthonormalising these product basis functions $\Qn\rightarrow \Rn$, we 
obtain a rapid and convenient method for calculating the relevant expansion coefficients $\alpha_{prs}$ in 
(\ref{eq:separablebasis}).  The subsequent discussion and implementation of general primordial and 
CMB bispectrum estimators, as well as map-making methods, is then built around these mode functions
$q_p, \,\Qn,$ and $\Rn$.   Here, we use bounded symmetric polynomials as a concrete and working 
implementation of this methodology, and we defer discussion about other possible 
basis mode functions which have been investigated to the end of the section.

\subsection{Tetrahedral domain and weight functions}

\begin{figure}[t]
\centering
\includegraphics[width=.45\linewidth]{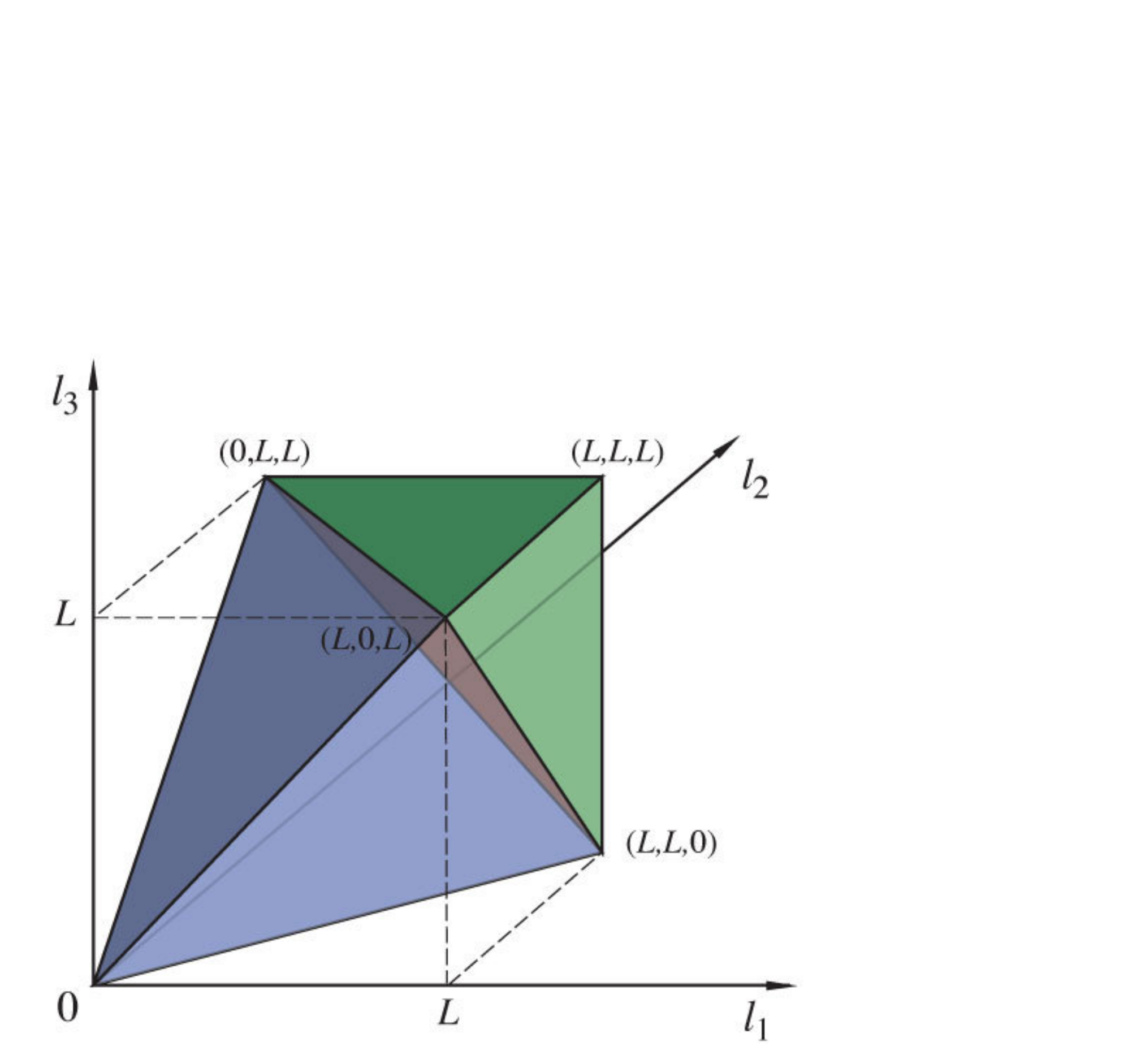}
\includegraphics[width=.45\linewidth]{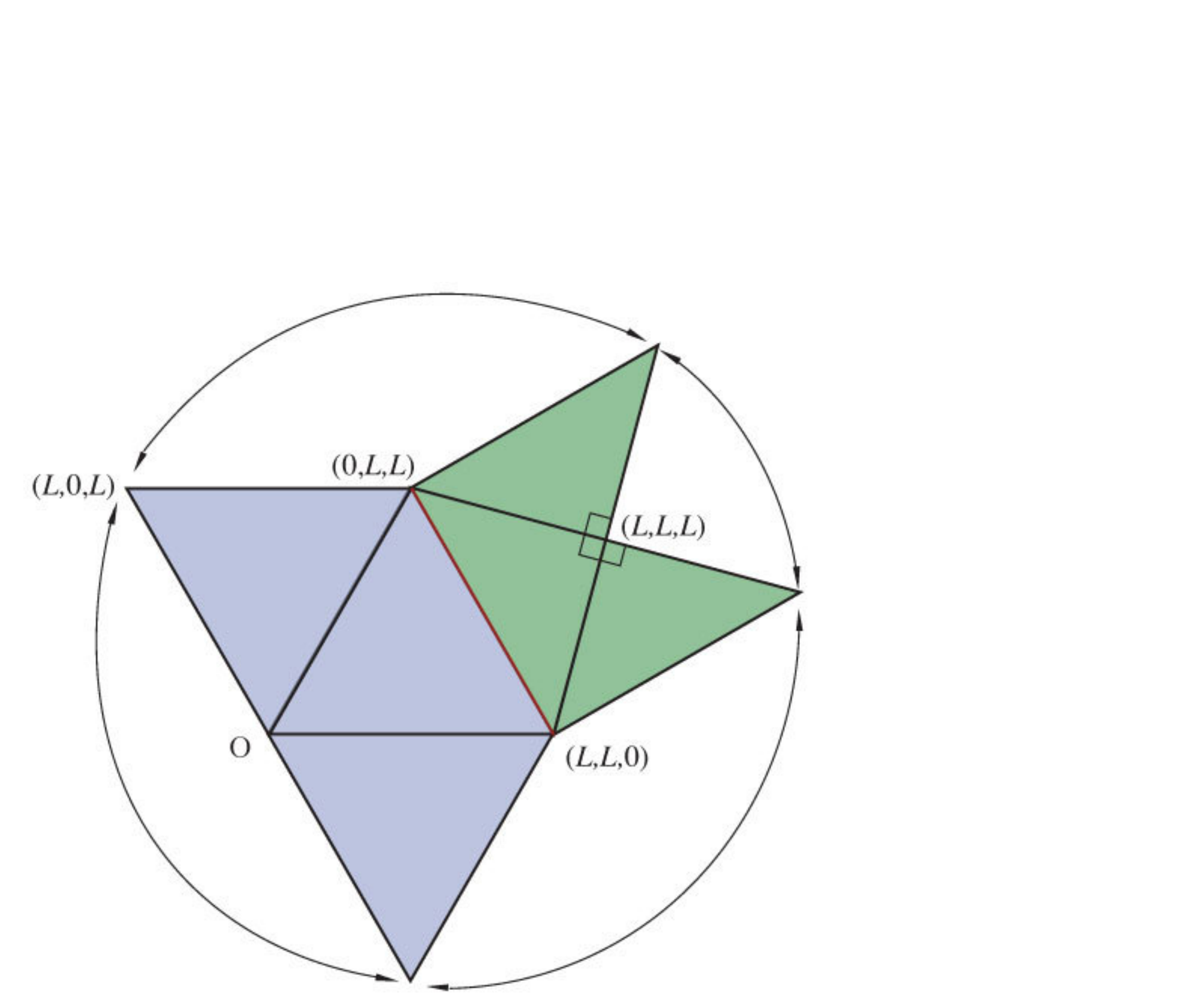}
\caption[Tetrahedral domain]{\small  Tetrahedral domain (`tetrapyd') for allowed multipole values $l$ for the CMB bispectrum $\blll$ or,
with wavenumbers $k$ for the primordial bispectrum $\Bkkk$).
The regular tetrahedral region defined up to the equilateral slice $\lsum \le 2\lmax\equiv 2L$ (shaded brown) contains two thirds of the overall volume.  The rest of the domain is given by the regular triangular pyramid on top which fills the volume to the corner of the encompassing cube defined by  $\lall \leq L$.  An origami tetrapyd is also shown (right) with folding instructions. } 
\label{fig:tetrapyd}
\end{figure}

In Fourier space, the primordial bispectrum $\Bkkk$ is defined when the three wavevectors
$\textbf{k}_1,\,\textbf{k}_2,\,\textbf{k}_3$ close to form a triangle 
$
\textbf{k}_1+\textbf{k}_2+\textbf{k}_3=0
$.
Since each such triangle
is uniquely defined by the lengths of its sides $k_1=|\textbf{k}_1|,\, k_2=|\textbf{k}_2|,\,k_3=|\textbf{k}_3|$,
we only require wavenumbers in the bispectrum argument.    In terms of these three
wavenumbers, the triangle condition restricts the allowed combinations into a 
tetrahedral region defined by 
\eq \label{eq:tetrapydk}
k_1 \leq k_2+k_3~\hbox{for}~k_1 \geq k_2,\,k_3,  ~~\hbox{or}~~ k_2 \leq k_1+k_3 ~\hbox{for}~k_2 \geq k_1,\,k_3,  ~~\hbox{or}~~ k_3 \leq k_1+k_2~\hbox{for}~
k_3 \geq k_1,\,k_2\,.
\qe
This region forms a regular tetrahedron if we impose the
restriction that $\ksum <2\kmax$, however, it is more natural to extend the domain out to values given 
by a maximum wavenumber in each direction $\kall \leq \kmax$.   This extension is motivated by issues both of separability 
and observation.   The allowed domain $\Vtetra$ is then a 
hexahedron formed by the intersection of a tetrahedron and a cube.   It can be obtained
from a regular tetrahedron (two-thirds of the total volume) by gluing on top
a regular triangular pyramid constructed from the corner of the cube (as illustrated in fig.~\ref{fig:tetrapyd}).   
For brevity, let us denote this asymmetric triangular bipyramid as a \textit{tetrapyd},  from the merger of a 
tetrahedron and a pyramid.  
Of course, bispectrum symmetries are such that it is only necessary to use one 
sixth of this domain, but aesthetics and intuition are helped by keeping the full domain while 
making a restriction to symmetrised functions. 

We will frequently need to integrate functions $f(k_1,k_2,k_3)$ over the tetrapyd domain (\ref{eq:tetrapydk}), which for brevity 
we will denote as $\Vtetra$ with the integration given explicitly by
\eq  \label{eq:tetrapydint}
{\cal T}[f] &\equiv& \int_\Vtetra f(k_1,k_2,k_3) \, w(k_1,k_2,k_3) \, d\Vtetra\\
 &=&K^3\left\{\textstyle{\int ^{1/2}_0 \int _y^{1-y}  \int ^{x+y}_{x-y} F\, W\, dz \,dx \,dy 
 \int ^{1/2}_0 \int _x^{1-x}  \int ^{x+y}_{y-x} F\,W\, dz \,dy \,dx +}\right.   \nonumber\\
 &&~~~~\left.+\textstyle{\int _{1/2}^1 \int _x^{1-x}  \int ^{1}_{x-y} F\, W\, dz \,dy \,dx +
 \int _{1/2}^1 \int _y^{1-y}  \int ^{1}_{y-x} F\,W\, dz \,dx \,dy} \right\}\,. \nonumber
 \qe
where $K = \kmax$,  $w(k_1,k_2,k_3)$ is an appropriate weight function, and we have made 
the transformation $x=k_1/K,\,y=k_2/K,\,x=k_3/K$
with $F(x,y,z) = f(Kx, Ky, Kz)$ and $W(x,y,z) = w(Kx, Ky, Kz)$.   For integrals over 
the product of two functions $f$ and $g$ we can define their inner product  $\langle f,\, g\rangle \equiv {\cal T}[fg]$, 
essentially defining a Hilbert space of possible shape functions in the domain (\ref{eq:tetrapydk}).  
The total volume of the tetrapyd domain is given by ${\cal T}[1] = K^3/2$. 
Initially, for the sake of simplicity, on the primordial wavenumber domain we will restrict attention to unit sidelength $K=1$ and weight
 $w =1$.

We note that it is important to incorporate a weight function for a variety of reasons. For example, the primordial shape function $\Skkk$ can be shown to possess a nearly linear scaling with respect to the CMB bispectrum 
estimator; on the multipole 
domain $w_{l_1 l_2 l_3}$ is non-trivial. A fairly close correspondence between the two can be obtained using $w(k_1,k_2,k_3) \approx 1/(k_1+k_2+k_3)$ \cite{Fergusson:2008ra} which explains its choice in the shape correlator (\ref{eq:shapecorrelator}).  The choice of weight function also affects mode expansion convergence and for certain shapes it may be convenient to eliminate dependencies by rescaling with a separable fuction. For the shapes we consider here however, this is not necessary.

\begin{figure}[t]
\centering
\includegraphics[width=.45\linewidth]{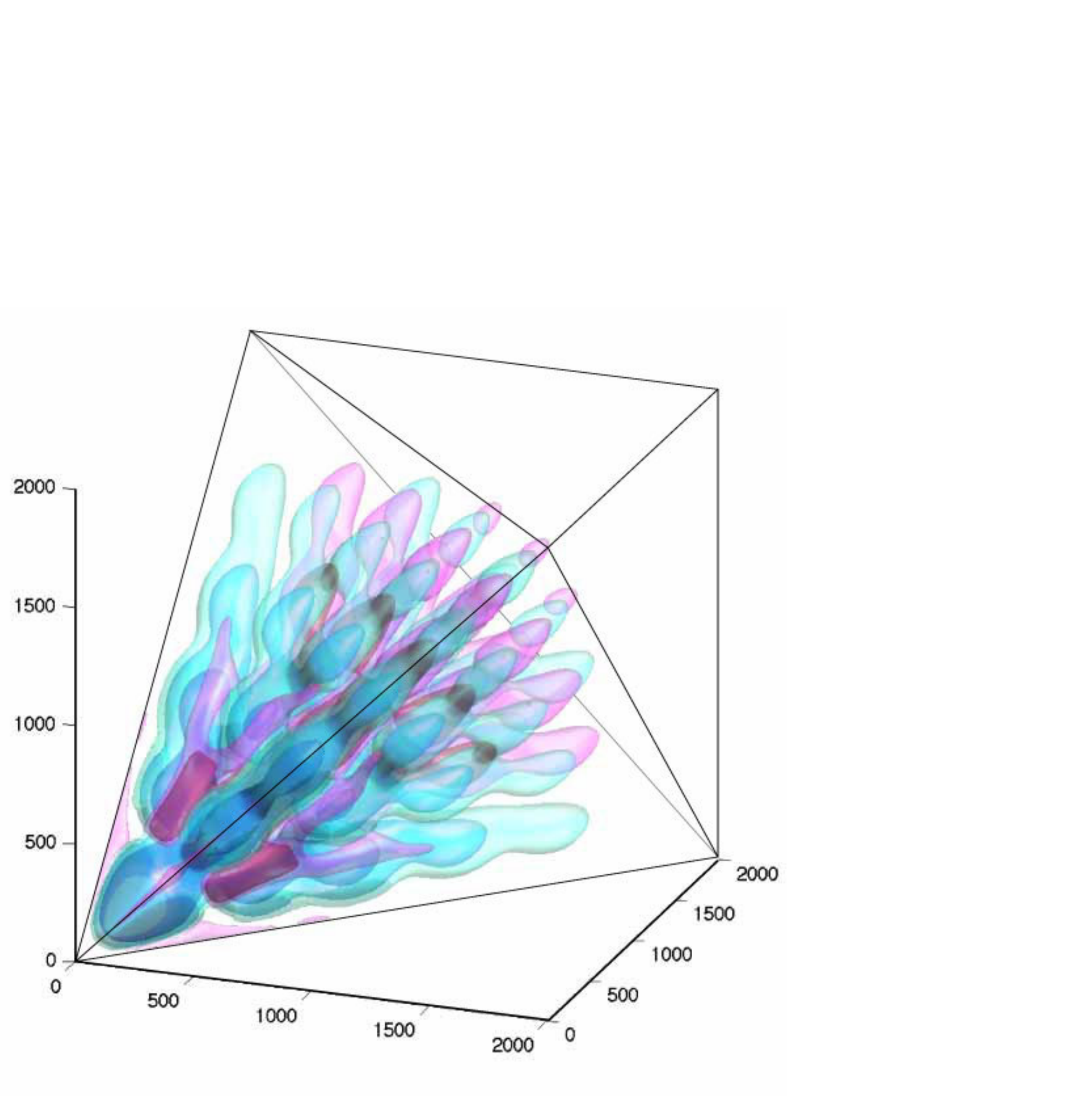}
\includegraphics[width=.45\linewidth]{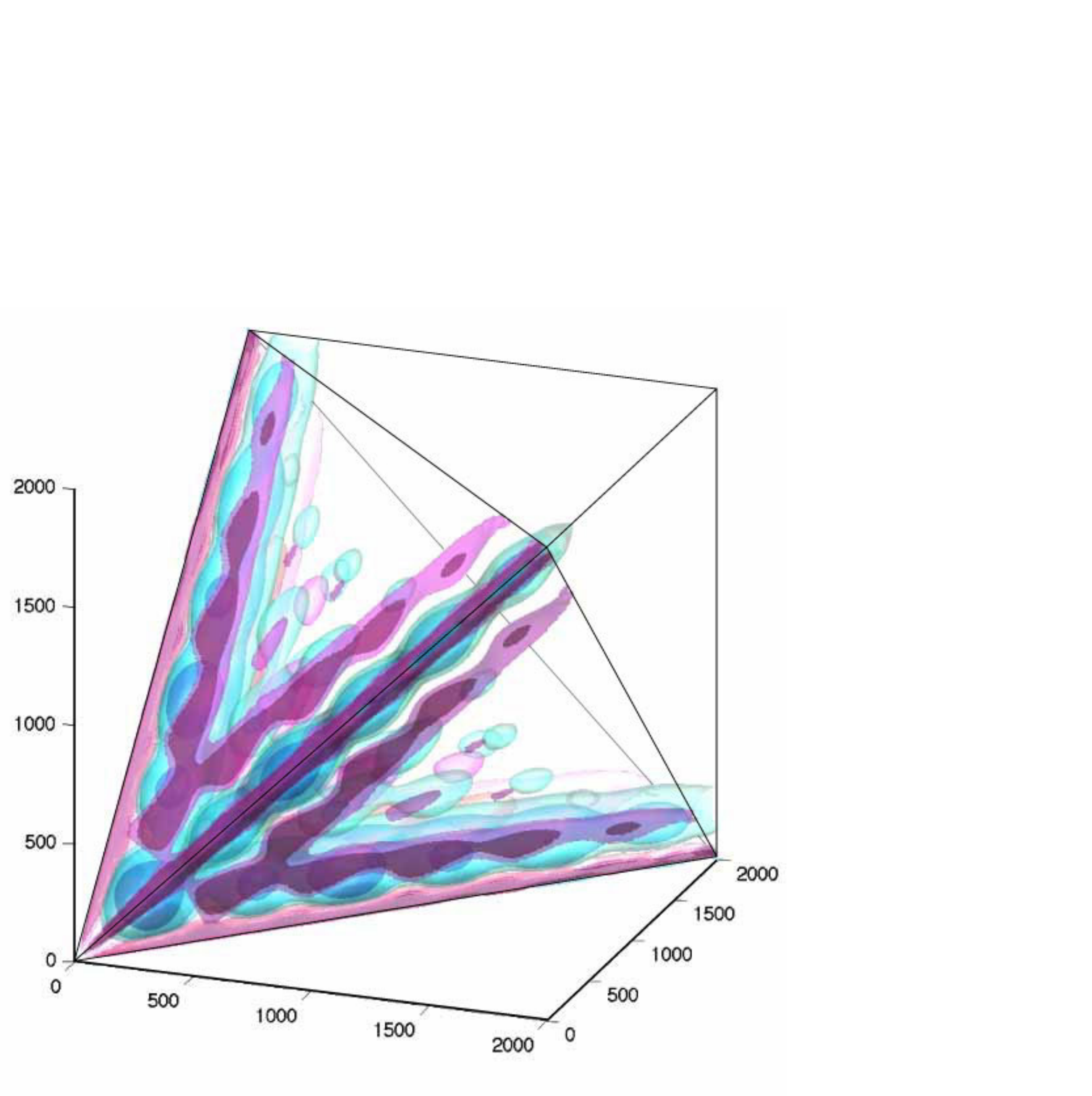}
\caption[Equilateral and local non-Gaussianity]{\small The reduced CMB bispectra for the equilateral 
model (left)
and the local model (right) plotted on the tetrahedral region shown in figure~\ref{eq:tetrapydl} (from 
\cite{Fergusson:2008ra}).  
Several density contours are illustrated (light blue positive and magenta negative) and $\blll$ is normalised 
by scaling relative to the constant Sachs-Wolfe solution (\ref{eq:constbispect}) $\blll^{\rm model}/\blll^{\rm const.}$.
Note the acoustic peaks induced by the transfer functions and the centre weighting for the equilateral model, contrasting with the corner-weighting for the local case \cite{Fergusson:2006pr}.
}
\label{fig:cmbbispectra}
\end{figure}

When analysing the CMB bispectrum it is particularly important to extend the tetrahedral domain to 
include multipoles in the top pyramidal region shown in fig.~\ref{fig:tetrapyd}.   In principle, this pyramid 
contains 33\% of the triple $l_1l_2l_3$ combinations available in the observational data, e.g.\ with Planck 
out to $\lall \leq 2000$.   The tetrapyd domain for the reduced bispectrum $\blll$ becomes the discrete $\{\lall\}$ combinations satisfying
\eq \label{eq:tetrapydl}
\nonumber && \lall \leq \lmax\,,\quad \lall \in  \mathbb{N}\,,\\
&&l_1 \leq l_2+l_3  ~~\hbox{for}~~ l_1 \geq l_2,\,l_3,  ~~  +~\hbox{cyclic}~\hbox{perms.}\,,\\
\nonumber &&\lsum = 2n\, ,~~~n\in\mathbb{N}\,.
\qe
In fig.~\ref{fig:cmbbispectra} we illustrate contrasting bispectra on this domain for the equilateral and local models (here with $\lmax=2000$).

In multipole space, we will be primarily dealing with a summation over all possible $\{\lall\}$ combinations in 
the estimator (\ref{eq:estimator}) or the 
closely related correlator (\ref{eq:cmbcor}).  The appropriate weight function in the sum is then 
\eq\label{eq:lweightdiscrete}
w_{l_1 l_2 l_3} ={\textstyle \frac{1}{4\pi}}\,(2l_1+1)(2l_2+1)(2l_3+1) \( \begin{array}{ccc} l_1 & l_2 & l_3 \\ 0 & 0 & 0 \end{array} \)^2\,,
\qe
where we note that the third condition in (\ref{eq:tetrapydl}) arises as a selection rule from the Wigner-3$j$ symbol.   
Despite the discrete origin of the function $w_{l_1 l_2 l_3}$, 
like the reduced bispectrum $b_{l_1 l_2 l_3}$,  it  varies smoothly.  It is particularly uniform on cross-sectional slices
$\lsum = 2L$, except for a finite rise very close to the boundaries.
While the Wigner-3$j$ symbols are easily calculable (especially in the $m_i=0$ case when performed
in advance for a look-up table), it is more convenient to work in the continuum limit $w_{l_1 l_2 l_3} \kern-4pt\rightarrow w(\lall)$ 
when considering 
domains with large $\lmax$.   To achieve this we take the exact expression in terms of factorials (for even combinations
with $\lsum = 2l\,, ~l\in  \mathbb{N}$),
\eq
 \( \begin{array}{ccc} l_1 & l_2 & l_3 \\ 0 & 0 & 0 \end{array} \) = (-1)^l \sqrt{ \frac {(2l-2l_1)!\,(2l-2l_2)!\, (2l-2l_3)!}
 {(2l +1)!} }\frac {l!}{(l-l_1)!\,(l-l_2)!\,(l-l_3)!}\,,
\qe 
and then we substitute  the Gosper approximation for all these factorials, that is, 
\eq
l! \approx \sqrt{(2l + {\textstyle \frac{1}{3}})\pi}\, \,l^l \,e^{-l}\,.
\qe
The discrete multipole weight function  (\ref{eq:lweightdiscrete}) then reduces to a straightforward continuum 
version
\eq\label{eq:lweight}
w(\lall) =  \frac{1}{2\pi^2} \frac{(2l_1+1)(2l_2+1)(2l_3+1)(2l+{\textstyle \frac{1}{3}})}{(2l-2l_1+{\textstyle \frac{1}{3}})  (2l-2l_2+{\textstyle \frac{1}{3}})  (2l-2l_3+{\textstyle \frac{1}{3}})  }\sqrt{\frac{ (2l-2l_1+{\textstyle \frac{1}{6}})  (2l-2l_2+{\textstyle \frac{1}{6}})  (2l-2l_3+{\textstyle \frac{1}{6}})         }{(2l +{\textstyle \frac{1}
{6}})}}\,.
\qe
This is a remarkably accurate representation for the exact discrete $w_{l_1 l_2 l_3}$ with the difference between the weight functions being less than 0.01\% (0.1\%)  for about 95\% (99\%) of the allowed triples $l_1l_2l_3$ on the domain (\ref{eq:tetrapydl}) with 
$2\le\lall\le2000$. The worst approximation by $w(\lall)$ never 
differs by more than 2.5\% and such points are exclusively located very near the boundaries, leaving an overall integrated error over the entire domain (\ref{eq:tetrapydl}) of less than 0.01\%.  Nevertheless, care must be exercised using this approximation for edge- or corner-weighted models.  With this caveat in mind, we can define the multipole sum 
equivalent to the wavenumber tetrapyd integration (\ref{eq:tetrapydint}) as
\eq\label{eq:tetrapydintk}
{\cal T}[f] = \sum_{\{ l_1l_2l_3\}\in \Vtetra} {\kern-10pt w_{l_1l_2l_3}\, f_{l_1l_2l_3}} \;=\; {\textstyle \frac{1}{2}}\int _\Vtetra w(\lall)\,f(\lall) \,d\Vtetra\,,
\qe
with the inner product again defined by $\langle f,\,g\rangle= {\cal T}[fg]$.  It will be clear from the context whether we are dealing 
with multipole or wavenumber integrations. 

\begin{figure}[t]
\centering
\includegraphics[width=.45\linewidth]{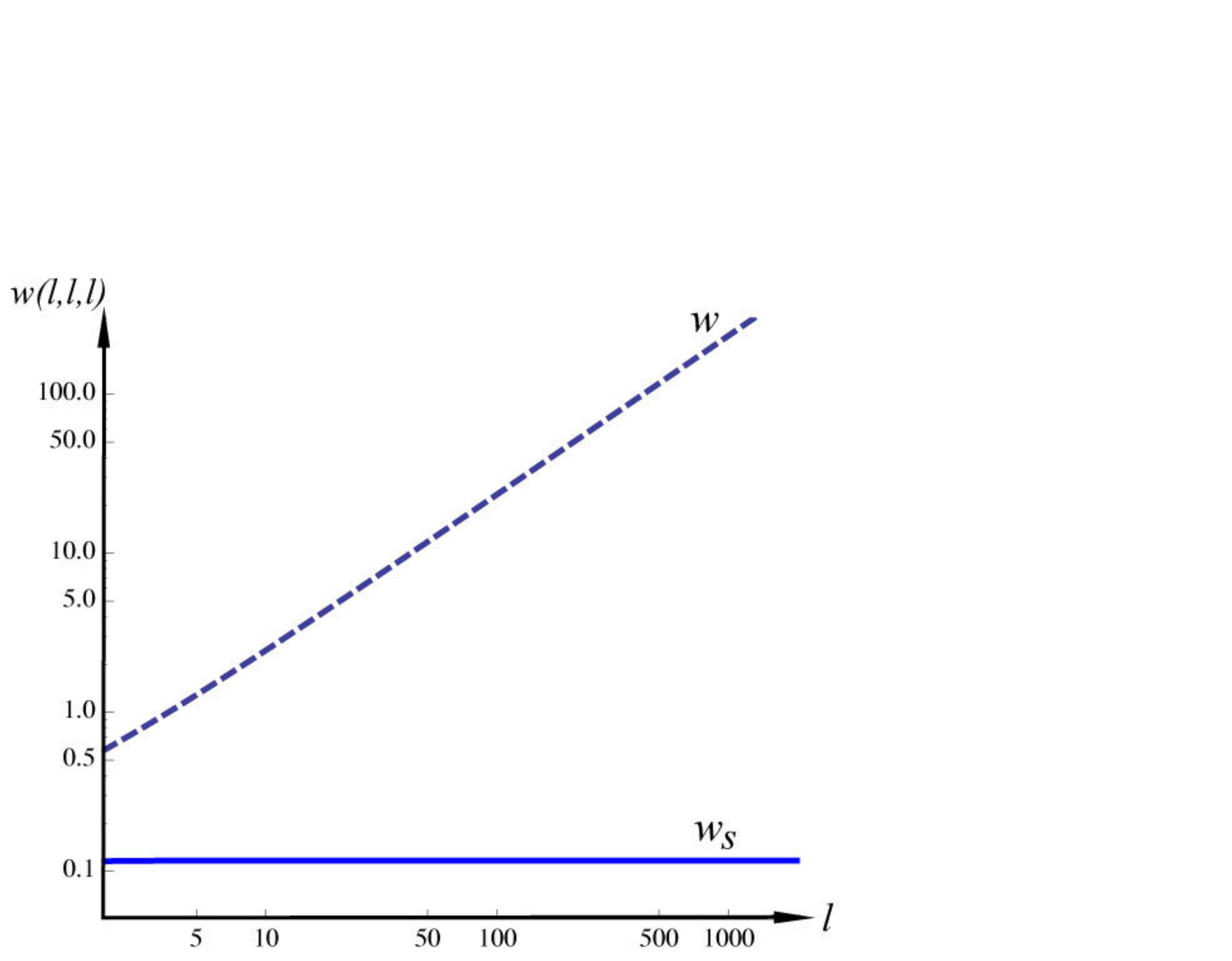}
\includegraphics[width=.40\linewidth]{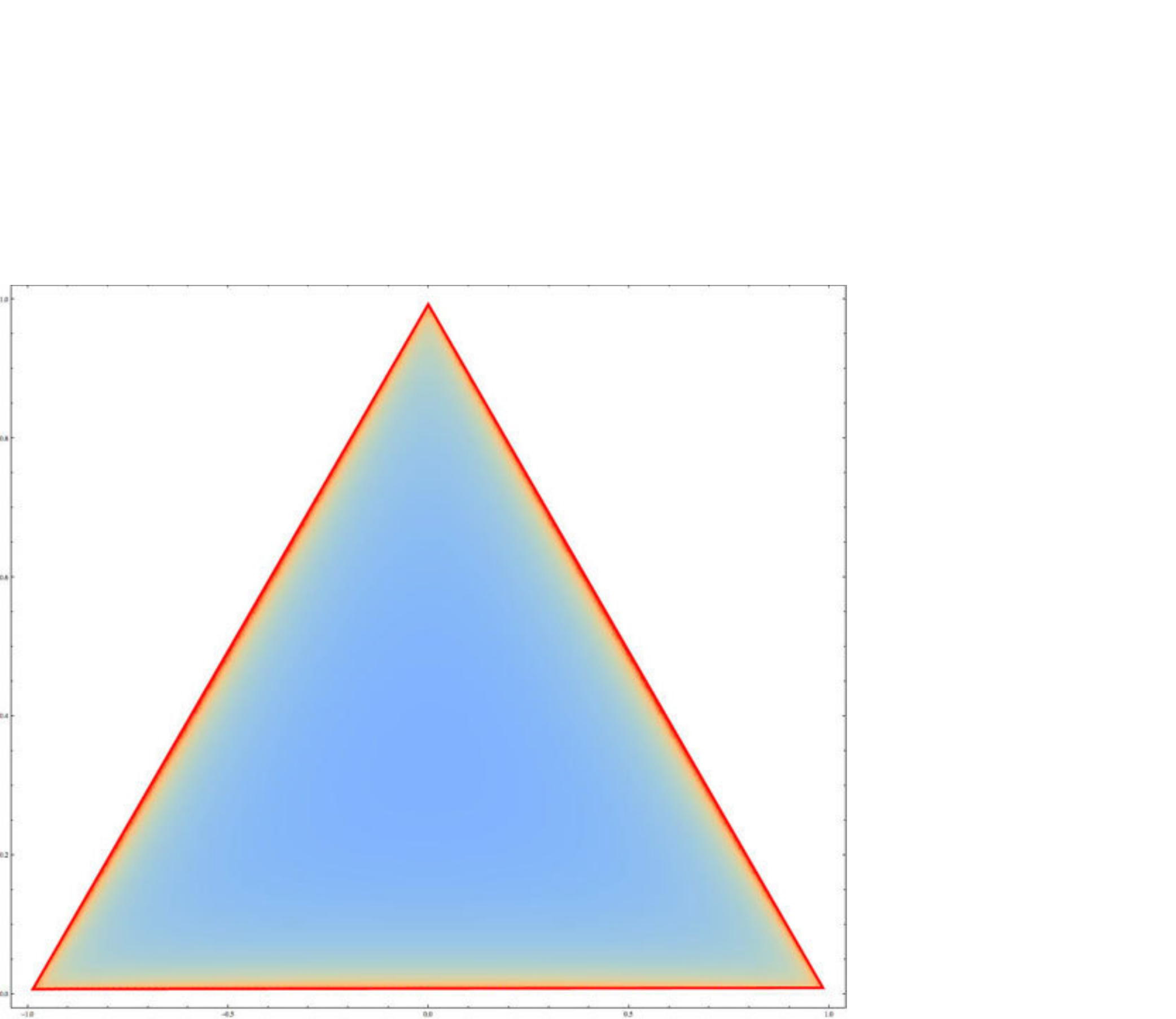}\hskip12pt
\includegraphics[height=6.0cm]{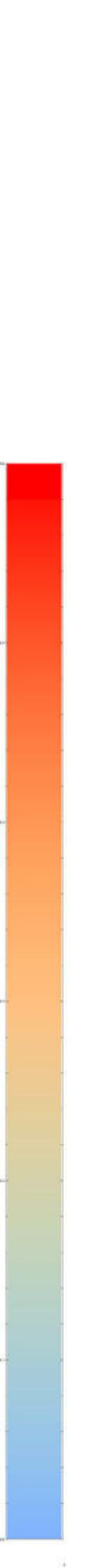}
\caption[Multipole weight functions]{\small Scaling comparison of the multipole domain weight function $w(\lall)$ (or $w_{l_1 l_2 l_3}$) 
given in (\ref{eq:lweight}) and the modified weight function $w_s(\lall)$ given in (\ref{eq:lweightsep}), which is rescaled by a separable function.  On the left, the equal-$l$ values are shown with the linear scaling of $w$ (dashed) contrasting with the constant $w_s$ (solid).  On the right, a density plot of $w_s$ is shown on the $\lsum = 2L$ slice with $L=2000$.  Note the
uniformity $w_s\approx {\rm const.}$, except very close to the edges where there is about 
a factor of 4 rise to the maximum value on the perimeter. 
}
\label{fig:lweightfns}
\end{figure}

The weight function  $w(\lall)$    (or $w_{l_1 l_2 l_3}$)  in (\ref{eq:lweight}) possesses an overall scaling which grows
linearly with $l$, as illustrated in 
fig.~\ref{fig:lweightfns}.  It can be convenient to eliminate this scaling, so that the weight function 
becomes very nearly constant.   We can achieve this by dividing $w(\lall)$ by a separable function as 
\eq  \label{eq:lweightsep}
w_s(\lall) = \frac{w(\lall)}{(2l_1+1)^{1/3}(2l_2+1)^{1/3}(2l_3+1)^{1/3}}\,.
\qe
The result is shown in fig.~\ref{fig:lweightfns} where it is evident that $w_s\approx {\rm const.}$ everywhere except 
on the boundaries.   For uniform or centre-weighted bispectrum models, such as the equilateral model, the multipole 
domain with weight $w_s(\lall)$ becomes essentially identical to that for the primordial wavenumbers (\ref{eq:tetrapydk}) 
with $w(\kall)=1$, so that it is a good approximation to proceed with the same polynomial expansions.

Finally, we comment on the freedom to absorb an arbitrary separable function $v_l$ into the weight functions 
$w(\kall)$ or $w_{l_1 l_2 l_3}$, such as in the example (\ref{eq:lweightsep}) above.  If we define a new weight $\bar w$
in the  estimator as
\eq\label{eq:lweightgeneral}
\bar w_{l_1 l_2 l_3} = w_{l_1 l_2 l_3}/\left(v_{l_1}v_{l_2}v_{l_2}\right)^2\,,
\qe
then we must similarly rescale the estimator functions as $\overline {(\blll /\Delta)}= v_{l_1}v_{l_2}v_{l_2}{(\blll /\Delta)}$.
This rescaling should be separable, otherwise it would compromise the separability of the methods we outline
here, undermining their efficiency.  As we have seen it can prove convenient to make the weight functions 
scale-invariant for practical purposes,  thus facilitating better convergence of mode expansions for
typical bispectra.   However, in principle, we can also exploit this separability in order to remove pathologies 
from singular shapes, such as the local model, using a mode expansion to describe the more regular deviations 
away from these shapes.    The important point is to consistently use both the new weight $\bar w$ and the estimator rescaling throughout the analysis pipeline, 
including the generation of appropriate orthonormal mode functions.

\subsection{Orthogonal polynomials on a tetrahedral domain}

We next construct some concrete realizations of mode functions which are orthogonal on the tetrahedral domain $\Vtetra$
and which have the form required for a separable expansion (\ref{eq:separablebasis}).    First, we will generate one-dimensional orthogonal polynomials $q_p(x)$ for unit weight $w=1$, before discussing their promotion to three-dimensions
and alternative weights.   These tetrahedral polynomials are analogues of the more 
familiar Legendre polynomials  $P_n(x)$ on the unit interval.  Considering functions $q_p(x)$ depending only on 
the $x$-coordinate, we integrate over the $y$- and $z$-directions to yield the reduced weight function $\tilde w(x)$ for $x \in [0,1]$ (we take $K=1$):
\eq\label{eq:tetrapyd1dweight}
\tilde w(x) ={\textstyle{\frac{1}{2}}}x(4  - 3x)\,, \quad \hbox{with} \quad {\cal T}[f] = \int _0^1 f(x)\,\tilde w(x)\,dx\,.
\qe
This simplifies our domain integration (\ref{eq:tetrapydint}) for functions of only $x$, and
 the moments for each power of $x$  become simply
\eq
w_n \equiv {\cal T}[x^n] = \frac{n + 6}{2(n + 3)(n + 2)}\,.
\qe
From these we can create  orthogonal polynomials  using the generating function,
\begin{align} \label{eq:Qgenerator}
q_n (x) = \frac{1}{{\cal N}} \left| \begin{array}{ccccc}
\hbox{\small1/2} & \hbox{\small 7/24} & \hbox{\small1/5} & ... &w_n \\
\hbox{\small7/24} &\hbox{\small1/5} &\hbox{\small3/20} & ... & w_{n+1} \\
... & ... & ... & ... & ... \\
w_{n-1} & w_n &w_{n+1} & ... & w_{2n-1} \\
1& x & x^2 & ... & x^n \end{array} \right|\,,
\end{align}
where we choose the normalisation factor ${\cal N}$ such that ${\cal T}[q_n]=1$ for all $n\in \mathbb{N}$,
that is,
so that the $q_n(x)$ are orthonormal 
\eq
\langle q_n,\, q_p\rangle \equiv{\cal T} [q_n q_p ] = \int_\Vtetra q_n(x) \,q_p(x) \, d\Vtetra = \delta_{np}\,.
\qe
The first few orthonormal polynomials on the tetrahedral domain (\ref{eq:tetrapydk}) are explicitly  
\begin{eqnarray}  \label{eq:Qpolys}
q_0(x) &=&\sqrt{2}\,,\nonumber\\
q_1(x) &=& 5.787 \,(-{\textstyle\frac{7}{12}} +x)\,,\nonumber\\
q_2(x) &=& 23.32 \left({\textstyle\frac{54}{215}} -{\textstyle\frac{48}{43}}x+ x^2\right)\,,\\
q_3(x) &=& 93.83 \left(-0.09337 + 0.7642 \,x - 1.631\,x^2 + x^3\right)\,,\nonumber\\
q_4(x) &=& 376.9 \left(0.03192 - 0.4126\,x + 1.531\,x^2 - 2.139\,x^3 + x^4\right)\,,\nonumber\\
q_5(x) &=& 1512 \left(-0.01033 + 0.1929\,x  - 1.084\,x ^2 + 2.549\,x ^3 - 2.644\,x ^4 + 
 x^5\right)\,,~ ...\nonumber
\end{eqnarray}
These can be obtained easily from the generating determinant (\ref{eq:Qgenerator}) in Mathematica or similar applications.  

\begin{figure}[t]
\centering
\includegraphics[width=.65\linewidth]{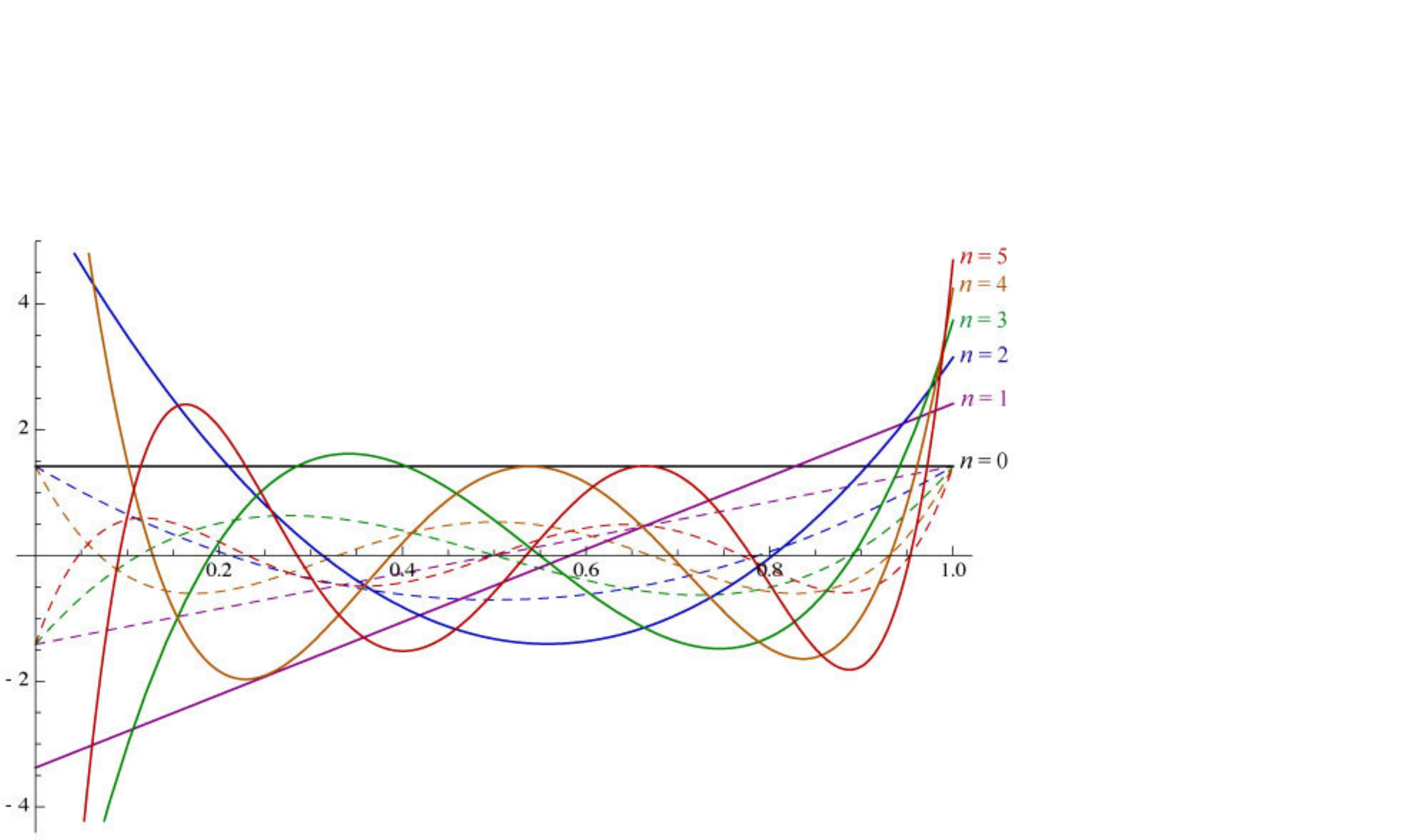}
\caption[Eigenfunctions in 1D on a tetrahedral domain]{\small The orthonormal one-dimensional tetrahedral
$q_n(x)$ plotted on the unit interval for $n=0$--$5$.     The behaviour is smooth and bounded across the domain even for high $n$, 
except where the weight function $w(x)$ vanishes at $x=0$. 
Also plotted for comparison are the rescaled  Legendre polynomials $P_n(2x-1)$ (dashed lines).  Despite $q_n$ and 
$P_n$ sharing qualitative features such as  $n$ nodal points, their properties and orthogonality on $\Vtetra$ are very different. 
}
\label{fig:Qpolynomials}
\end{figure}

\begin{figure}[t]
\centering
\includegraphics[width=.65\linewidth]{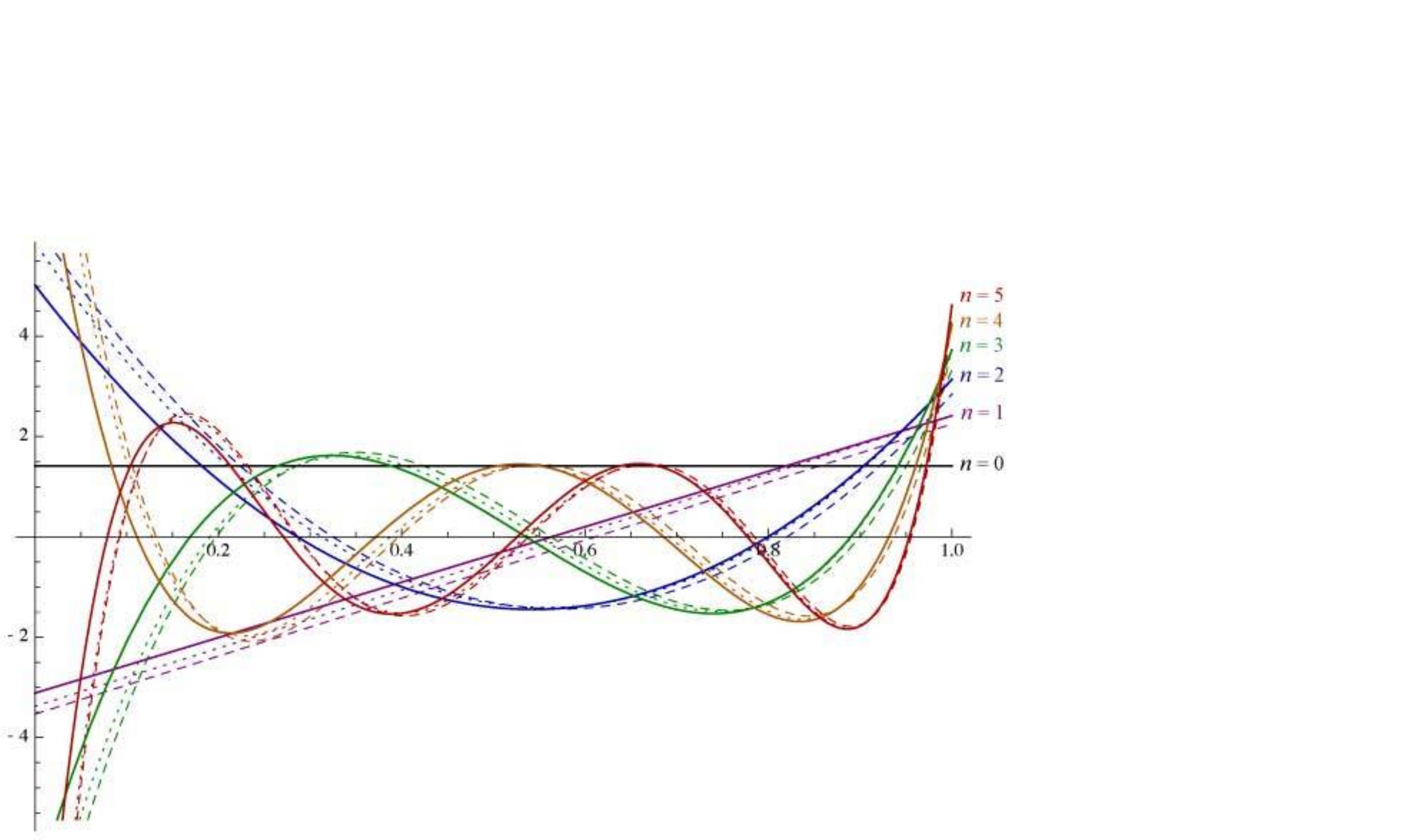}
\caption[Multipole eigenfunctions in 1D on a tetrahedral domain]{\small Orthonormal polynomials
$\bar q_n(x)$ for the multipole domain (\ref{eq:tetrapydl}) with weight functions $w$ given in (\ref{eq:lweight}) [solid] and
$w_s$ given in  (\ref{eq:lweightsep}) [dashed], as well as the previous $q_n(x)$ for unit weight [dashed] (shown already in 
fig.~\ref{fig:Qpolynomials}).   Despite
the different scaling of $w$, these tetrahedral polynomials are very similar,  particularly the latter two with flattened weight functions. }
\label{fig:Qcomparison}
\end{figure}

We note that the $q_n$'s are only orthogonal in one dimension
(e.g.\ we have ${\cal T}[q_n(x) \,q_p(y)] \ne \delta_{np}$ in general).  However, 
as product functions of $x$, $y$ and $z$ they form an independent and well-behaved basis 
which we will use to construct orthonormal three-dimensional eigenfunctions.  
In  practice, these $q_n$'s will remain the primary calculation tools throughout, notably 
when performing separable integrations.   Where they differ from the separable functions 
used to represent bispectra in the literature, they generally have a number of distinct advantages,
as we shall detail at the end of this section.   Finally, we point out that for a regular tetrahedron 
(in contrast to the tetrapyd domain  (\ref{eq:tetrapydk})), 
the volume weight function is $\tilde w(x)= 2x(1-x)$ and so the behaviour is different at $x=1$ where the weight vanishes,
unlike (\ref{eq:tetrapyd1dweight}). The first orthonormal polynomials in this case are 
$q_0(x) = \sqrt{3},\; q_1(x) = 0.387(2y-1), \;  q_2 = 32.4 (y-0.724 ) ( y-0.276)\,, ...\,$.

ow let us turn to the polynomials $\bar q(x)$ which are orthonormal on the multipole domain (\ref{eq:tetrapydl}),
using the weight functions $w$ given in (\ref{eq:lweight}) and $w_s$ given in  (\ref{eq:lweightsep}).   For definiteness
we take $L\equiv\lmax=2000$, so that $ x=l_1/L$, $y = l_2/L$ and $z=l_3/L$.   The generating function 
(\ref{eq:Qgenerator}) can be obtained
as above but now using the moments $w_n \equiv {\cal T}[x^n] = \int w(x,y,z) \,x^n d\Vtetra$ (or by undertaking 
Gram-Schmidt orthogonalisation from $\bar q_0 = {\rm const.}$).   The resulting 
first few polynomials for the multipole domain are then  
\begin{eqnarray}  \label{eq:Qlpolys}
\bar q_0(x) &=&0.07378\,,\nonumber\\
\bar q_1(x) &=& 0.3017 \,(-0.6110 + x)\,,\nonumber\\
\bar q_2(x) &=& 1.223 \left(0.2665 - 1.145 \,x + x^2\right)\,,\\
\bar q_3(x) &=& 4.933 \left(-0.1000+ 0.7951\,x  - 1.659\,x^2 + x^3\right)\,,\nonumber\\
\bar q_4(x) &=&19.85 \left(0.0345- 0.4342\,x  + 1.578\,x ^2 - 2.169\,x ^3 + x^4\right)\,,\nonumber\\
\bar q_5(x) &=& 79.55 \left(-0.0106 + 0.1975\,x  - 1.103\,x ^2 + 2.576\,x ^3 - 2.657\,x ^4 + 
 x^5\right)\,,~ ...\nonumber
\end{eqnarray}
A cursory comparison with $q_n$ given above for the flat wavenumber domain will show that 
these polynomials are very similar for low $n$, despite the linear scaling behaviour of $w$.   However, 
if we remove this scaling as in the flatter weight $w_s$ in (\ref{eq:lweightsep}s),  the polynomials
become near identical as illustrated in fig.~\ref{fig:Qcomparison}.   It is clear that each of these polynomial sets would 
suffice as independent basis functions on the multipole domain. However, using the correctly weighted
versions leads to improvements in the immediate orthogonality of the three-dimensional polynomials 
we shall construct in the following discussion.

\subsection{Bispectrum symmetries and three-dimensional basis functions}

 We can represent arbitrary bispectra on the tetrahedral domain (\ref{eq:tetrapydk}) 
using a suitable set of independent basis functions  formed from products $q_p(x)\,q_r(y)\,q_s(z)$ 
of the orthogonal polynomials (\ref{eq:Qpolys}) (or with different weight functions, such as (\ref{eq:Qlpolys}).    
(Here, we again take $x = k_1/\kmax$, 
$y = k_2/\kmax$, $z = k_3/\kmax$ or $x=\l_1/\lmax$, etc.)
Both primordial bispectra $\Bkkk$ and CMB bispectra $\blll$ on (\ref{eq:tetrapydk}) possess 
six symmetries made from combinations of discrete $\pi/3$ rotations around the line $x=y=z$ and/or 
reflections which interchange the axes.  We can impose these six symmetries on our products 
by summing the relevant permutations  and defining the 3D basis function 
\begin{align}\label{eq:Qdefn}
\curl{Q}_n(x,y,z) &={\textstyle{ \frac{1}{6{\cal N}}}}\left [q_p(x) q_r(y) q_s(z) + q_r(x) q_s(y) q_p(z) +q_s(x) q_p(y) q_r(z)\right. \nonumber\\
&\left.~~~~~~~ + ~q_p(x) q_s(y) q_r(z) + q_s(x) q_r(y) q_p(z) +q_r(x) q_p(y) q_s(z)\right ]\nn\\
&\equiv q_{\{p}\,q_r\,q_{s\}}  ~~~\hbox{with}~~ n \leftrightarrow \{prs\}\,,
\end{align}
where we use the notation $\{prs\}$ to denote the six permutations of $prs$.  Here, for convenience, we have specified
 a one-to-one mapping 
 $n \leftrightarrow \{prs\}$ ordering the permuted indices into a list labelled by $n$ (see below).
Alternatively, we could directly represent bispectra in a power series using sums of 
monomial symmetric polynomials which like (\ref{eq:Qdefn}) are also separable; that is, we could identify our 
set of  basis functions with the following 
\begin{align}\label{eq:monsympoly}
1,\,~~x+y+z\,,~~ xy+yz+zx\,,~~x^2+y^2+z^2,~~ xyz\,,~~ x^3 +y^3 + z^3, ~~etc.
\end{align}
The $\Qnxyz$ we defined in (\ref{eq:Qdefn}) are themselves ultimately constructed from these through the $q_p$ products.  
However, the $\Qn$ have two distinct advantages which are, first, they already have partial orthogonality 
built in which improves their convenience and convergence and, secondly, unlike the elements of (\ref{eq:monsympoly}), the $q_p$ polynomials 
remain bounded and well-behaved when convolved with transfer functions, as we shall emphasise in the map-making
discussion.  

Since we will be dealing with relatively small numbers of basis functions, it is convenient to order the symmetric products $\Qn = q_{\{p}\,q_r\,q_{s\}} $ linearly with a single index $n$;  here we offer two comparable alternatives for achieving this.   The first is by `slicing' such that 
triples are ordered by the sum $p+r+s$ and the second is by `distance' from 
the origin, that is,  $p^2+r^2+s^2$.   

Slicing the $prs$ naturally groups the  $\Qn$ by the overall order of the polynomials 
from which they are made.   The subscript $n$, with a specific choice of sub-ordering, relates to 
the $prs$ via
\begin{align}\label{eq:slicingorder}
\nn &\underline{0  \rightarrow 000}   \qquad \nn 4  \rightarrow 111  \qquad~\,  \nn 8  \rightarrow 022\qquad  \nn 12  \rightarrow 113 \\
\nn  & \underline{1\rightarrow 001}   \qquad  \nn 5  \rightarrow 012 \qquad~\,  \nn 9  \rightarrow 013  \qquad \nn 13  \rightarrow 023 \\
& 2 \rightarrow 011  \qquad \underline{6  \rightarrow 003}\qquad \underline{10  \rightarrow 004}  \qquad  14  \rightarrow 014 \\   \nn
 &\underline{ 3 \rightarrow 002}  \qquad  7  \rightarrow 112 \qquad   11  \rightarrow 122 \qquad   \underline{15  \rightarrow 005} ~\cdots\,,
\end{align}
where we have underlined the transitions between polynomial order.
The number $d_N$ of independent symmetric polynomial products $Q_nQ_pQ_r$ which can be formed at each polynomial order $N$ is a combinatorial 
problem but the sequence begins as follows and we give a recurrence relation for any further elements:
\eq
\{d_N\} = \left\{1, 1, 2, 3, 4, 5, 7, 8, 10, 12, ... \right\},\qquad
d_N=1+d_{N-2}+d_{N-3}-d_{N-5}\,.
\qe
For consistency when using slicing we will usually decompose functions with polynomials up to a specific order $N$.

The distance ordering of the $\Qn$ is more straightforward with 
\begin{align}\label{eq:distanceorder}
\nn &{0  \rightarrow 000}   \qquad \nn 2 \rightarrow 011  \qquad~\,  \nn4 \rightarrow 002\qquad  \nn 6  \rightarrow 112 \qquad  8  \rightarrow 122\\
 & {1\rightarrow 001}   \qquad   3  \rightarrow 111 \qquad~\, 5  \rightarrow 012  \qquad  7  \rightarrow 022 \qquad 9  \rightarrow 003~\cdots\,.
\end{align}
  This approach is the analogue of state counting over spherical 
shells in the continuum limit and the basis functions can be grouped accordingly.   
Distance ordering has some advantage by reshuffling to higher $n$ the pure states 
$00p$ which turn out to be most affected by masking.

\begin{figure}[t]
\centering
\includegraphics[width=.425\linewidth]{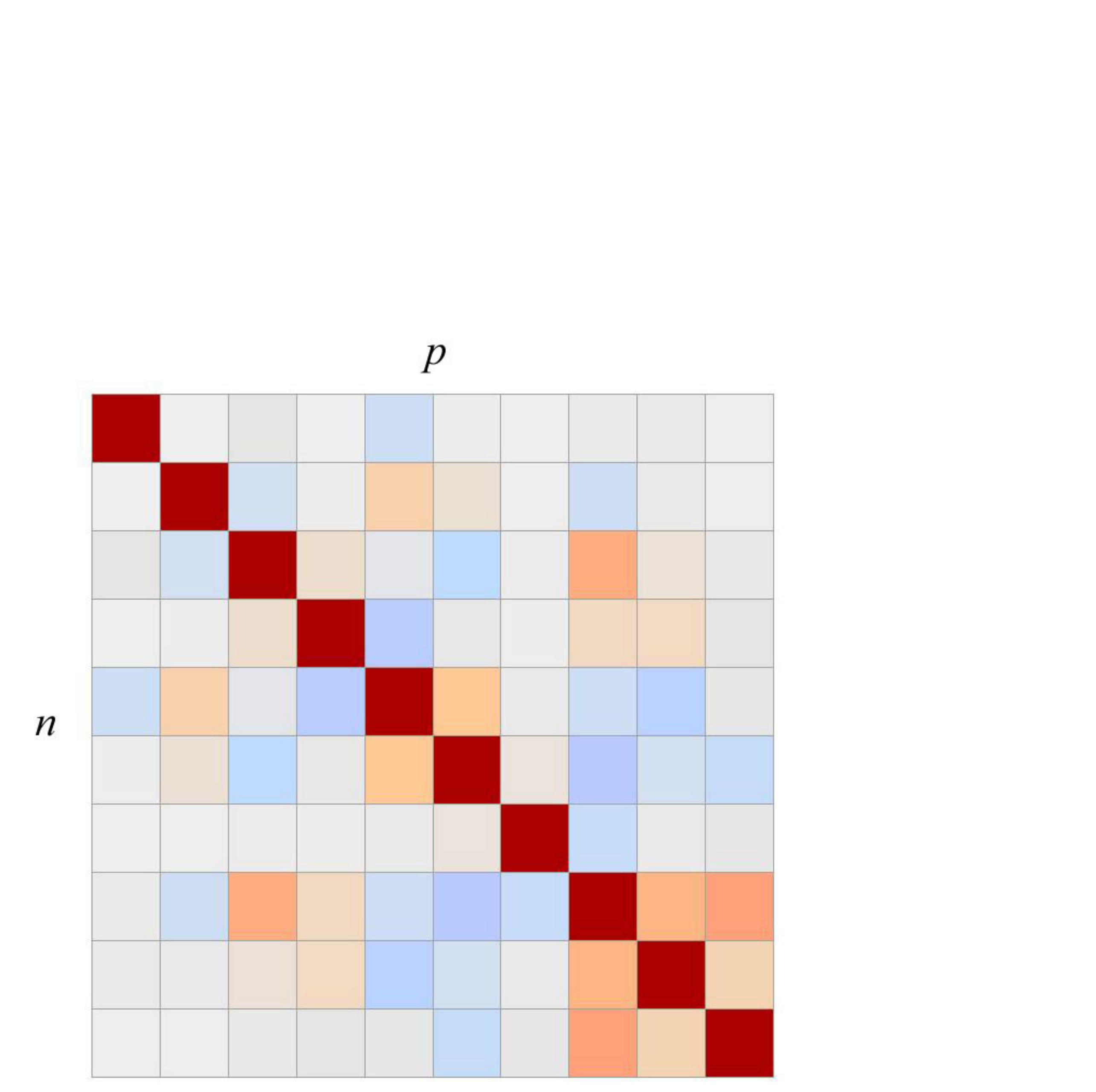}
\includegraphics[width=.425\linewidth]{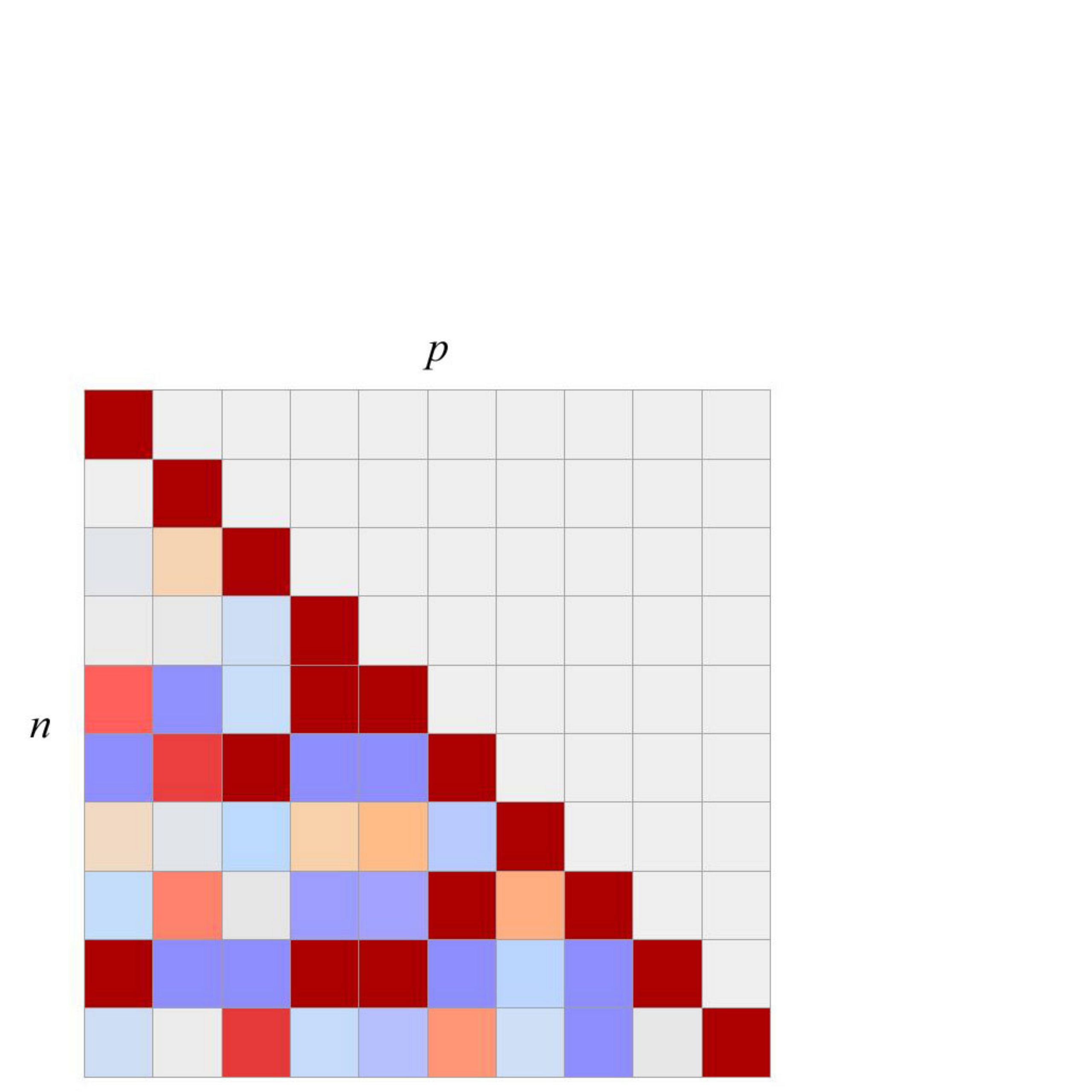}
\hskip8pt
\includegraphics[height = 6.75cm]{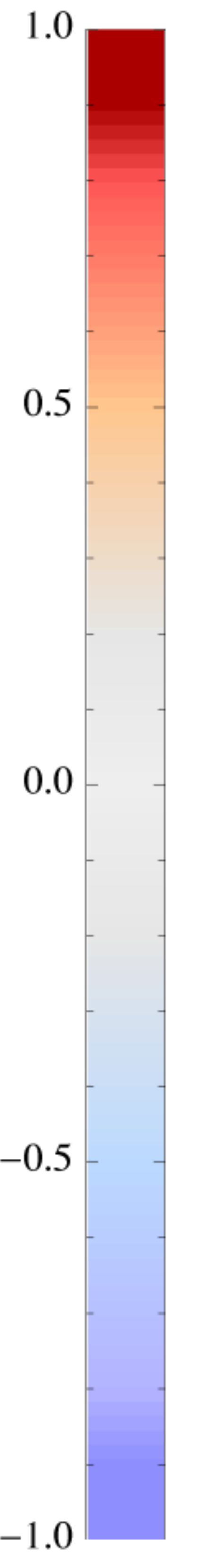}
\caption[QR relations]{\small  Partial orthogonality of the symmetric product polynomials $\Qn$  illustrated through 
the inner product matrix $\langle\Qn,\,\Qp\rangle$ for $0\le n,p < 10$ (left panel).  Lower triangular matrix $\lambda_{np}$ 
in (\ref{eq:RQinverse}) illustrating the decomposition of the orthonormal $\Rn$ into the  $\Qp$ arising through the Gram-Schmidt 
process (right panel); this is the inverse of $\langle\Qn,\,\Rp\rangle$.    To improve comparison, the 
$\Qn$'s have been unit normalised.   
}
\label{fig:QRrelation}
\end{figure}

\begin{figure}[p]
\centering
\includegraphics[width=.45\linewidth]{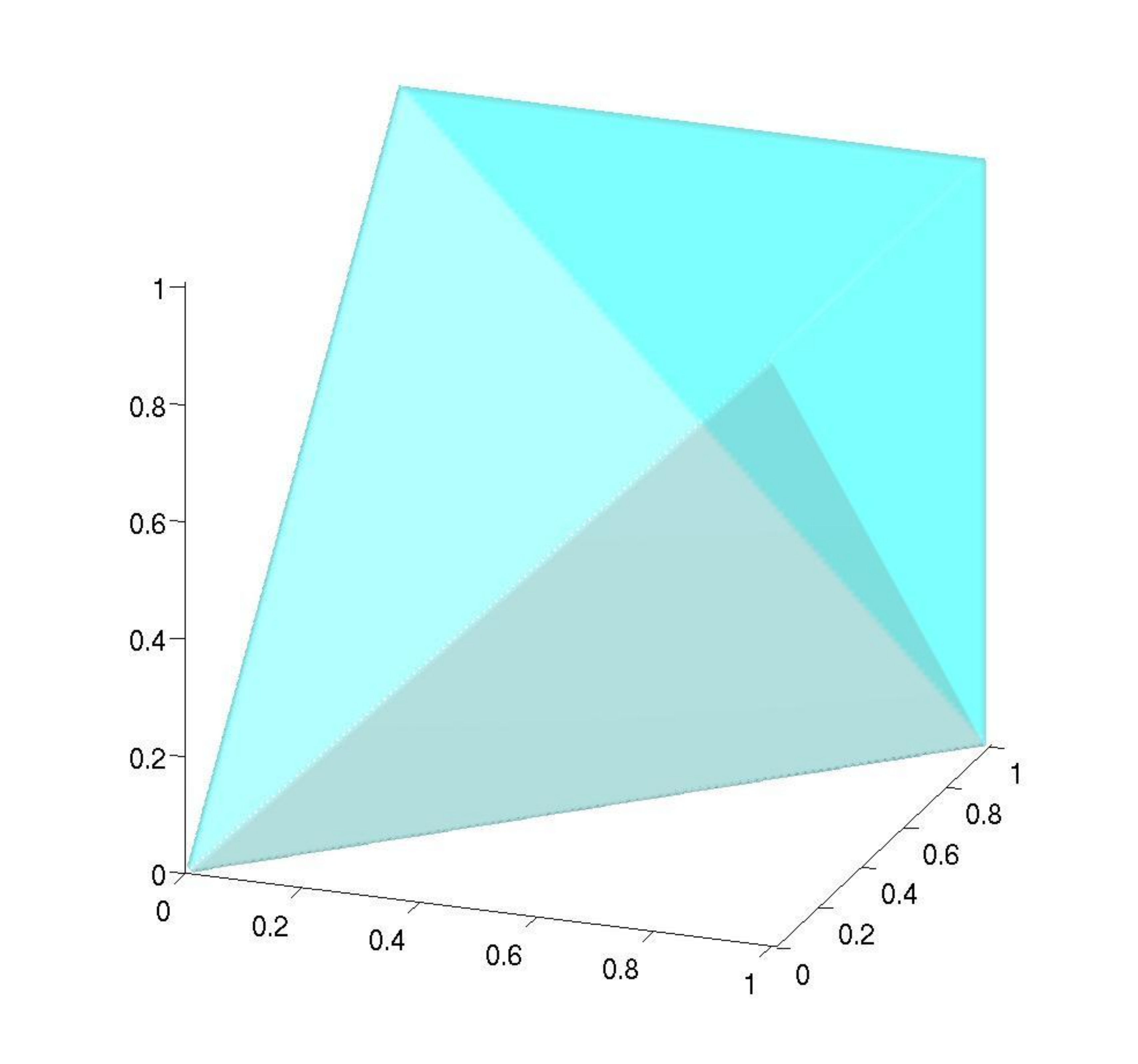}
\includegraphics[width=.45\linewidth]{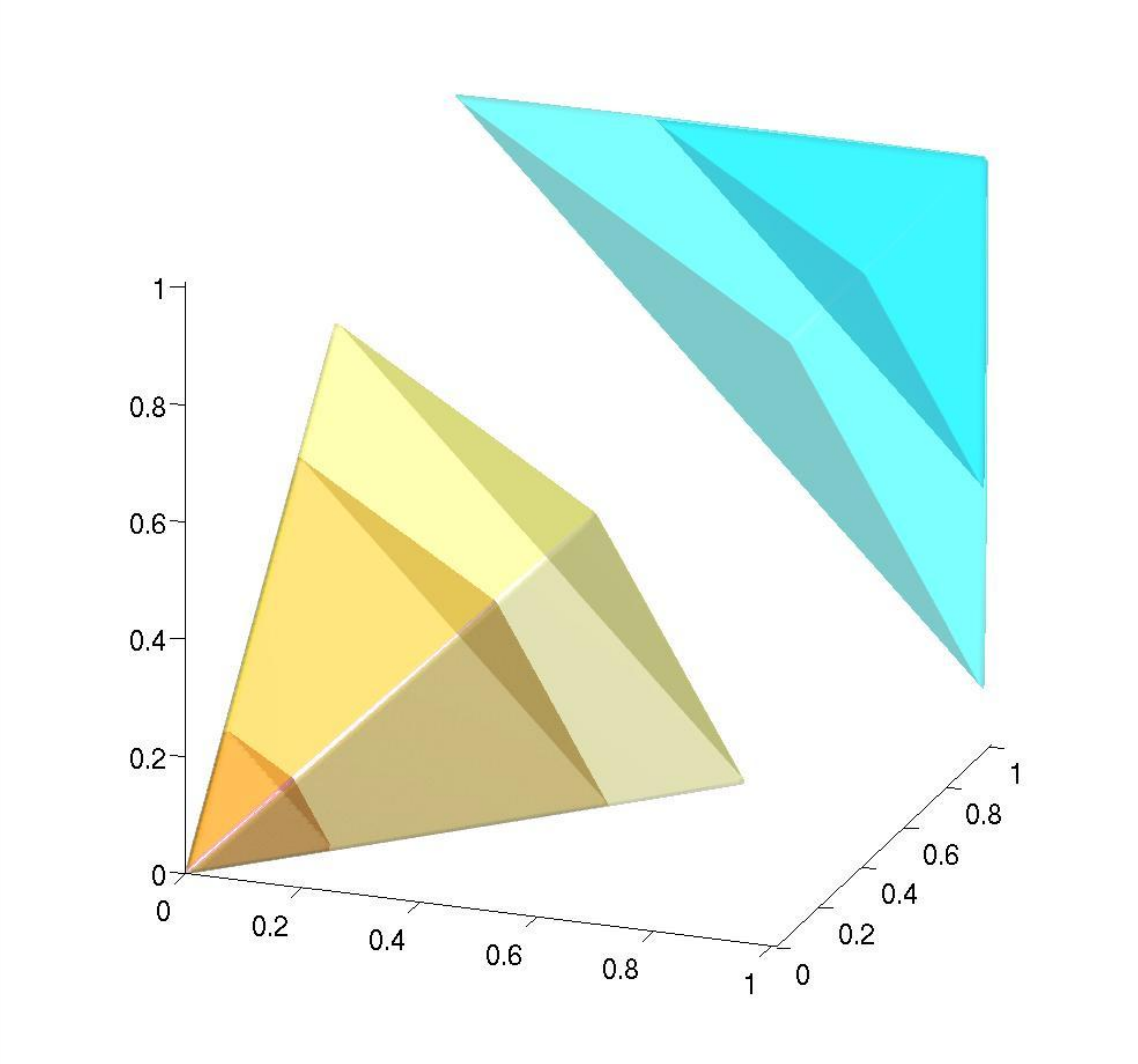}
\includegraphics[width=.45\linewidth]{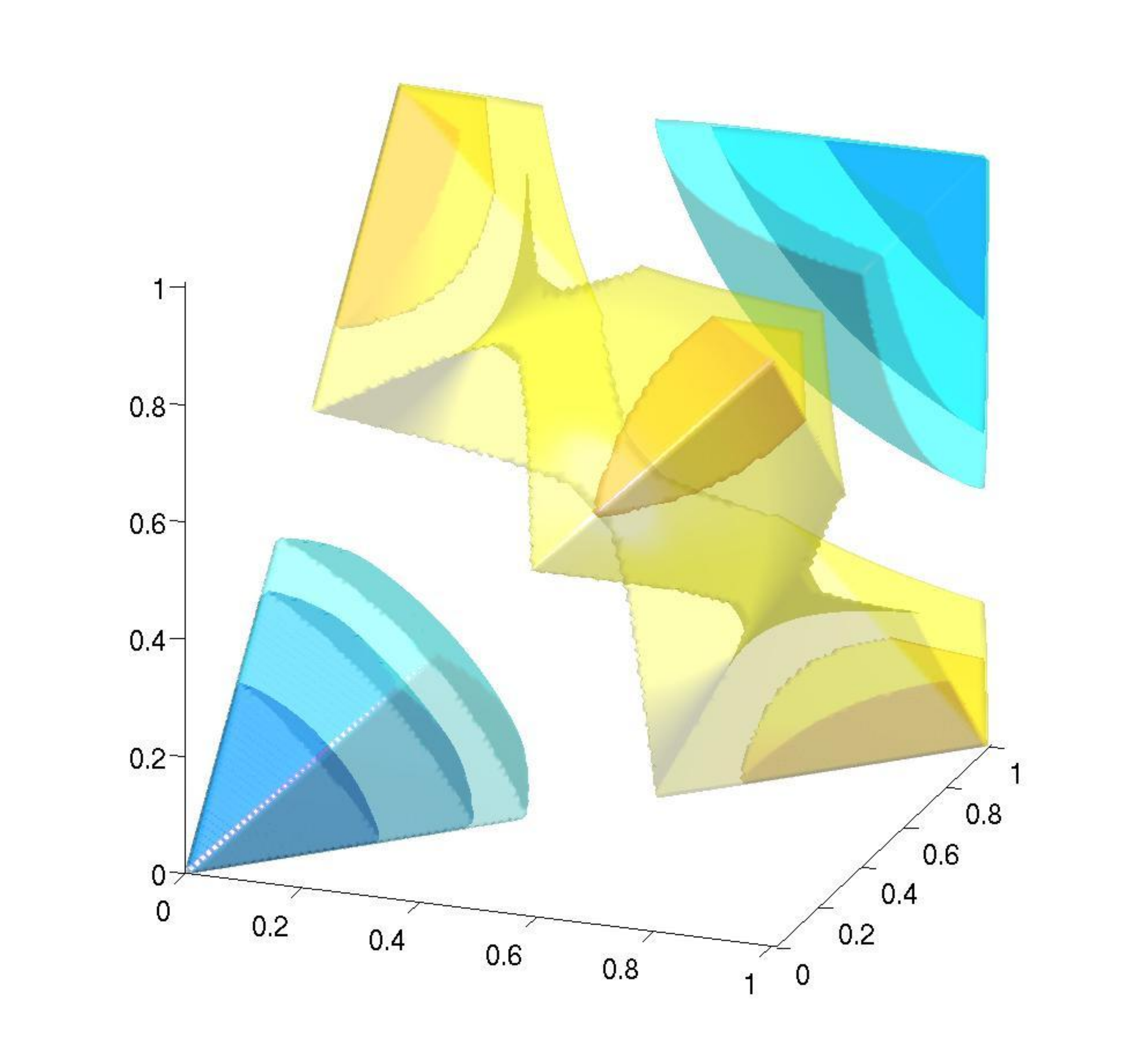}
\includegraphics[width=.45\linewidth]{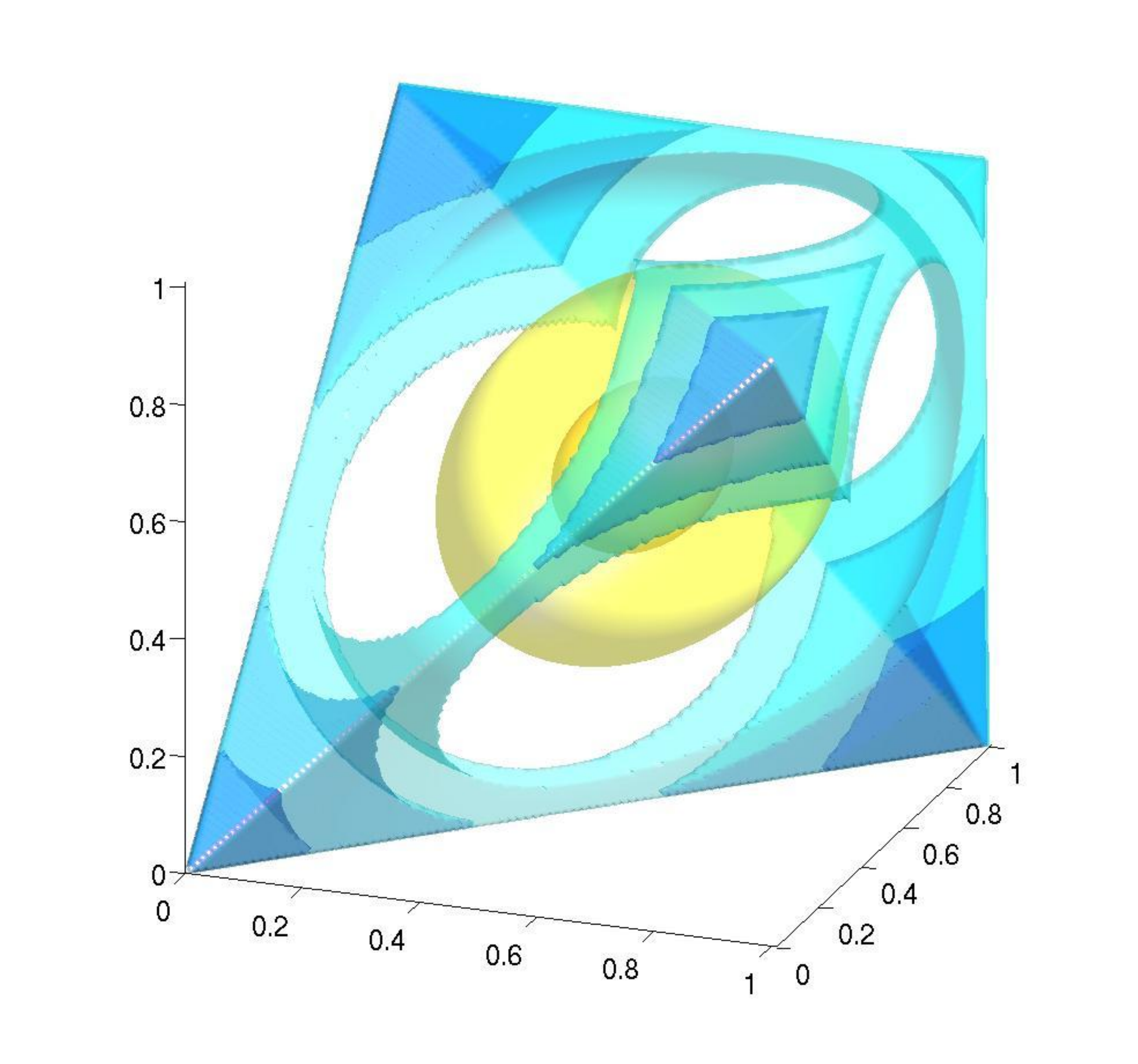}
\includegraphics[width=.45\linewidth]{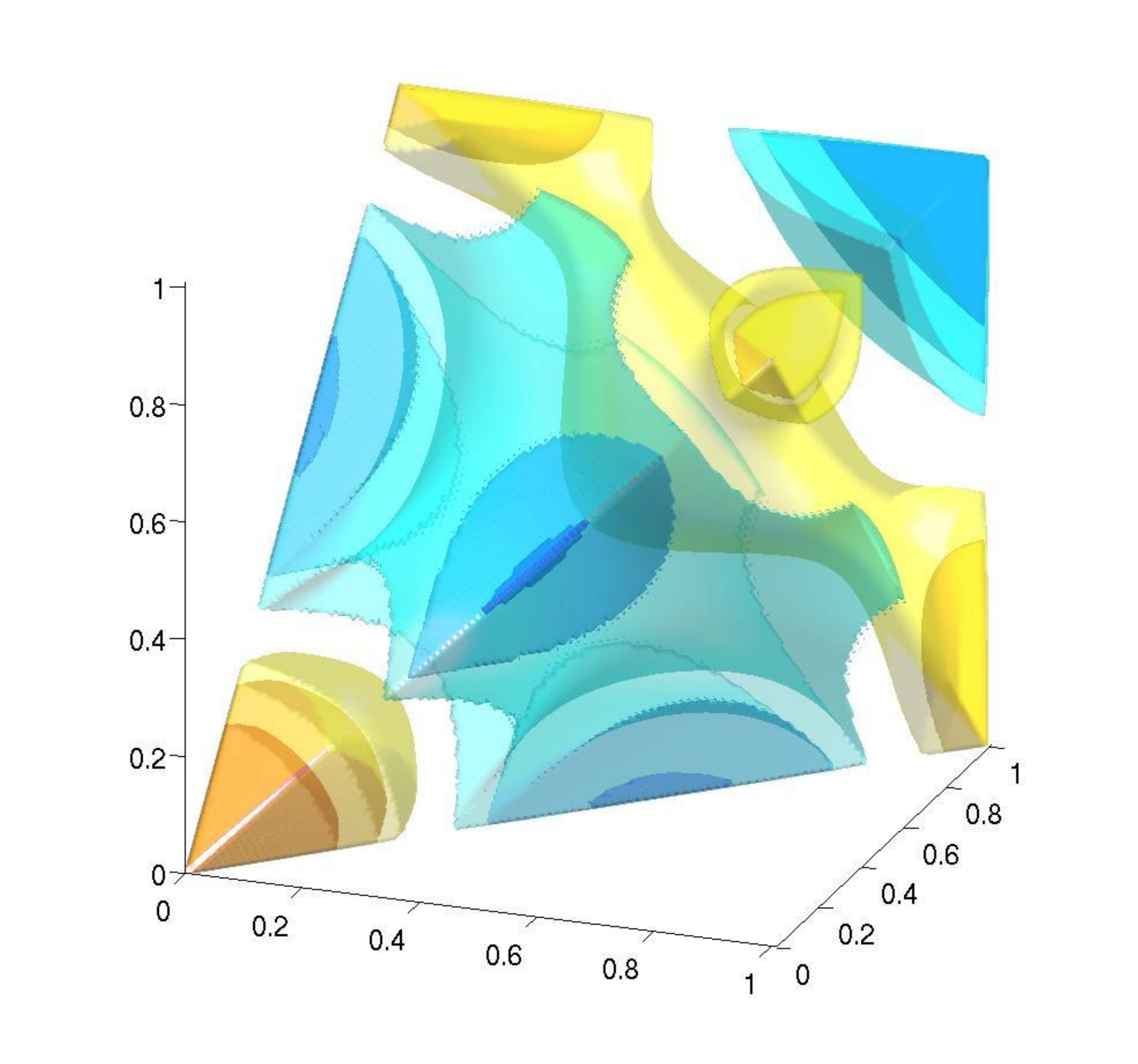}
\includegraphics[width=.45\linewidth]{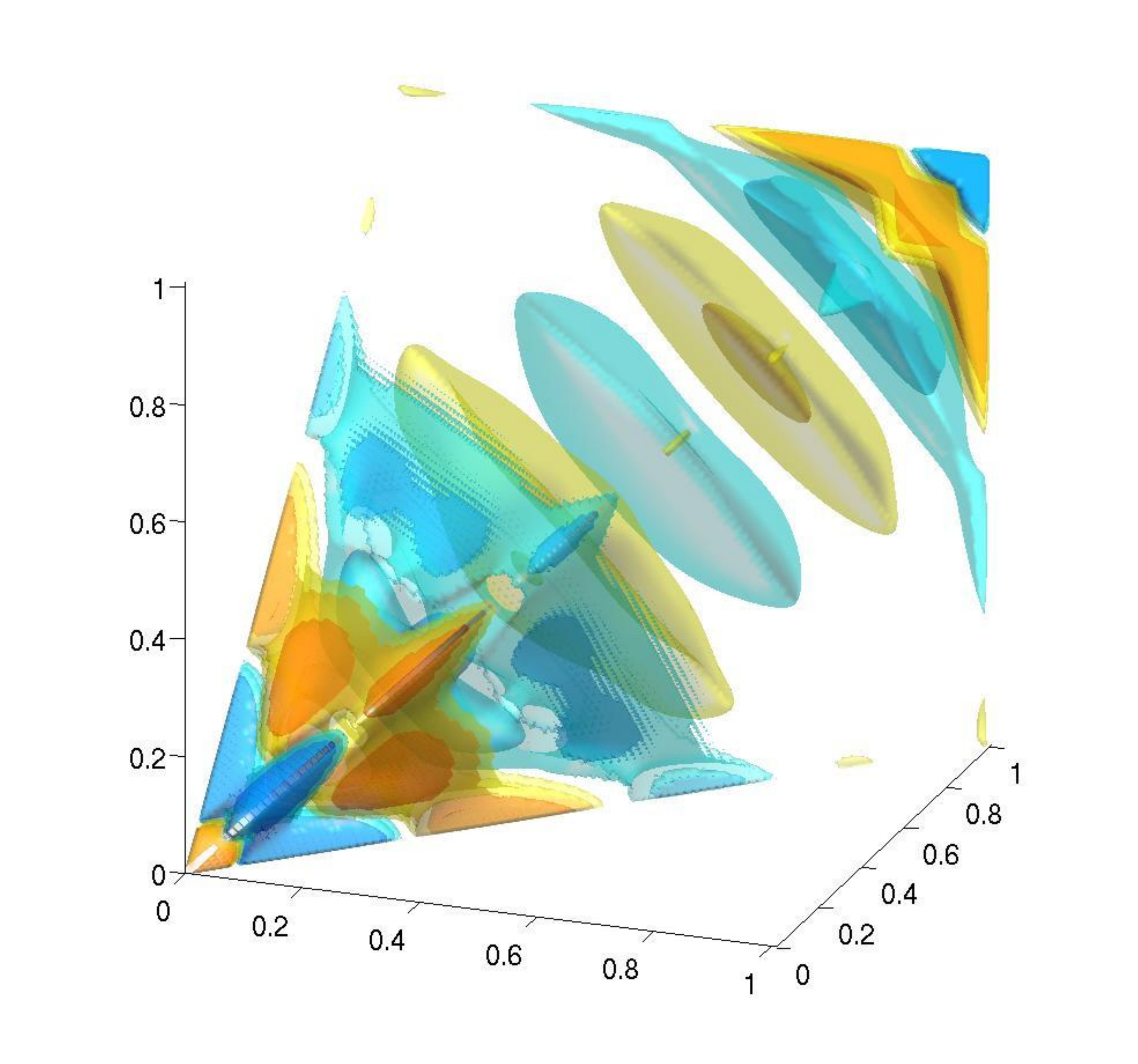}
\caption[Primordial shape function modes]{\small  Three-dimensional orthonormal 
polynomials $\Rn$ on the tetrahedral domain (\ref{eq:tetrapydk}).   Taken 
from top left (and moving across and then down) these are ${\cal R}_0$,  ${\cal R}_1$,  
${\cal R}_2$,
${\cal R}_3$,  ${\cal R}_4$,  and ${\cal R}_{41}$ (bottom right). 
}
\label{fig:shapemodes}
\end{figure}

While the $\Qn$'s by construction are an independent set of three-dimensional basis functions on the domain (\ref{eq:tetrapydk}), they 
are not in general orthogonal.  In fig.~\ref{fig:QRrelation},  we illustrate the inner product matrix $\gamma_{np} = \langle \Qn,\,\Qp\rangle$, showing 
partial orthogonality (nearly diagonal $\gamma_{np}$) because of their origin as products of orthogonal $q_r$'s.   
However, this is not sufficient because we need the convenience of a fully orthonormal basis to efficiently decompose 
arbitrary bispectra.  For this reason,  we undertake an iterative Gram-Schmidt orthogonalisation process  to construct an 
orthonormal set $\curl{R}_n$ from the $\curl{Q}_n$, that is, satisfying 
\eq \label{Rorthog}
\langle \Rn,\, \R_p \rangle = \delta_{np}\,.
\qe 
Formally, we have a Gram matrix $\Gamma= \left(\langle\Qn,\,\Qp\rangle\right)$ made from the
independent functions $\Qn$, and therefore positive definite, which needs to be factorised as 
$\Gamma = \Lambda^{\top}\Lambda$ where $\Lambda= \left(\langle\Qn,\,\Rp\rangle\right)$ is triangular (i.e.\ an LU or Cholesky decomposition).
As we require explicit relationships between $\Qn$ and 
$\Rn$, we run through the main steps in the Gram-Schmidt process.   

Let us assume that we have achieved this orthonormalisation 
up to $n$, that is, such that $\langle\Rn,\,\Rm\rangle =\delta_{nm}$, $\forall \,m\le n$.  This means we can represent any basis function 
$\Q_p$ in terms of the $\Rm$ and vice versa by inversion, so we can write 
\begin{align}\label{eq:RQinverse}
\curl{R}_m = \sum_{p=0}^m \l_{mp} \curl{Q}_p \quad \hbox{for}~~ m,p\le n\,,
\end{align}
where $ \l_{mp}$ is a lower triangular matrix with $(\l^{-1})_{np} ^{\top} = \langle\Qn,\,\Rp\rangle$. 
We wish by induction to construct the next orthonormal polynomial $\R_{n+1}$ and infer from this 
the sum over basis functions up to $\Q_{n+1}$.   We achieve this by taking the next independent basis function, $\Q_{n+1}$, as 
a first approximation to an unnormalised $\R^\pr_{n+1}$ and then we project out all components dependent on the $\Rm$ ($m\le n$), 
\begin{eqnarray}\label{eq:RQorthog}
\curl{R}^\pr_{n+1}&\equiv&  \sum_{p=0}^{n+1} \l^\pr_{n+1\, p} \curl{Q}_p ~ = ~\curl{Q}_{n+1} - \sum_{m=0}^{n}\,\R_m\int_{\Vtetra} \curl{Q}_{n+1}\, \curl{R}_m \,w\,d\Vtetra\nn\\
&=&\Q_{n+1} - \sum_{m=0}^{n} \sum_{r=0}^{m}\sum_{s=0}^m\l_{mr} \l_{ms} \g_{n+1\, s}\Q_r
\end{eqnarray}
where in the second line we have substituted  (\ref{eq:RQinverse}) and the $\gamma_{n+1\,s}$ are determined 
from the relative orthogonality of the $\Qn$'s,
\begin{align}\label{eq:QQorthog}
\g_{n+1\,s} = \langle \curl{Q}_{n+1},\,\curl{Q}_s\rangle = \int_\Vtetra \curl{Q}_{n+1} \,\curl{Q}_s\,w \,d\Vtetra\,.
\end{align}
By equating coefficients in the expression (\ref{eq:RQorthog}) we can determine that 
\begin{align}\label{eq:RQorthog2}
\l^\pr_{n+1\, p} = \d_{n+1\, p} - \sum_{r=p}^{n}\sum_{s=0}^r\l_{rp}\, \l_{rs}\, \g_{n+1\, s}\,.
\end{align}
Unit normalising appropriately, we obtain the coefficients $\lambda_{n+1\,p}$ which 
define the new orthonormal $\R_{n+1}$ we are seeking, that is, we have
\begin{align}
\l_{n+1\, p} = {\l^\pr_{n+1\,p}}~\hbox{\huge/}\left({\sum_{r,s=0}^{n+1} \l^\pr_{n+1\,r}\, \l^\pr_{n+1\,s}\, \g_{rs}}\right)\,.
\end{align}

In fig.~\ref{fig:QRrelation}, we see the orthogonalisation process at work for the first 10 modes by plotting 
the matrix coefficients for $ \langle \Qn,\,\Qp\rangle$ and $ \langle \Qn,\,\Rp\rangle$. 
At each order $n$,  the independent component in $\Rn$ is provided by $\Qn$, as 
indicated by the dominant diagonal term.  This is a good approximation at low order, but the mixing increases with 
$n$.   We also illustrate several of the orthogonal polynomials $\Rn$ on the 
tetrapyd domain  fig.~\ref{fig:shapemodes} for the slicing ordering (\ref{eq:slicingorder}).   These  are primarily the lowest modes 
and demonstrate the build up of the number of  
 nodal points and lines as the order increases.   As an aside, we note Gram-Schmidt orthogonalisation in the form
 given above is  inherently unstable numerically, though this can be easily corrected by using the modified Gram-Schmidt process.  However, we do not iterate to sufficiently high $n$ to notice
 any significant degradation in accuracy, as verified by determining orthogonality. 

\subsection{Mode decomposition of the bispectrum}

We have constructed examples of an orthonormal basis $\{\Rn\}$ out of monomial symmetric 
polynomials (\ref{eq:monsympoly}) which span the set of symmetric functions on the tetrahedral 
domain (\ref{eq:tetrapydk}).    The $\Rn$ polynomials will possess
the properties of more familiar  orthonormal eigenmodes in other contexts, notably 
completeness and the convergence of  mode expansions for 
well-behaved functions.    We proceed by considering an arbitrary primoridal 
bispectrum (\ref{eq:shapefn}) described by the shape function 
$S(\kall)$ and decomposing it as follows
\begin{align}\label{eq:Rshapedecomp}
S(\kall) = \sum_{n=0}^{\infty} \aRn\, \Rn(x,y,z)\,,
\end{align}
where the expansion coefficients $\aRn$ are given by
\begin{align}\label{eq:Rshapecoeff}
\aRn = \langle \Rn, \, S\rangle = \int_{\Vtetra} \curl{R}_n \, S\,w\, d\Vtetra \,,
\end{align}
and $K = \kmax$ and $k_1 = Kx$ etc on the domain $\Vtetra$ defined in (\ref{eq:tetrapydk}).   For practical purposes, 
we shall always work with partial sums up to a given $N=\nmax$ with 
\eq\label{eq:RshapeN}
S_N =  \sum_{n=0}^{N} \aRn\, \Rn(x,y,z)\,, \qquad S = \lim_{N\rightarrow\infty}S_N\,.
\qe
We shall assume that the expansion (\ref{eq:RshapeN}) is the best fit mode expansion of degree $N$
(for this particular mode ordering).   Given the complete orthonormal basis $\Rn$, Parseval's theorem for the
integrated product of two functions implies
\eq \label{eq:Parseval}
\langle S, \,S'\rangle  = \int _\Vtetra S \,S'\, w\,d\Vtetra = \lim_{N \rightarrow \infty} \sum_{n=0}^{N} \aRn\,{\aRn}'\,,
\qe  
which, for the square of a function $S$, yields the sum of the squares of the 
expansion coefficients, ${\cal T}[S^2] = \sum_n{\aRn}^2$. 

In order to accomplish our original goal of a general separable expansion (\ref{eq:separablebasis}), 
we must now transform backwards from the 
orthonormal $\Rn$ sum (\ref{eq:RshapeN}) 
into an expansion over  the separable product  functions $\Qn = q_{\{p}q_rq_{s\}}$ through 
\begin{align}\label{eq:Qshape}
S_N = \sum_{n=0}^{N} \aQn\, \Qn(x,y,z)\,, 
\end{align}
where the $\aQn$ can be obtained from the $\aRn$ as
\begin{align}\label{eq:Qshapecoeff}
\aQn = \sum_{p=0}^{N}(\l^\top)_{np}\, \aRp\,,
\end{align}
with the transformation matrix  $\l_{np}$ defined in (\ref{eq:RQinverse}) (this is triangular and not 
orthogonal in general).    Note the complication that $\aQn$ also contains contributions from $\Rp$ components with $
n< p\le N$, since $(\l^\top)_{np}$ is upper triangular.
The inverse transformation
\begin{align}\label{eq:aRaQ}
\aRn = \sum_{p}^N(\l^{-1})^T_{np}\, \aQp\,,
\end{align}
has coefficients given by $(\lambda^{-1})_{np} = \langle \Qn ,\,\Rn\rangle$.  We have already noted that the degree of 
non-orthogonality of the $\Qn$ basis is described by $\gamma_{np} = \langle \Qn,\,\Qp\rangle$ in (\ref{eq:QQorthog})
which is in turn related to $\l_{np}$ through
\begin{align}\label{eq:gammalambda}
(\gamma^{-1})_{np} = \sum_{r}^N(\l^\top)_{nr}\l_{rp}\, .
\end{align}
When substituted into Parseval's theorem (\ref{eq:Parseval}) in the $\Qn$ basis, we see that the coefficients of different degrees become mixed as 
\eq \label{eq:Parsevalmixed}
\langle S_N, \,S_N'\rangle =  \sum_n^N{\aRn}^2 = \sum_n^N\sum_p^N{\aQn}\gamma_{np}{\aQp}
\qe

\begin{figure}[t]
\centering
\includegraphics[width=.85\linewidth,height = 6.5cm]{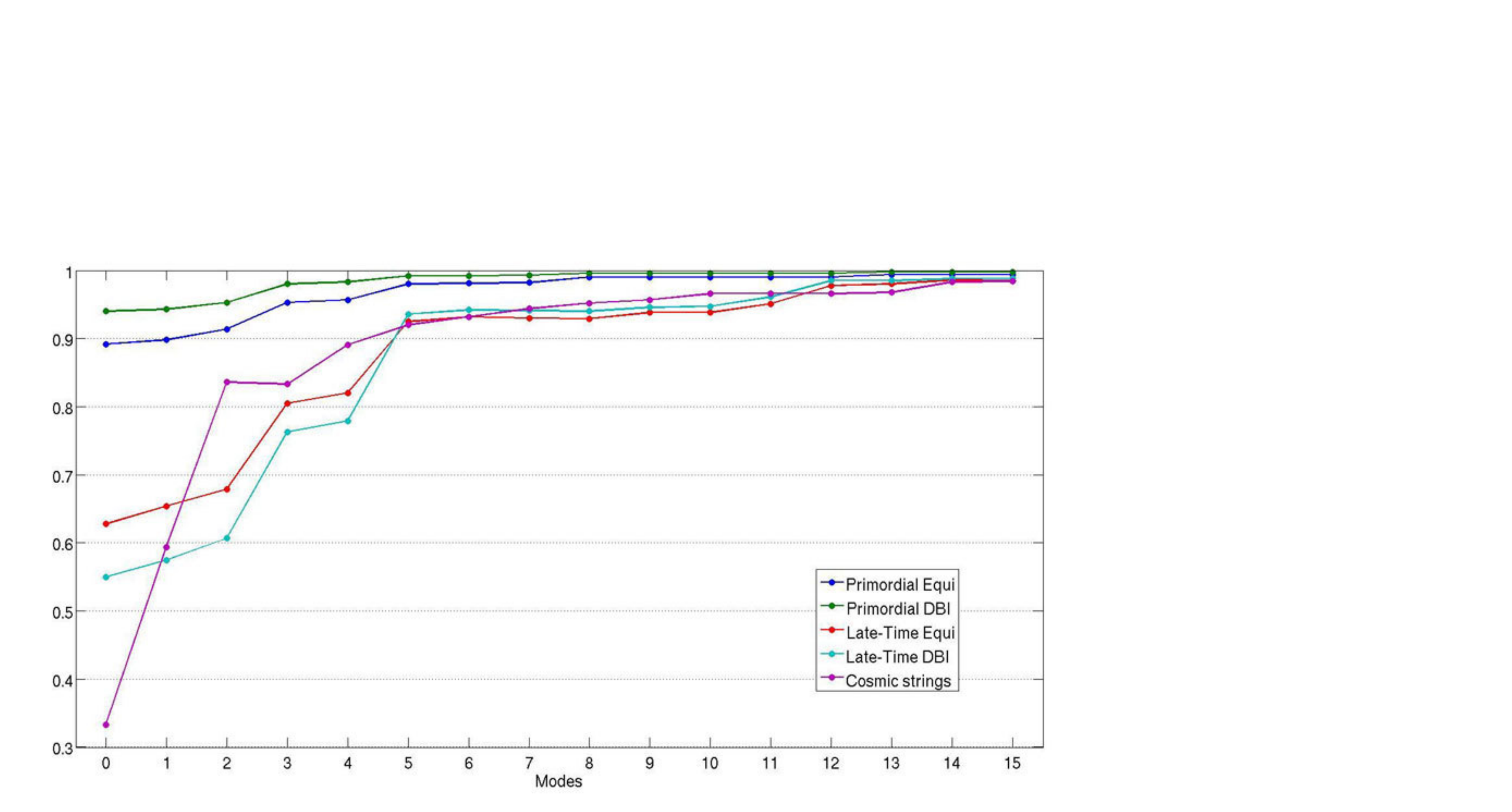}
\caption[Convergence]{\small Correlation of the reconstructed bispectra to the original for partial sums of the decomposition up to a given mode $n$. The plot includes the primordial bispectra for the equilateral and DBI models,   the CMB bispectrum for the equilateral and DBI models and the CMB bispectrum produced at late times by cosmic strings.  In all cases, we find that with 15 three-dimensional modes we have a correlation greater than 98\%, thus demonstrating very rapid convergence. For the CMB bispectra, convergence is limited by matching the acoustic peaks introduced by the transfer functions, whereas the primordial models converge at 98\% accuracy with only 6 modes. 
}
\label{fig:convergence}
\end{figure}

\begin{figure}[t]
\centering
\includegraphics[width=.85\linewidth, height = 6.5cm]{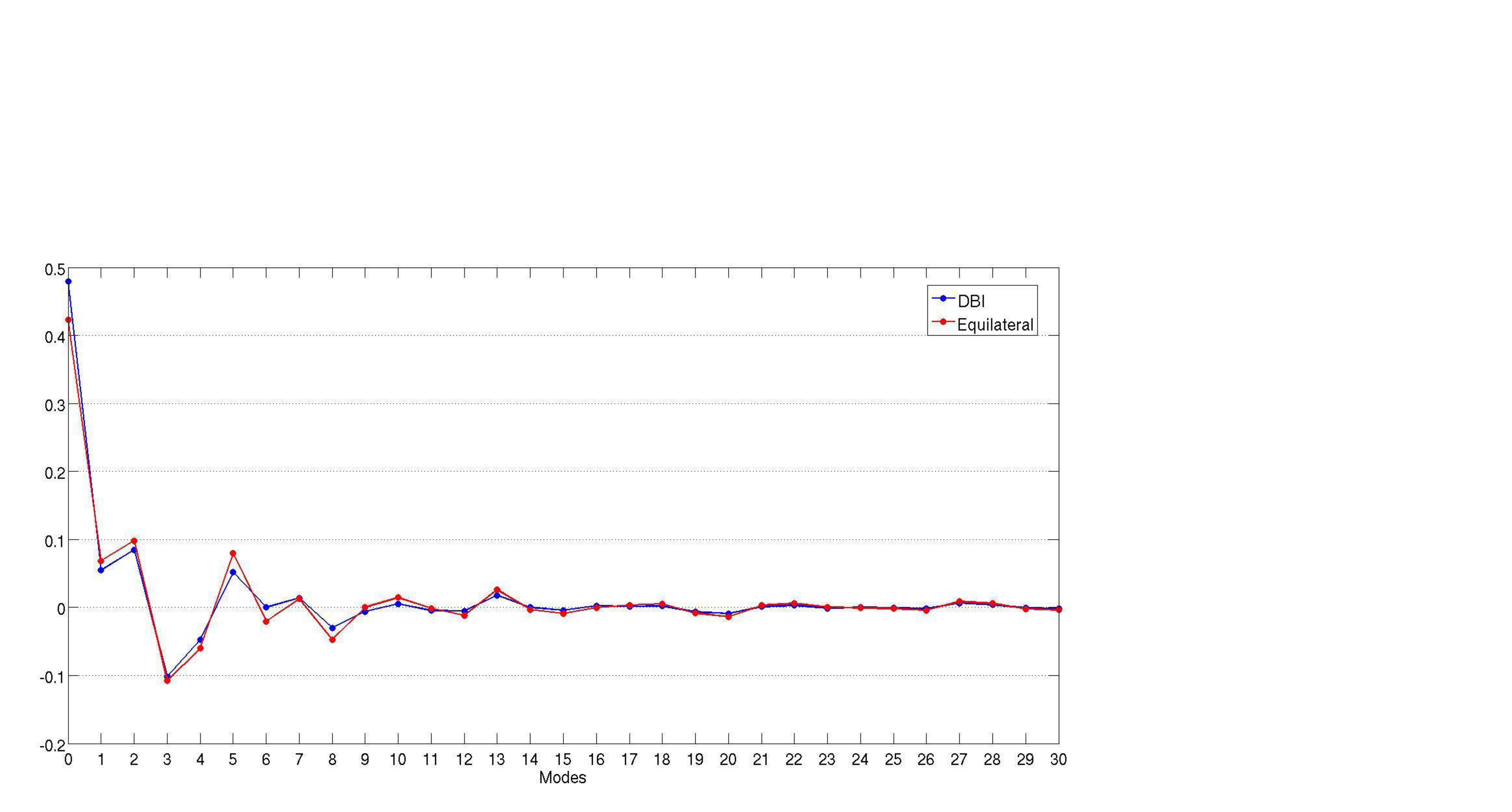}
\includegraphics[width=.85\linewidth, height = 6.5cm]{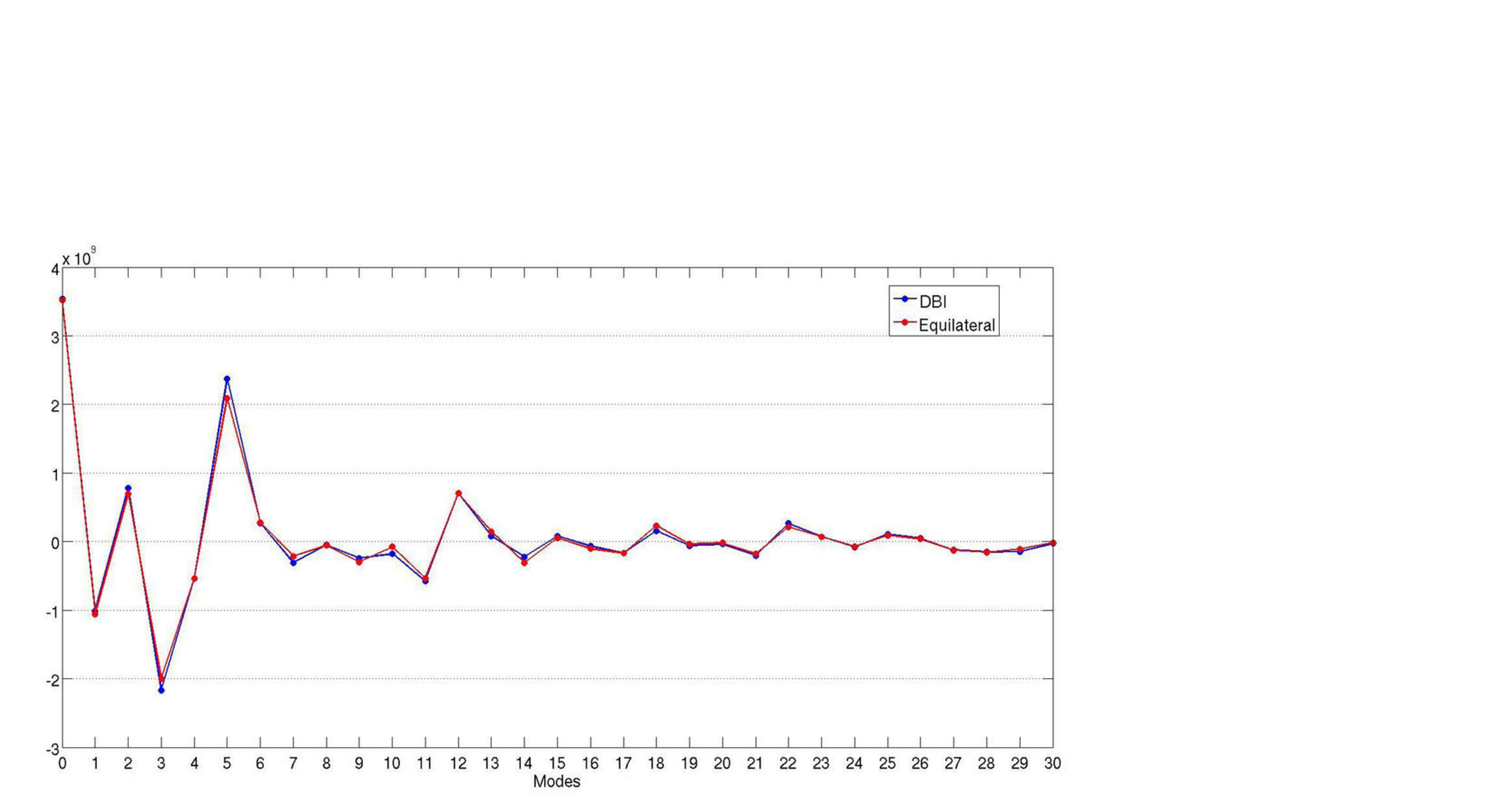}
\caption[QR relations]{\small  Decomposition into orthonormal polynomials $\Rn$ for both the primordial shape function (\ref{eq:Rshapedecomp}) (above)
and the CMB bispectrum estimator (\ref{eq:cmbestmodes}) (below) for the equilateral (red) and DBI (blue)
models.  In both cases, these results are for `slicing' polynomial ordering given in (\ref{eq:slicingorder}).  The peak in the CMB 
bispectrum estimator modes (here, for $\lmax =500$ at $n\approx 5$) arises because of the power shifted into the coherent 
acoustic peaks observed in fig.~\ref{fig:cmbbispectra}; this is a distinguishing feature of the CMB bispectrum for most primordial models \cite{Fergusson:2006pr}. 
}
\label{fig:Requildecomp}
\end{figure}

The separable $\Qn$ expansion (\ref{eq:Qshape}) is 
important for most practical calculational purposes but its coefficients are constructed at the outset 
using the orthonormal $\Rn$.  For interpreting results from the estimator it is helpful to 
transform back to the $\Rn$ basis in order to understand the normalised spectrum $\aRn$ using Parseval's
theorem (\ref{eq:Parseval}).  
We finally note that all the transformation matrices, $\l_{np}$ and
$\gamma_{np}$ in (\ref{eq:QQorthog}), need only be calculated once,
at the same time as the $\Rn$ polynomials are generated, and then stored for later reference.

In fig.~\ref{fig:convergence}, we demonstrate polynomial convergence for the 
DBI model and its separable equilateral approximation by showing the cross-correlation between 
the shape function and the partial sum (\ref{eq:RshapeN}).  We also provide the actual expansion coefficients 
$\aRn$ for the primordial shape functions in fig.~\ref{fig:Requildecomp} (along with the $\baRn$ for the CMB bispectra).  
Using only 6 three-dimensional $\Rn$ polynomials we achieve a better than 98\% cross-correlation with the original 
analytic expressions in both cases (i.e.\ using symmetric products of at most quadratic $q_p$ polynomials from (\ref{eq:Qpolys})). 
Here, we undertake the full Fisher matrix analysis between the theoretical CMB bispectrum and its approximation using 
the methods described in ref.~\cite{Fergusson:2008ra}.   More generally, we note that for all well-behaved bispectra the polynomial 
expansion has proved to be rapidly convergent. The decomposition of the primordial bispectra is also numericaly efficient using the orthogonal $\Rn$ modes with 
each $\aRn$ coefficient taking an average of 7 seconds to calculate.

We can equally well expand the CMB bispectrum $\blll$ at late times, using the same polynomials $\blll= \sum_n \baRn \Rn(x,y,z)$.   However, our aim is to represent the bispectrum estimator ${\cal E}$ given in (\ref{eq:normestimator}), rather than $\blll$ itself.  We, therefore, consider expanding a separable product which approximates ${\cal E}$ with the same weight and scaling (schematically, $\sqrt{l} \, \blll /\Cl^{3/2}$).   We discuss this in the next section, but in the lower half of fig.~\ref{fig:Requildecomp} we show the corresponding late-time expansion coefficients $\baRn$ for the equilateral and DBI CMB bispectra.   Once again, convergence is rapid, see figure~\ref{fig:convergence}, with a 95\% correlation achieved with only 12 $\barRn$ polynomials at $\lmax =500$, improving to 98\% with 15 polynomials for both CMB bispectra.  This is despite the fact that  the expansion must incorporate additional features induced by the transfer functions.   We emphasise the power shift from the low modes in the primordial bispectrum to a peak at higher modes $n\approx 5$ in the CMB bispectrum (for this slicing and $\lmax$).  This is a common characteristic of the polynomial expansion for almost all bispectra of primordial origin and is a manifestation of the 
pattern of coherent acoustic peaks observed in ref.~\cite{Fergusson:2006pr}. The decomposition of the CMB bispectra is also numericaly efficient using the orthogonal $\barRn$ modes with each $\baRn$ coefficient taking an average of 8 seconds to calculate.

\subsection{Utility of the tetrahedral polynomials $\Qn$ and other alternatives}

The three-dimensional polynomials $\Qn$ we have presented are just one possible 
set of basis functions which can be used as bispectrum eigenmodes for the methodology we present 
in the next section.   They are built from products of the one-dimensional $q_p$'s which are 
orthonormal on the tetrahedral region (\ref{eq:tetrapydk}) with given weight functions. These 
are the analogues of 
Legendre polynomials $P_n$.  Unfortunately, unlike the $P_n$'s on a cube, they do 
not retain full orthogonality as separable products on the tetrahedral domain, though 
there is a substantial remnant.   There are significant advantages to using the $q_p$'s, rather
than the monomial symmetric polynomials in (\ref{eq:monsympoly}), in the same way 
that Legendre polynomials are more efficient than simple power series representations.  
As we shall discuss subsequently, there are further important benefits which arise 
when the $\Qn$'s are decomposed into separable integrals over the $q_p$'s.   Given 
the bounded and well-behaved nature of the $q_p$'s on their domain, these integrals
reflect these properties, eliminating diverging artifacts which are known for other separable
approximations to bispectra in the literature (including difficulties for simple powers $x^n$).

There are other alternatives to expansions using the tetrahedral polynomials
$\Qn$ and $\Rn$ which we have considered.  
It is possible, for example, to expand an arbitrary bispectrum using 
separable products of more familiar orthonormal functions such as Legendre the $P_n$ and 
Chebyshev $T_n$ polynomials,  as discussed in ref.~\cite{Fergusson:2006pr}.   This entails using 
shifted polynomials on the full cubic domain  $\lall \le \lmax$. The shortcoming of this approach 
is that the bispectrum is only defined on the tetrahedral region (\ref{eq:tetrapydk}), so it 
has to be zero elsewhere or arbitrarily extended in some manner to fill the cube.  
This leads to generic overshooting of the expansion near the boundaries (the 
analogue of the Gibbs phenomena for Fourier series).     Extensive 
experiments yielded very poor convergence with Legendre and Chebyshev 
polynomial expansions, as well as Fourier series, especially relative 
to that achieved with the tetrahedral $\Qn$ and $\Rn$ polynomials.  A further
simple alternative is to transform the tetrahedral region into a cube (see ref.~\cite{Fergusson:2008ra}).  
This allows the bispectrum to be defined everywhere on the standard domain using the 
more familiar eigenmodes and thus yielding more rapid convergence.  However, this 
compromises separability which is essential for the estimators we discuss below.    

Were the rate of convergence to become a primary issue when representing the bispectrum, 
then there are further alternatives to polynomials.   There is a 
significant literature on eigenmodes on the regular tetrahedron or simplex because of its importance 
in crystallography and other contexts.  For example, it is possible to define generalised sine and cosine 
functions on the simplex, as well as Koornwilder and generalised Chebyshev polynomials of the 
first and second kind (see, for example, ref.~\cite{0803.0508}).    Such generalised eigenfunctions 
could, in principle, improve convergence, however, two significant developments are required. First, it is 
more natural to define the observational
data on the tetrahedral domain with $\lall \leq \lmax$  (the tetrapyd), rather than the simplex
$\lsum\leq 2\lmax$, so generalised eigenfunctions must be derived explicitly for this domain (\ref{eq:tetrapydl}).
Secondly, these should be able to conveniently represent functions in separable form.   
The present tetrahedral polynomials $\Rn$ and $\Qn$ do converge satisfactorily for all the primordial 
models studied to date, but more efficient mode expansions will continue to be investigated \cite{inprep}.

\section{Measures of $\Fnl$}

\subsection{Primordial $\Fnl$ estimator}

We have  obtained two related mode expansions for a general primordial shape function (\ref{eq:shapefn}), one for an orthonormal basis $\Rn$  (\ref{eq:RshapeN}) and the other for separable basis functions $\Qn$ (\ref{eq:Qshape}). 
Substitution of the 
separable form into the expression for the reduced bispectrum (\ref{eq:redbispect2}) offers an 
efficient route to its direct calculation through \begin{eqnarray} \label{eq:sepblllprim}
\nn b_{l_1 l_2 l_3} &=& {\textstyle\(\frac{2}{\pi}\)^3} \D_\O^2 f_{NL}\int  x^2 dx\, d k_1 d k_2 d k_3\, \,6 \sum_n \aQn \,\Qn(\kall)
\,\D_{l_1}(k_1)\, \D_{l_2}(k_2)\, \D_{l_3}(k_3)\, j_{l_1}(k_1 x) \,j_{l_2}(k_2 x)\, j_{l_3}(k_3 x) \\
\nn & =& \D_\O^2 f_{NL} \sum_{n} \aQn \int  x^2 dx 
\left[\({ \frac{2}{\pi} }\int d k_1\,  q_p(k_1) \, \D_{l_1}(k_1) \, j_{l_1}(k_1x) \)
\( {\frac{2}{\pi}} \int d k_2 \,  q_r(k_2) \, \D_{l_2}(k_2) \, j_{l_2}(k_2x) \)\right.\\
&& \qquad \qquad \qquad \qquad \qquad ~~\left.\times \( {\frac{2}{\pi}} \int d k_3 \,  q_s(k_3) \, \D_{l_3}(k_3) \, j_{l_3}(k_3x) \) ~+ ~\mbox{5~ permutations}\right]\\
\nn &=& \D_\O^2 f_{NL} \sum_{n} \aQn \int x^2dx \, q_{\{p}^{\,l_1} q_{r}^{\,l_2} q_{s\}}^{\,l_3} ~=~\D_\O^2 f_{NL} \sum_{n} \aQn\int x^2dx \, \Qn^{l_1l_2l_3}\,,
\end{eqnarray}
where here we implicitly assume the mapping $n\leftrightarrow {prs}$ between indices for the $\Qn$ and the product basis
functions  from which they are  formed, that is, $\Qn=\Qndefn$ (e.g.\ see the ordering in (\ref{eq:slicingorder})).  For brevity we 
have also denoted as $q_p^l$ the convolution of the basis function $q_p(k)$ with the 
transfer functions 
\eq\label{eq:qmultipole}
q_p^{\,l}(x) = \frac{2}{\pi} \int d k\,  q_p(k) \, \D_{l}(k) \, j_{l}(kx)\,,\quad \hbox{with}\quad \Qn^{l_1l_2l_3}(x) = q_{\{p}^{\,l_1}(x)\,q_r^{\,l_2}(x)\,q_{s\}}^{\,l_3}(x)\,.
\qe
(These $q_p^{\,l}$ are the primordial counterparts of the $\bar q$ defined in multipole space (\ref{eq:Qlpolys}).) 
Here, in (\ref{eq:sepblllprim}),  the previously intractable three-dimensional wavenumber integral separates into the product of three one-dimensional integrals
which  are relatively easy to to evaluate.  This has been achieved because the triangle condition has been enforced through the product of Bessel functions, giving a manifestly separable form and allowing us to interchange the orders of integration with $x$; it is the basis for the analytic local (\ref{eq:localbispect}) and constant (\ref{eq:constbispect}) solutions on large angles, as well as all the analysis of separable shape functions to date (see, for example, ref.~\cite{Komatsu:2003iq}).  With this mode expansion, all  non-separable theoretical CMB bispectra $\blll$ become calculable provided there is a convergent expansion for the shape function.  Accurate hierarchical schemes already exist against which to benchmark this method \cite{Fergusson:2006pr} but, in principle, it is more efficient.

Now consider the implications of this mode expansion for  $\fnl$ by substituting the decomposed $\blll$ 
 (\ref{eq:Qshape}) into the estimator expression (\ref{eq:estimator}) to obtain 
 \eq\label{eq:primestsep}
 {\cal E} &=& \D_\O^2 f_{NL}  \sum_{n} \aQn \sum_{l_i,m_i}  \int x^2 dx\,q_{\{p}^{\,l_1}(x)\, q_{r}^{\,l_2}(x)\, q_{s\}}^{\,l_3}(x) \int d^2\hat {\bf n}\,  Y_{l_1m_1}(\hat {\bf n})\,Y_{l_2m_2}(\hat {\bf n})\, Y_{l_3m_3} (\hat {\bf n})\,\frac{a_{l_1m_1}a_{l_2m_2}a_{l_3m_3}}{C_{l_1}C_{l_2}C_{l_3}}\\
&=&  \D_\O^2 f_{NL} \sum_{n} \aQn \int d^2\hat {\bf n}  \int x^2 dx\, \left[\sum_{l_1,m_1}q_p^{\,l_1}\, \frac { a_{l_1m_1} Y_{l_1m_1} }{C_{l_1}}\sum_{l_1,m_1}q_r^{\,l_2}\, \frac { a_{l_2m_2} Y_{l_2m_2} }{C_{l_2}}\sum_{l_3,m_3}q_s^{\,l_3}\, \frac { a_{l_3m_3} Y_{l_3m_3} }{C_{l_3}}\right]\,.
 \qe
The break up of the wavenumber integration now extends also to the separation of the  
sum over the multipoles $\lall$.  The summation between the $\alm$'s and each $q_p$ integral creates a filtered
map of the original data, which we can define in the above as 
\eq\label{eq:mapfilter}
M_p(\un,x) =  \sum_{lm} q_p^{\,l}\,\, \frac { a_{lm} Y_{lm} }{C_{l}} ~=~ \sum_{lm}  \[\frac{2}{\pi} \int q_p(k) \D_l(k) j_l(kx) dk \] \frac{a_{lm} Y_{lm}(\un)}{C_l}\,.
\qe 
From these we can efficiently calculate product maps which essentially extract the $\Qn$ basis function contribution from the observational data, 
\begin{align}\label{eq:productmap}
\curl{M}^{\cal Q}_n(\un,x) = M_p(\un,x)M_r(\un,x)M_s(\un,x)
\end{align}
where again we exploit the correspondence $n\leftrightarrow {prs}$. 
Note, that in this case, there is no need to symmetrise the product map
because it is implicit in the estimator expression.   Integrating over directions and shells we 
can now obtain for the observational maps, 
the analogue of the primordial mode expansion coefficients $\aQn$, 
\begin{align}\label{eq:betadefn}
\bQn = \int d^2\un \int x^2 dx \, \curl{M}^{\cal Q}_n(\un,x)\,.
\end{align}
In common with the analysis of simple separable shapes, the shell integral over $x$ is where the most significant computational effort is required.

Substituting into (\ref{eq:primestsep}), the bispectrum estimator then collapses into a compact diagonal form
\begin{align} \label{eq:primestimator}
\curl{E} = \frac{6 \D_\O^2}{N} \sum_{n} \aQn\, \bQn\,.
\end{align}
The estimator has been reduced entirely to tractable integrals and sums which can be performed
rapidly even at $\lmax =2000$.   We will demonstrate how efficiently it can 
recover $\fnl$ from simulated maps in subsequent sections. 

The form of the estimator in (\ref{eq:primestimator}) suggests that further information  can be 
extracted about the observed bispectrum beyond the $\fnl$ for one specific 
theoretical model.   This is because, through the coefficients $\bQn$,
we have obtained some sort of mode decomposition of the 
 bispectrum of the observational map.   However, the non-orthogonal and primordial nature of the $\Qn$ basis 
functions means that these $\bQn$ require some effort in their interpretation.   Consider 
the expectation value of $\bQn$ obtained from an ensemble of maps generated for a particular
theoretical model with shape function $S = \sum_n \aQn\,\Qn$. 
Noting that the relation $ \langle a_{l_1m_1}a_{l_2m_2}a_{l_3m_3}\rangle= \curl{G}^{\,l_1 ~l_2~ l_3}_{m_1 m_2 m_3} \blll $,
the average over the product maps (\ref{eq:productmap}) becomes 
\eq\label{eq:betaprim}
\langle\bQn\rangle &=& \int d^2\un \int x^2 dx \, \langle \curl{M}^{\cal Q}_n(\un,x) \rangle
~= ~\sum_{l_i,m_i} \left(\int x^2 dx\,  q_{\{p}^{l_1} q_{r}^{l_2} q_{s\}}^{l_3}\right)\left(   \curl{G}^{\,l_1 ~l_2~ l_3}_{m_1 m_2 m_3}  \right)^2    \frac{\blll}{C_{l_1}C_{l_2}C_{l_3}}\\
&=& \sum_{l_1l_2l_3}{\frac{1}{4\pi}}\,\frac{(2l_1+1)(2l_2+1)(2l_3+1)}{C_{l_1}C_{l_2}C_{l_3}} \( \begin{array}{ccc} l_1 & l_2 & l_3 \\ 0 & 0 & 0 \end{array} \)^2
\int x^2 dx\, \Qn^{l_1l_2l_3}\sum_p \aQ_p{\int x^2 dx\,\Q_p^{l_1l_2l_3}}\\
&=& \sum_p \aQ_p\sum_{l_1l_2l_3}\frac{w_{l_1l_2l_3}}{C_{l_1}C_{l_2}C_{l_3}} 
\int x^2 dx\, \Qn^{l_1l_2l_3}\int x^2 dx\,\Q_p^{l_1l_2l_3}~~\equiv~~ \sum_p \Gamma^{\cal Q} _{np} \aQ_p\,,
\qe
where we have substituted the expression (\ref{eq:sepblllprim}) for the reduced bispectrum and the weight $w_{l_1l_2l_3}$ is described in (\ref{eq:lweightdiscrete}).  Here, the matrix $\Gamma^Q_{np}$ represents a late time inner product 
$\Gamma^Q_{np} = \langle\kern-2pt\langle\Qn,\,\Q_p\rangle\kern-2pt\rangle$ analogous to $\gamma_{np}= \langle\Qn,\,\Q_p\rangle$
in (\ref{eq:QQorthog}) (but with a different weight so that $\langle\kern-2pt\langle\Rn,\,\R_p\rangle\kern-2pt\rangle\ne \delta_{np}$). 
Determining the transformation matrix $\Gamma^{\cal Q} _{np}$  relating the $\aQn$ and $\bQn$ appears to be a complicated task but, in fact, it reduces to separable sums and integrals over the one-dimensional products $q_pq_r$ convolved with Bessel
and transfer functions, together with the final sum over the multipole domain (\ref{eq:tetrapydl}).  The latter is straightforward, 
especially in the continuum limit (\ref{eq:lweight}).   It need only be evaluated once, given a robust prior estimate for the power spectrum $C_l$'s.   

This discussion demonstrates that we can recover spectral information about the primordial shape function from 
the observational data through the relation
\eq
\aQn = \sum_p\({\Gamma^{\cal Q}}^{-1}\)_{np}\langle\bQp\rangle\,,
\qe
which extends to the orthonormal coefficients $\aRn$ using (\ref{eq:aRaQ}).   If the decomposition coefficients $\bQn$ are found
with adequate significance, we can reconstruct the shape function from a single realization through the expansion 
\eq
S(\kall) = \sum_{n,p}\({\Gamma^{\cal Q}}^{-1}\)_{np}\bQp\,\Qn\,.
\qe
We will discuss this prospect in more detail in the next section about the late-time CMB estimator where the relation 
between the $\aQn$ and $\bQn$ is more transparent.  

\subsection{CMB $\Fnl$ estimator}

\begin{figure}[p]
\centering
\includegraphics[width=.45\linewidth]{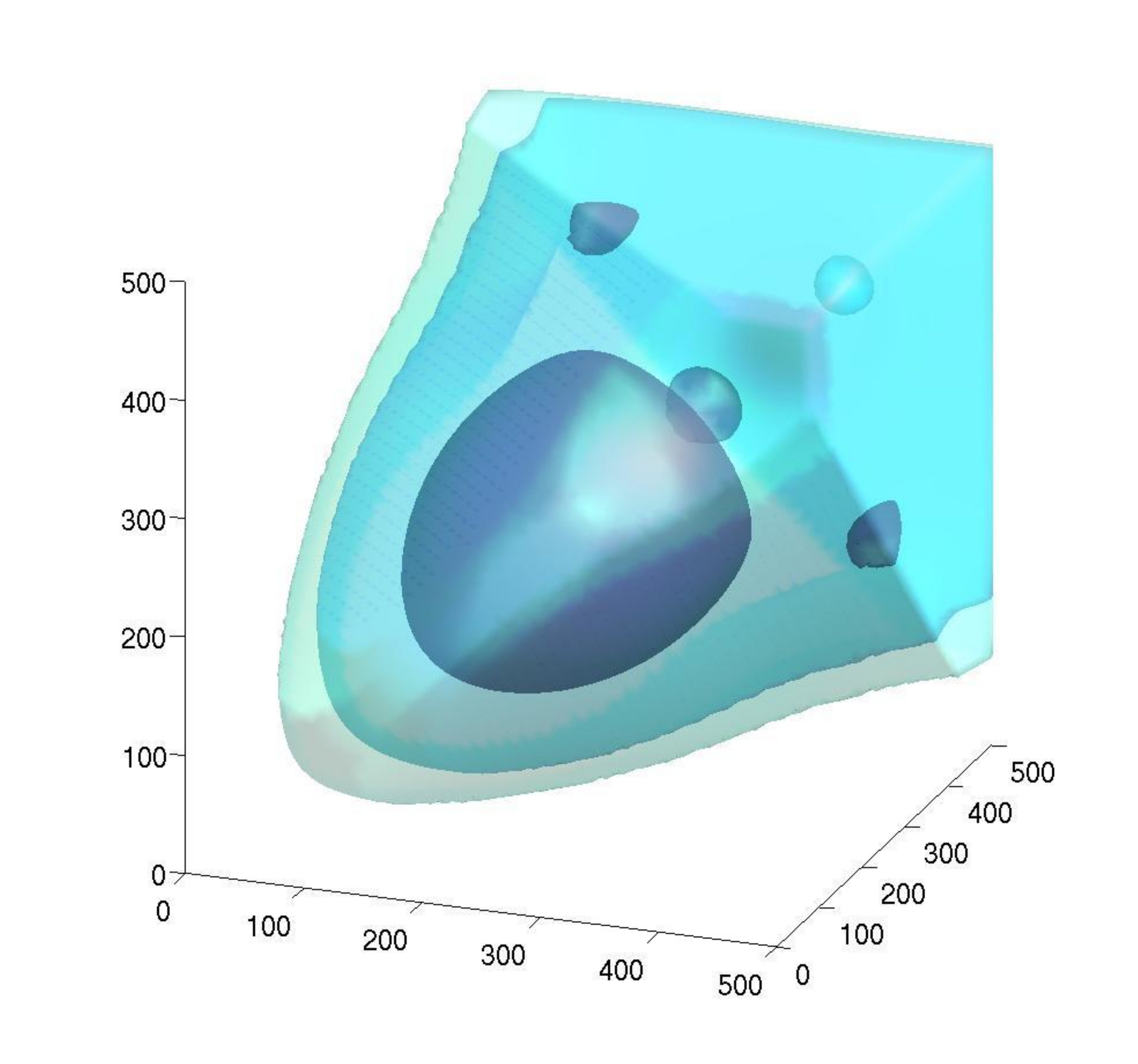}
\includegraphics[width=.45\linewidth]{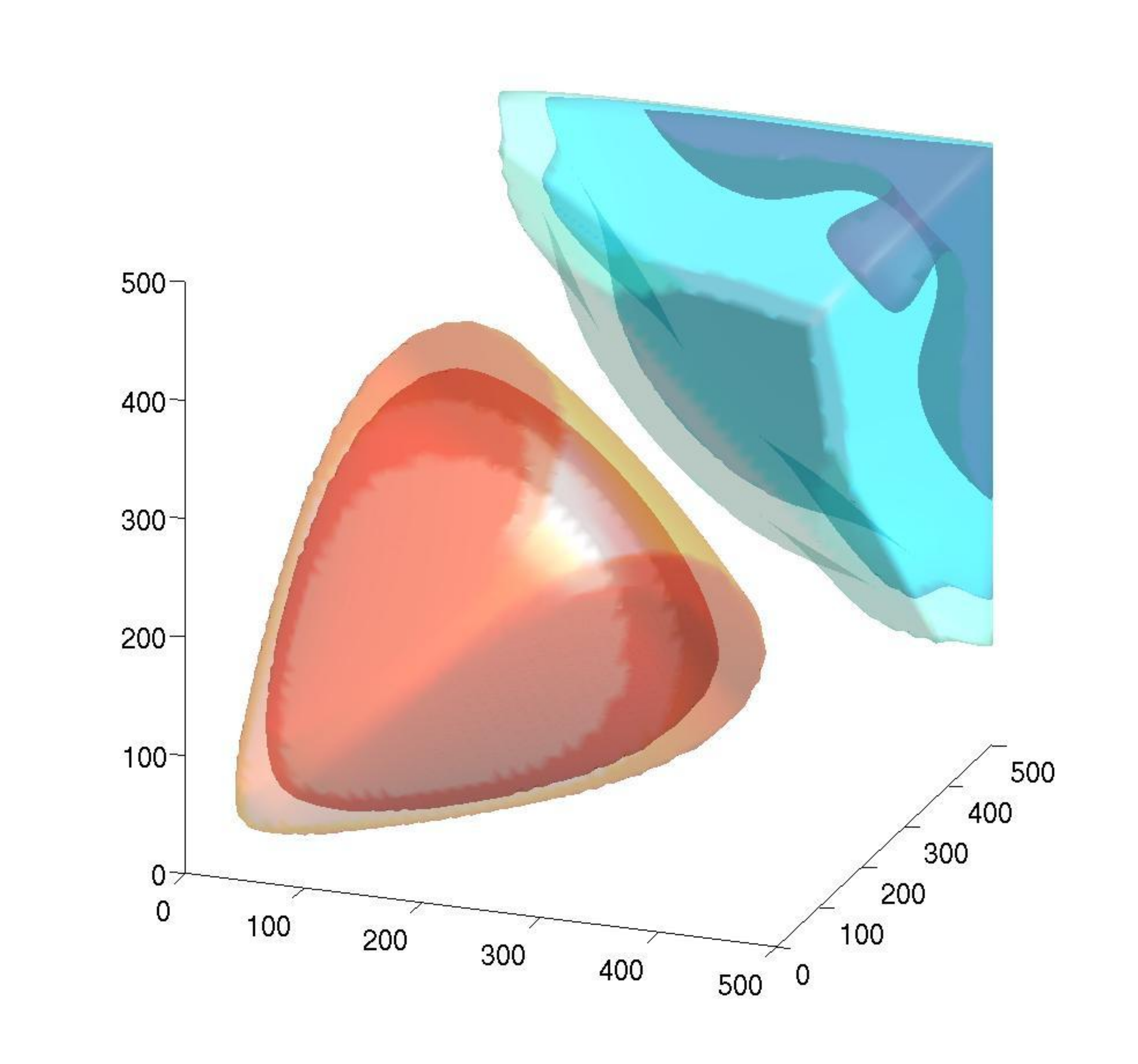}
\includegraphics[width=.45\linewidth]{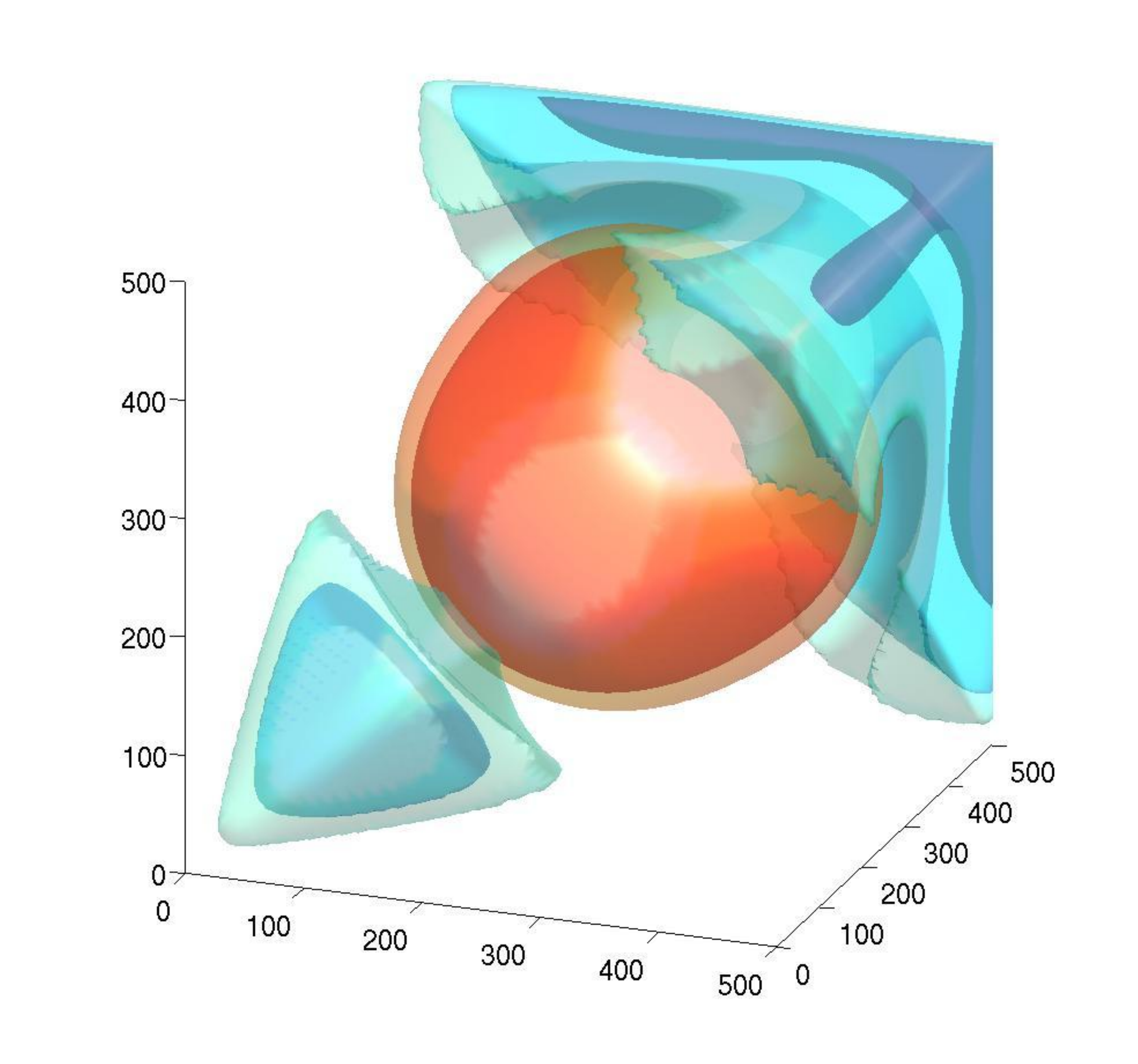}
\includegraphics[width=.45\linewidth]{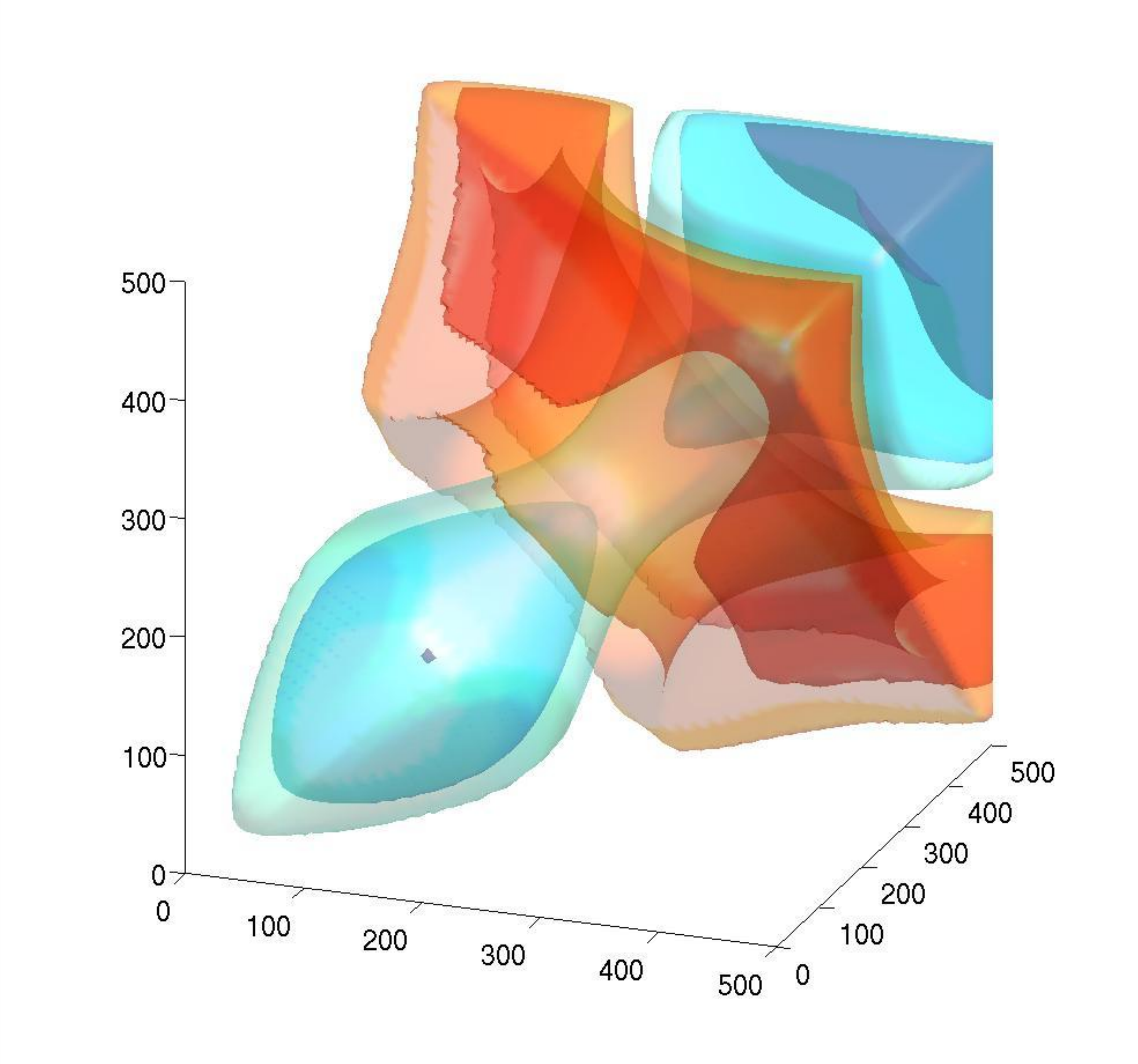}
\includegraphics[width=.45\linewidth]{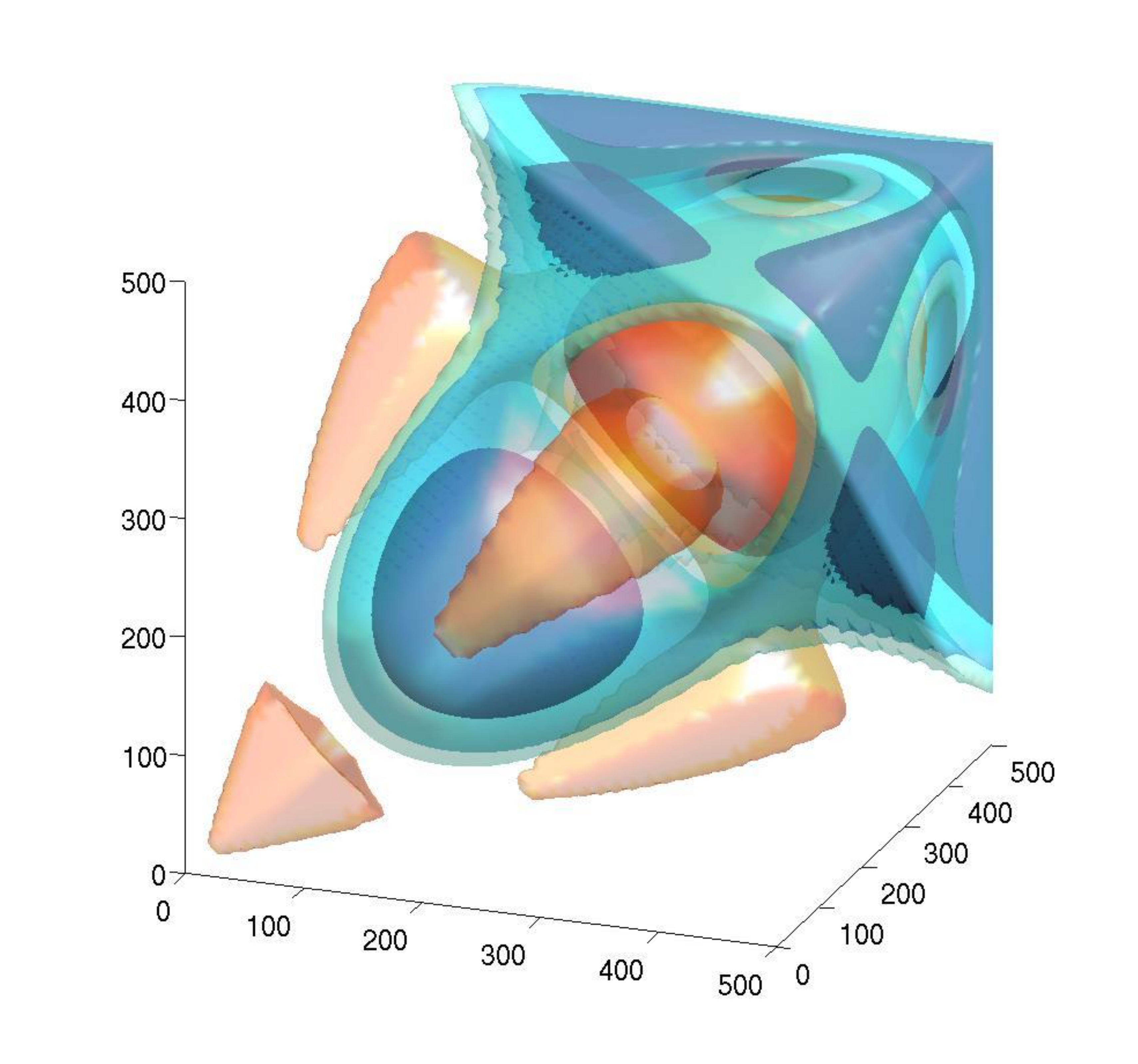}
\includegraphics[width=.45\linewidth]{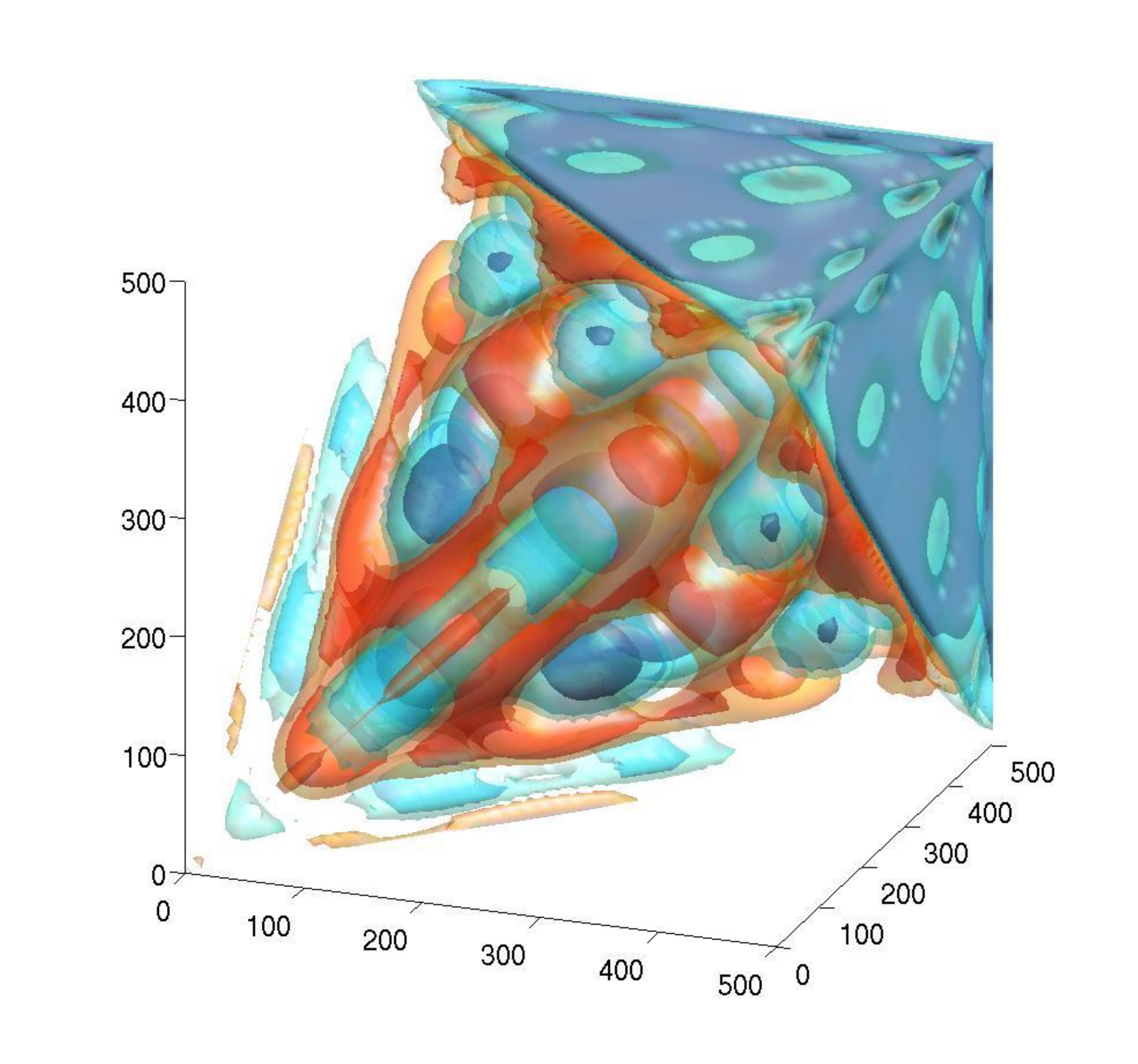}
\caption[CMB modes]{\small  Polynomials $\barRn$ on the tetrahedral domain (\ref{eq:tetrapydl}) used for
representing modes in the CMB estimator multiplied by  the 
weight function given in (\ref{eq:cmbestmodes}).   These are ordered just as in fig.~\ref{fig:shapemodes} with $\bar{\cal R}_0$ (top left),  $\bar{\cal R}_1$,  
$\bar{\cal R}_2$,
$\bar{\cal R}_3$,  $\bar{\cal R}_4$,  and $\bar{\cal R}_{41}$ (bottom right).  The last 
higher mode bears a superficial resemblance 
to the equilateral bispectrum in fig.~\ref{fig:cmbbispectra}.}
\label{fig:cmbmodes}
\end{figure}

We turn now to the implementation details of the late-time CMB estimator originally proposed in \cite{Fergusson:2006pr}.
Here, we presume that the CMB bispectrum $\blll$ for our non-separable primordial model is precomputed using 
eqn~(\ref{eq:sepblllprim}) or a robust hierarchical scheme \cite{Fergusson:2006pr,Fergusson:2008ra}.  In addition, this approach
can accommodate any late-time source of non-Gaussianity in the CMB, including secondary anisotropies, gravitational 
lensing, active models such as cosmic strings, and even systematic experimental effects. For the late-time analysis
we wish to expand the estimator functions using the  orthonormal $\barRn(\lall)$ and separable $\barQn(\lall)$ mode functions 
created out of products of the $\bar q_p(l)$ polynomials, for which we gave a concrete example (\ref{eq:Qlpolys}).  (Note
that we denote the multipole modes with a bar, distinguishing them from the primordial $q_p,\, \Qn,\,\Rn$ which are 
functions of wavenumber $k$).   Convergence of mode expansions on the multipole domain (\ref{eq:tetrapydl}) 
has been found to be poor for quantities as scale-dependent 
as $\blll$, so we choose to decompose the estimator functions directly as 
\eq \label{eq:cmbestmodes}
\frac{v_{l_1}v_{l_2}v_{l_3}}{\sqrt{C_{l_1}C_{l_2}C_{l_3}}} \, \blll = \sum_n \baQn \barQn\,,
\qe 
where the separable $v_l$ incorporates the freedom to make the weight function $w_{l_1l_2l_3}$ given in 
(\ref{eq:lweightdiscrete}) even more scale invariant (typically we shall use $v_l = (2l+1)^{1/6}$ as defined in (\ref{eq:lweightsep})).
The expression (\ref{eq:cmbestmodes}) means that we are effectively expanding in mode functions modulated 
by the $C_l$'s, that is, $\barQn \rightarrow \sqrt{C_l} \barQn/v_l$.   These more closely mimic the acoustic peaks observed
in the $\blll$ as illustrated in fig.~\ref{fig:cmbmodes}.    We shall see that the estimator expansion with $C_l$ in (\ref{eq:cmbestmodes})
is appropriate for primordial models, but different flatter choices will be more suitable for late-time anisotropy, such as that
from cosmic strings. 

We determine the implications for $\fnl$ of our mode expansion (\ref{eq:cmbestmodes}) by substituting into the 
estimator  (\ref{eq:estimator}), 
\eq\label{eq:cmbestsep}
{\cal E} &=& \sum_{l_i,m_i}\sum _{n\leftrightarrow prs}\kern-6pt \baQn\bar q_{\{p}\bar q_r \bar q_{s\}} \int d^2\hat {\bf n} \,Y_{l_2m_2}(\hat {\bf n})Y_{l_1m_1}(\hat {\bf n})\, Y_{l_3m_3} (\hat {\bf n})\,\frac{a_{l_1m_1}a_{l_2m_2}a_{l_3m_3}}{{v_{l_1}v_{l_2}v_{l_3}}\sqrt{C_{l_1}C_{l_2}C_{l_3}}}\\
&=& \sum _{n\leftrightarrow prs}\kern-6pt \baQn  \int d^2\hat {\bf n}\(\sum_{l_1,m_1} \bar q_{\{p} \, \frac{a_{l_1m_1} Y_{l_1m_1}}{v_{l_1} 
\sqrt{C_{l_1}}}\)\( \sum_{l_2,m_2} 
\bar q_{r} \, \frac{a_{l_2m_2} Y_{l_2m_2}}{v_{l_2} \sqrt{C_{l_2}}}\)\(
 \sum_{l_3,m_3} \bar q_{s\}} \, \frac{a_{l_3m_3} Y_{l_3m_3}}{v_{l_3} 
\sqrt{C_{l_3}}}\) \,,
\qe 
where again we assume the correspondence between the label $n$ and an ordered list of permuted triples $\{prs\}$, 
through $\barQn= \bar q_{\{p}\bar q_r \bar q_{s\}}$.   As previously for the primordial estimator (\ref{eq:primestsep}), we note 
that the sum between the $\bar q_p(l)$ and the $a_{lm}$ creates filtered versions of the original CMB map defined by 
\begin{align}\label{eq:barmapfilter}
\bar M_p(\un) = \sum_{lm} q_p(l)\frac{a_{lm}}{v_l\sqrt{C_l}} Y_{lm}(\un)\,,
\end{align}
which are multiplied together in (\ref{eq:cmbestsep}) to form the product map
\begin{align}
\bar{ \curl{M}}_n(\un) = \bar M_p(\un)\bar M_r(\un)\bar M_s(\un)\,.
\end{align}
Integrating over directions, we can obtain the  map mode expansion coefficient
\begin{align}
\b_n = \int d^2\un \curl{M}_n(\un)\,.
\end{align}
Thus the estimator reduces again to diagonal form 
\begin{align}\label{eq:cmbestimator}
\curl{E} = \frac{1}{N} \sum^{\nmax}_{n=0} \baQn \bbQn\,.
\end{align}
Like (\ref{eq:primestimator}), it consists entirely of separable sums and tractable integrals which can 
be performed rapidly.

As before, the separation of the estimator into two $\baQn$ and $\bbQn$ halves indicates that this approach 
could offer more direct information about the bispectrum than just $\fnl$ for one model.   Consider the expectation 
value of $\bbQn$ from an ensemble of maps with a given $\blll$ (and $\Cl$) expanded as (\ref{eq:cmbestmodes}).   
Following the steps used to derive (\ref{eq:betaprim}), we find a considerably simpler expression after substituting 
(\ref{eq:cmbestmodes}): 
\eq
\langle \bbQn\rangle &=& \int d^2\un \int x^2 dx \, \langle\bar {\curl{M}}^{\cal Q}_n(\un,x) \rangle
~= ~\sum_{l_i,m_i}  \frac{\bar q_{\{p}(l_1)\bar q_r(l_2) \bar q_{s\}}(l_3)}{{v_{l_1}v_{l_2}v_{l_3}}\sqrt{C_{l_1}C_{l_2}C_{l_3}}}\left(   \curl{G}^{\,l_1 ~l_2~ l_3}_{m_1 m_2 m_3}  \right)^2   {\blll}\\
&=&\sum_{l_i,m_i}  \frac{w_{l_1l_2l_3} \barQn(\lall)}{{v_{l_1}v_{l_2}v_{l_3}}\sqrt{C_{l_1}C_{l_2}C_{l_3}}}
\sum _p \baQp \barQp(\lall) \frac{\sqrt{C_{l_1}C_{l_2}C_{l_3}}}{{v_{l_1}v_{l_2}v_{l_3}}}\\
&=& \sum _p \baQp \sum_{l_1l_2l_3}\bar w_{l_1l_2l_3}\barQn\barQp ~= ~ \sum_p \bar\Gamma_{np}\baQp\,,
\qe
where the modified weight function $\bar w_{l_1l_2l_3}$ is given in (\ref{eq:lweightgeneral}) and $\Gamma _{np} = 
\langle \barQn,\,\barQp\rangle$ as discussed previously.  Hence, the estimator, when applied to a map containing the bispectrum defined by $\baQn$, should have the expectation value
\eq
\langle {\cal E}\rangle = {1\over N} \sum_n\sum_p \baQn \bar\Gamma_{np}\baQp\,.
\qe
Now rotating to our orthonomal basis $\barRn$, we note that from the relation (\ref{eq;Parsevalmixed}) we can 
deduce the simple and elegant form
\eq
\langle {\cal E}\rangle =  \frac{1}{ N} \sum_n {\bar{\aRn}}^2\,.
\qe
That is, we expect the best fit $\bbRn$'s for a particular realization to be the $\baRn$'s themselves.    
The simplicity of this result is not unexpected, 
since it would be obtained by correlating a bispectrum decomposed into the $\barRn$ with itself.   The advance here 
is that extracting the spectrum $\bbRn$ from the observed map would be intractable for large $\lmax$, were it not for
the transformation made to a non-orthogonal separable frame.   Assuming the coefficients $\bbRn$ are measured with 
some significance from a particular experiment, we can go further and reconstruct the map bispectrum using (\ref{eq:cmbestmodes})
\eq \label{eq:laterecon}
\blll = \frac {\sqrt{C_{l_1}C_{l_2}C_{l_3}}}{v_{l_1}v_{l_2}v_{l_3}} \sum_n \bbRn \barRn\,.
\qe
We reiterate that the viability of this fast and general reconstruction scheme \cite{Fergusson:2006pr} depends on two key factors, 
first, the smoothness of the reduced bispectrum $\blll$, requiring few modes to characterise it, and, secondly, on the 
completeness of the orthonormal basis from which the separable expansion was obtained.  We note that this methodology
can be applied using any complete mode expansions, beyond the polynomial examples given here, as well as 
with over-complete decompositions, such as wavelets, or with binning.  In the next section, we will demonstrate 
the efficacy of this method with simulated maps (for a sufficiently large $\fnl$), recovering the expected $\baRn$ spectrum and 
the main distinguishing features of the bispectrum $\blll$.

\subsection{Observable $\Fnl$ normalization}

In previous work  \cite{Fergusson:2008ra}, we pointed out the shortcomings of normalising the quantity $\fnl$ using the conventions employed to date in the literature (see also \cite{0509029}). At present, the central point in the primordial shape function defined in (\ref{eq:shapefn})  is normalised to unity assuming scale invariance, that is, $S(k,k,k)=1$ with no $k$-dependence.  This produces inconsistent results between models peaking or dipping at this central point (actually along this line); contrast the factor of 7 between the quoted variances of the equilateral and local models for exactly this reason.  Furthermore, the definition is not well-defined for models which are not  scale-invariant, such as feature models, and it is simply not applicable nonGaussian signals created at late times, such as those induced by cosmic strings or secondary 
anisotropies.    

We, therefore, propose a universally 
defined bispectrum non-Gaussianity parameter $\Fnl$ which (i) is a measure of the 
total observational signal expected for the bispectrum of the model in question 
and (ii) is normalised for direct comparison with the canonical local model 
(in particular, with $\Fnllocal = \fnllocal$ for a given $\lmax$).  
We presume that we have an unnormalised CMB bispectrum $\blll$ accurately 
calculated for a specific theoretical model over the whole observationally relevant domain $l\le \lmax$.   This 
can be achieved for any model using the separable mode expansion (\ref{eq:sepblllprim}) 
or hierarchical methods \cite{Fergusson:2008ra}.   We then define $\Fnl$ from an adapted version 
of the estimator (\ref{eq:estimator}) with 
\begin{align} \label{eq:normestimator}
\Fnl = \frac{1}{\,N{\bar N}_\textrm{loc}} \sum_{l_i m_i} \curl{G}^{l_1 l_2 l_3}_{m_1 m_2 m_3}
\blll \frac{a_{l_1 m_1} a_{l_2 m_2} a_{l_3 m_3}}{C_{l_1} C_{l_2} C_{l_3}}\,,
\end{align}
where $N$ is the appropriate normalisation factor for the given model,
\begin{align}\label{eq:normmodel}   
N^{\,2} = \sum_{l_i} \frac{B_{l_1 l_2 l_3}^2}{C_{l_1}C_{l_2}C_{l_3}}\,,
\end{align}
and  $\bar{N}_\textrm{loc}$ is the normalisation for the local model \textit{with} $\fnl = 1$, 
\begin{align}\label{eq:normlocal}
{\bar N}_\textrm{loc} ^{\,2}= \sum_{l_i} \frac{{B_{l_1 l_2 l_3}^{\textrm{loc}\,(\fnl\textrm{=}1)}}^2}{C_{l_1}C_{l_2}C_{l_3}}\,.
\end{align}
This $\Fnl$ estimator will certainly recover the usual $\fnl$ for the local model, but it is also clear that it will also equitably 
compare the total integrated observational bispectrum with that obtained from the $\fnl=1$ local model.    Of course, these definitions 
presume a sum to a given $l=\lmax$ (which should be quoted) but results for primordial models should not depend strongly on this cut-off, 
unless scale-invariance is broken. In any case, diffusion from  the transfer functions means that the primordial signal is dying out 
beyond $l \gtrsim 2000$, so we propose a canonical cut-off at $\lmax =2000$ (which is also relevant in the medium term for the 
Planck experiment).   Late-time anisotropies,  such as cosmic strings, do not generically fall-off exponentially 
for $l\gtrsim 2000$, but meaningful comparisons to the local $\fnl$$=$$1$ model can be made with the same definition (\ref{eq:normestimator})
on this domain, and alternative measures can be proposed elsewhere.   In principle, the normalised estimator (\ref{eq:normestimator}) can also be adapted as
a gross measure of the total bispectral signal over the given domain, irrespective of the possible 
underlying physical model.  For example,  using the reconstruction from 
Parseval's theorem (\ref{eq:Parseval}), the estimator provides a measure of $\Fnl^2$ which should then be normalised
relative to the total expectation for the local model with $N = N_{\rm loc}$ in (\ref{eq:normestimator}).  

If the CMB bispectrum $\blll$ is not known precisely for the primordial model under study, then the normalisation 
factor $N$ in (\ref{eq:normmodel}) can still be estimated using the shape function $S(\kall)$.   
Primordial and CMB correlators are closely related, so one can obtain a fairly accurate approximation to 
 the relative normalisations above (\ref{eq:normmodel}-\ref{eq:normlocal}) from  \cite{Fergusson:2008ra}
\begin{align}\label{eq:normalise}
\tilde N ^{\,2} = \int_{\curl{V}_k} S^{\,2}(k_1,k_2,k_3)\,w(k_1,k_2,k_3)\, d{\curl V}_k\,,
\end{align}
where the appropriate weight function was found to be 
$w(k_1,k_2,k_3) \approx 1/(k_1+k_2+k_3)$ and the domain ${\curl{V}_k}$ is given by $\kall \le\kmax(\lmax)$ (refer
to the discussion before (\ref{eq:shapecorrelator}) in section II).  Here, we note that  $N/\bar N_{loc\, \fnl=1} \approx 
\tilde N/\tilde N_{loc\, \fnl=1}$.
Using this primordial shape function normalisation $\tilde N$   in ref.~\cite{Fergusson:2008ra} led to a comparable definition of  $\barfnl\approx \Fnl$, 
which can be useful for making fairly accurate projections of nonGaussianity or for 
renormalising $\fnl$ constraints for different models into more compatible $\Fnl$ constraints.

Below we renormalise published and model-dependent constraints on $\fnl$ into the integrated bispectral measure $\Fnl$, using the expression (\ref{eq:normestimator}) together with accurate calculations of $\Blll$ for each model: 
\begin{align} 
\qquad\qquad-4 &< \fnl^{\rm local} < 80  \quad\,~~\mbox{\cite{Smith:2009jr}} & \&\Rightarrow & -4 &< \Fnl^{\rm local}  < 80\qquad\qquad \\
\qquad\qquad-125 &< \fnl^{\rm equil} < 435  \quad~\mbox{\cite{Senatore:2009gt}} & \&\Rightarrow & -24 &< \Fnl^{\rm equil} < 83 \qquad\qquad\\
\qquad\qquad-375 &< \fnl^{\rm warm} < 37 \quad~~\mbox{\cite{0701302}} & \&\Rightarrow & -93 &<\Fnl^{\rm warm} < 9 \qquad\qquad\\
\qquad\qquad-369 &< \fnl^{\rm ortho} < 71 \quad\,~~\mbox{\cite{Senatore:2009gt}} & \&\Rightarrow & -114 &<\Fnl^{\rm ortho}< 22\qquad\qquad
\end{align}
Note the much more consistent variance found for the different models with $\Fnl$, thus aiding direct comparison, 
as well as the exact correspondence for the local model to which it  is normalised.

\section{CMB map simulations for general bispectra}

\subsection{Map-making with separable shape functions and its limitations}

In the limit of weak non-Gaussianity, an algorithm to produce
non-Gaussian CMB simulations with a given power spectrum and bispectrum  for separable primordial shapes 
was described in ref.~\cite{0612571}.   Here, we present it in a more transparent notation, generalising the method
to non-separable shapes using the mode decompositions of the previous sections.  A byproduct is that the generalised
approach is more robust and reliable, because the polynomial mode functions are better behaved than the separable
approximations which have been previously employed. In this algorithm the non-Gaussian components of the CMB multipoles 
are obtained using the following formula:
\begin{align}\label{eq:almNG}
a_{l m}^{\rm NG} = 
\frac{1}{6} \sum_{l_i m_i} B_{l\, l_2 l_3} 
\(  \begin{array}{ccc} l & l_2 & l_3 \\ m & m_2 & m_3 \end{array} \)
  \frac{a_{l_2 m_2}^{\rm G *}}{C_{l_2}} \frac{a_{l_3 m_3}^{\rm G *}
  }{C_{l_3}} \; ,
\end{align}
where $a_{l m}^G$ is the Gaussian part of the CMB multipoles,
generated using the angular power spectrum $C_l$, while $B_{l\, l_2 l_3}$ is the given 
bispectrum of the theoretical model for which simulations are required. 
Although equation (\ref{eq:almNG})  is completely general, as before,  its numerical evaluation is
only computationally affordable for bispectra that can be
written in separable form. We have emphasised already that separability results in a
reduction of the computational cost of the estimator (\ref{eq:estimator}) from $O(l_{max}^5)$ to $O(l_{max}^3)$ 
operations; the same argument applies here allowing a rewriting of (\ref{eq:almNG}) into  
an equivalent form in pixel space (see below). 

The limitation dictated by separability is clearly overcome by using
our eigenfunction representation for the bispectrum (\ref{eq:Qshape}). The basic idea 
is to start by expanding an arbitrary bispectrum shape $S$ using the
separable polynomial decomposition $S_N$ until a good level of convergence is
achieved and then to substitute the mode decomposition into (\ref{eq:almNG}).
The accuracy of convergence is parametrized in terms of the
correlation $\bar {\cal C}(S,S_N)$ between the original non-separable shape and the eigenmode
expansion, as defined previously (\ref{eq:shapecorrelator}).   Note that this convergence can also be checked
more accurately using the full Fisher matrix correlation on the CMB bispectra ${\cal C}(\blll,\blll^N)$, 
calculated using the separable approach (\ref{eq:sepblllprim}) or else accurate hierarchical 
approaches \cite{Fergusson:2008ra}.   In previous sections (see fig.~\ref{fig:convergence}), we have noted 
how rapid this convergence 
is for well-behaved non-separable shapes, such as DBI inflation (or at late times with cosmic strings).   

In addition to the bispectrum separability requirement, there is an important further caveat which 
can prevent the straightforward implementation of the algorithm  (\ref{eq:almNG}).
By construction, terms $\curl{O}(f^2_{\rm NL})$ and higher are not explicitly
controlled. Following the discussion in \cite{Hanson:2009kg} we can
write the connected N-point functions as:
\begin{eqnarray}
\langle a^{*}_{l_1 m_1} a^{*}_{l_2 m_2} \rangle  & = & \left[ C_{l_1}
  + f_{\rm NL}^2 C_{l_1}^{NG} \right] \label{eq:clsim} \\ 
\langle a^{*}_{l_1 m_1} a^{l_2 m_2} a^{l_3 m_3} \rangle & = &
\left[f_{\rm NL} B_{l_1 l_2 l_3} + \curl{O} (f_{\rm NL}^3) \right] \\
\langle a^{l_1 m_1} a^{l_2 m_2} a^{l_3 m_3} \dots
a^{l_N m_N} \rangle & = & \curl{O} \left( f_{\rm NL}^3 \right) \; .
\end{eqnarray} 
Thus the condition that the map has the power spectrum $C_l$ specified
in the input will only be satisified if the power spectrum of the
non-Gaussian component in (\ref{eq:clsim}) remains small. Since 
this method does not control $\curl{O}(f^2_{\rm NL})$ terms, 
one has to ascertain that spuriously large $C_l^{NG}$ contributions do not affect the overall
power spectrum significantly. It turns out that this effect plagues current map
simulations if the standard separable expressions for the local and equilateral bispectra are
directly substituted into (\ref{eq:almNG}), as we now demonstrate. 

In section II, we showed how the reduced bispectrum could be written explicitly in separable
form for the local model $\bllllocal$  in (\ref{eq:localbispectsep}) and for the equilateral 
model $\blllequil$ in (\ref{eq:equilbispectsep}).   These were expressed in terms of one-dimensional 
convolution integrals between the transfer functions $\Delta_l(k)$ and powers of the power spectrum
$P(k)$, with $\balpha,\,\bbeta,\,\bgamma,\,\bdelta$ corresponding to $\hbox{const.},\, P(k),\,P(k)^{1/3},\,
P(k)^{2/3}$ respectively (refer to eqns (\ref{eq:ABterm}) and (\ref{eq:CDterm}).)
Just as we did with the $\fnl$ estimator (\ref{eq:mapfilter}), we can sum the $\alm^G$'s from the Gaussian 
maps with these functions 
to create filtered maps,
\eq\label{eq:lemapfilter}\textstyle
\nn \Malpha & \equiv  \sum_{lm} \balpha(x)\,a_{lm}^G
\frac{Y_{lm}(\hat{\mathbf{n}})}{C_l}\,,\qquad 
\Mgamma& \equiv  \sum_{lm} \bgamma(x) \,a_{lm}^G
{\textstyle\frac{Y_{lm}(\hat{\mathbf{n}})}{C_l}}\,, \\
\Mbeta & \equiv  \sum_{lm} \bbeta(x) \,a_{lm}^G
\frac{Y_{lm}(\hat{\mathbf{n}})}{C_l}\,,\qquad
\Mdelta& \equiv  \sum_{lm} \bdelta (x)\, a_{lm}^G
{\textstyle \frac{Y_{lm}(\hat{\mathbf{n}})}{C_l}}
\qe
From products of these maps in pixel space,  we can now obtain explicit expressions 
for the nonGaussian $\alm^{NG}$'s in these two separable cases (compare with the
bispectrum expressions (\ref{eq:localbispectsep}) and (\ref{eq:equilbispectsep})):
\begin{eqnarray}
  a_{l m}^{\rm local} & = &{\textstyle}{  \int dx x^2 \left[ \frac{2}{3} \bbeta(x)
    \int d^2 \hat{\mathbf{n}} \,Y^*_{lm}(\hat{\mathbf{n}})
      \Malpha \, \Mbeta + \frac{1}{3}\balpha(x)\int d^2 \hat{\mathbf{n}}\,
      Y^*_{lm}(\hat{\mathbf{n}}) \Mbeta\,\Mbeta
        \right] \,,}\label{eq:almNGlocal} \nn\\
  a_{l m}^{\rm equil} & = & {\textstyle}{2 \int dx x^2 \left[
      -2 \bbeta(x)\int d^2 \hat{\mathbf{n}}\,
             Y^*_{lm}(\hat{\mathbf{n}}) \,\Malpha\,\Mbeta
      -\balpha(x) \int d^2 \hat{\mathbf{n}}\,
            Y^*_{lm}(\hat{\mathbf{n}}) \,\Mbeta \,\Mbeta \right.}  \nonumber \\
      & & {\textstyle}{-2 \bdelta(x) \int d^2 \hat{\mathbf{n}}\,
            Y^*_{lm}(\hat{\mathbf{n}}) \, \Mdelta\,\Mdelta
      +2 \bgamma(x) \int d^2 \hat{\mathbf{n}}\,
            Y^*_{lm}(\hat{\mathbf{n}}) \,\Mbeta\,\Mdelta} \nonumber \\   
      & & {\textstyle}{+2 \bbeta(x)\left.\int d^2 \hat{\mathbf{n}} \,
         Y^*_{lm}(\hat{\mathbf{n}}) \, \Mgamma\,\Mdelta       
      +2 \bdelta(x) \int d^2 \hat{\mathbf{n}}\,
            Y^*_{lm}(\hat{\mathbf{n}}) \, \Mbeta\,\Mgamma
         \right] \,.}\label{eq:almNGequil} 
\end{eqnarray}

\begin{figure}[t]
\centering
\includegraphics[width=.65\linewidth]{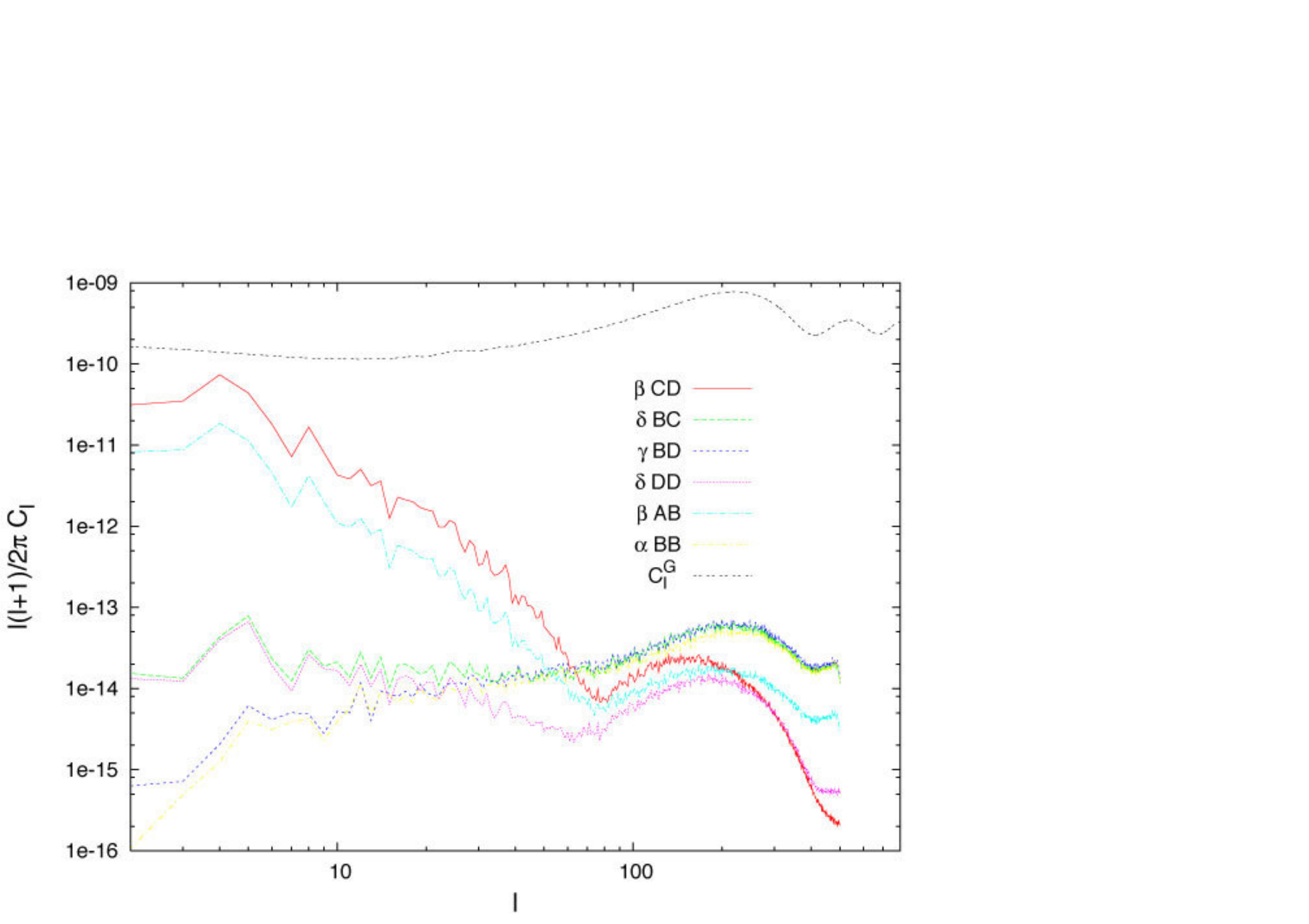}
\includegraphics[width=.65\linewidth]{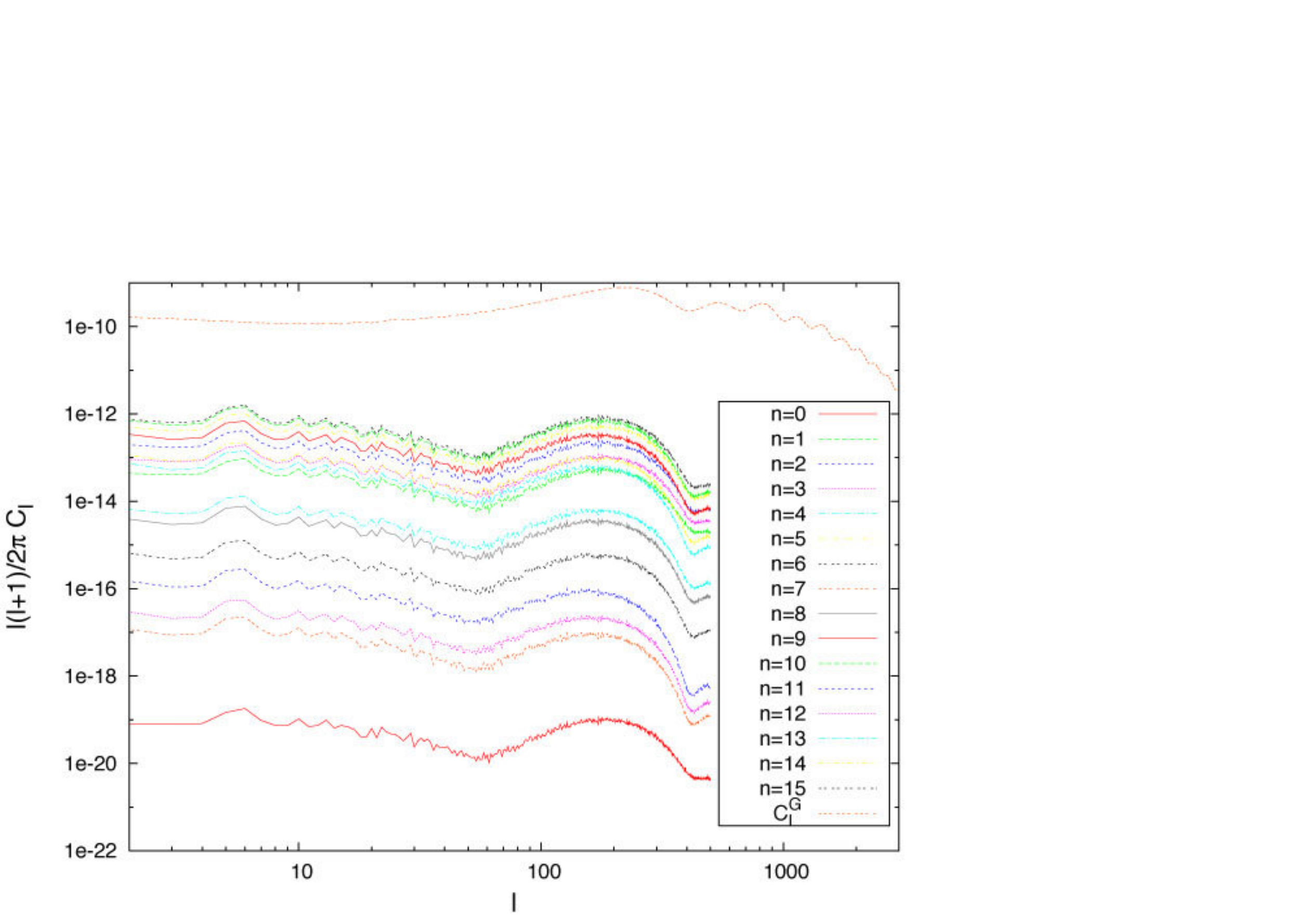}
\caption[Transfer functions]{\small Convergence properties of the 
standard separable functions used in the literature to represent local and 
equilateral models (top panel).   Here the functions have been convolved with 
the transfer functions $\Delta _l$  required in the bispectrum estimator or map-making 
algorithms.   Note the poor scaling and divergence at low $l$ for two of the separable 
combinations with the resulting power spectrum from non-Gaussianity rising to 
compete with the CMB power spectrum $\Cl$'s ($\fnl=1$).  This poor scaling is 
contrasted with results for the tetrahedral polynomials $q_n(k)$  (lower panel).  These remain 
bounded and roughly scale-invariant over the full multipole range, even for very high order polynomials. }
\label{fig:Qconvergence}
\end{figure}

In the top panel of fig.~\ref{fig:Qconvergence}  we consider the contribution to the final $C_l^{NG}$
from the various terms appearing in equation (\ref{eq:almNGequil})
taken separately.  For example,  we build a set of multipoles from
the term 
$\int dx x^2 \frac{2}{3} \bbeta(x) \int d^2 \hat{\mathbf{n}} \,Y^*_{lm}(\hat{\mathbf{n}})\,
\Malpha\,\Mbeta $ and compute the resulting power spectrum, neglecting all other terms, 
and so forth. We then compare the power spectra of the nonGaussian
part to the input power spectrum of the Gaussian part for $f_{\rm NL}
= 100$. Our procedure underlines what was pointed out in
\cite{Hanson:2009kg}: some terms in the separable approximations to both
the local and equilateral shapes produce spurious
divergences at low l's that are large enough to affect the final power
spectrum of the map. More precisely, as can be seen in fig.~\ref{fig:Qconvergence},
the biggest problems come from the terms 
$\int dx x^2  \frac{2}{3} \bbeta(x) \int d^2 \hat{\mathbf{n}} \,Y^*_{lm}(\hat{\mathbf{n}})\,
\Malpha\,\Mbeta$ and  
$\int dx x^2 \bbeta(x) \int d^2 \hat{\mathbf{n}}\,
Y^*_{lm}(\hat{\mathbf{n}}) \,\Mgamma\,\Mdelta$. In ref.~\cite{Hanson:2009kg}, it was
pointed out that the problem can be circumvented for the local model by modifying the
expression (\ref{eq:almNGlocal}) of $a_{lm}^{\rm local}$ so as to eliminate 
the pathological term, while leaving the final bispectrum of the map preserved with a change of
weight for the remaining term. The same approach can also be applied to the equilateral case, 
leaving new tailored expressions for the nonGaussian parts:
\begin{eqnarray}
  a_{l m}^{\rm local} & = & \int dx x^2 \, \balpha(x)\int d^2 \hat{\mathbf{n}}\,
      Y^*_{lm}(\hat{\mathbf{n}}) \Mbeta\,\Mbeta
        \,,\label{eq:newalmNGlocal} \nn\\
  a_{l m}^{\rm equil} & = & 2 \int dx x^2 \left[
      -3\balpha(x) \int d^2 \hat{\mathbf{n}}\,
            Y^*_{lm}(\hat{\mathbf{n}}) \,\Mbeta \,\Mbeta \right.
      -2 \bdelta(x) \int d^2 \hat{\mathbf{n}}\,
            Y^*_{lm}(\hat{\mathbf{n}}) \, \Mdelta\,\Mdelta\nonumber \\
      & &\left.\qquad\qquad\quad+2 \bgamma(x) \int d^2 \hat{\mathbf{n}}\,
            Y^*_{lm}(\hat{\mathbf{n}}) \,\Mbeta\,\Mdelta
               \right] \,.\label{eq:newalmNGequil} 
\end{eqnarray}
It is easy to verify that these modified expressions produce the
correct bispectra in the final maps, they are
numerically stable and so allow the simulation of non-Gaussian maps of
the local and equilateral type with given power spectrum and
bispectrum. However, one can see how the necessity of looking at all 
the individual terms in the equations defining $a_{lm}^{NG}$, and the need
to produce suitable modifications of the original formulae, means that the 
algorithm loses its generality. If additional shapes are considered then, in principle, 
different separation schemes could well encounter the problems outlined
above. The good news is that the full generality of this approach is
regained when the separation of the original shape is done using the
eigenmode expansion introduced in this paper. 

\subsection{Map-making from arbitrary primordial shape functions}

In order to see why this
happens, it
is useful to write down the equation for $a_{lm}^{NG}$ in terms of our
polynomial expression. Since we can decompose the three-point
functions both at early and late times it is actually possible to
generate a map in two different way. The closest method to the
``standard'' one, just outlined above, is the one that start from the
early time decomposition. In this case the primordial shape
$S(k_1,k_2,k_3)$ is
written as:
\begin{align}
S(k_1,k_2,k_3) = \sum_n \aQn \Qn = \sum_{pqr} \alpha^{\scriptstyle\cal Q}_{pqr} q_p(k_2)
q_q(k_1) q_r(k_1) \; ,
\end{align}
where the $\Qn(\kall)$ are formed from products of the tetrahedral polynomials $q_p(k)$ given in (\ref{eq:Qpolys})  and the $\aQn \leftrightarrow \alpha^{\scriptstyle\cal Q}_{pqr}$ are the coefficients 
of the eigenmode expansion for a given shape (recall the convenience of ordering the $pqr$ with a single label $n$).  
The reduced angular bispectrum $\blll$ is obtained, as was shown in (\ref{eq:sepblllprim}), by linearly projecting the
primordial shape on the sphere using  radiation
transfer functions:
\begin{eqnarray} \label{eq:sepblllprim2}
\nn b_{l_1 l_2 l_3} & =& \D_\O^2 f_{NL} \sum_{n} \aQn \int  x^2 dx 
\left[\({ \frac{2}{\pi} }\int d k_1\,  q_p(k_1) \, \D_{l_1}(k_1) \, j_{l_1}(k_1x) \)
\( {\frac{2}{\pi}} \int d k_2 \,  q_r(k_2) \, \D_{l_2}(k_2) \, j_{l_2}(k_2x) \)\right.\\
&& \qquad \qquad \qquad \qquad \qquad ~~\left.\times \( {\frac{2}{\pi}} \int d k_3 \,  q_s(k_3) \, \D_{l_3}(k_3) \, j_{l_3}(k_3x) \) ~+ ~\mbox{5~ permutations}\right]\\
\nn &=& \D_\O^2 f_{NL} \sum_{n} \aQn \int x^2dx \, q_{\{p}^{\,l_1} q_{r}^{\,l_2} q_{s\}}^{\,l_3} \,, \qquad \hbox{where} \quad
q_{p}^{\,l}(x) = \frac{2}{\pi} \int d k\,  q_p(k) \, \D_{l}(k) \, j_{l}(kx)\,,
\end{eqnarray}
Substituting equation (\ref{eq:sepblllprim2}) into \ref{eq:almNG}, and using the standard technique of decomposing
the integrals into tractable products of one-dimensional integrals, after some algebra, we obtain the general
expression for $a_{lm}^{NG}$:
\begin{align}\label{eq:almNGsplit}
a_{lm}^{NG} = \frac{1}{18} \sum_{pqr\leftrightarrow n}\kern-6pt \aQn \int dx x^2 q_{p}^l(x)  \int d^2\hat{\mathbf{n}} 
Y^{m*}_l(\hat{\mathbf{n}}) \,M^G_q(\un,x) \, M^G_{r}(\un,x)\,,
\end{align}
where the $M_p^G(\un,x)$ are filtered maps found by summing a set of Gaussian $\alm^G$'s with the convolved
tetrahedral polynomial $q^l_p$ (refer to eqn (\ref{eq:mapfilter})
\eq\label{eq:Gmapfilter}
M^G_p(\un,x) =  \sum_{lm} q_p^{\,l}\,\, \frac { a^G_{lm} Y_{lm} }{C_{l}} ~=~ \sum_{lm}  \[\frac{2}{\pi} \int q_p(k) \D_l(k) j_l(kx) dk \] \frac{a^G_{lm} Y_{lm}(\un)}{C_l}\,.
\qe
The $M^G_p(\un,x)$ and $q_p^l(x)$ are now the analogues of the $\Malpha,\,\Mbeta,\,\Mgamma,\,\Mdelta$ 
and $\balpha,\bbeta,\bgamma,\bdelta$ defined above.  Note that it is not strictly necessary here to include 
cyclic permutations in (\ref{eq:almNGsplit}) running over the indices $\{p,q,r\}$, as these are incorporated automatically.  Further efficiencies can be achieved by exploiting the freedom to reorder terms in(\ref{eq:almNGsplit}), taking out the polynomial $q_p$ of highest order
and convolving the maps with the two lower order polynomials; this is not necessitated by stability requirements (see below).

\begin{figure}[p]
\centering
\includegraphics[width=.75\linewidth]{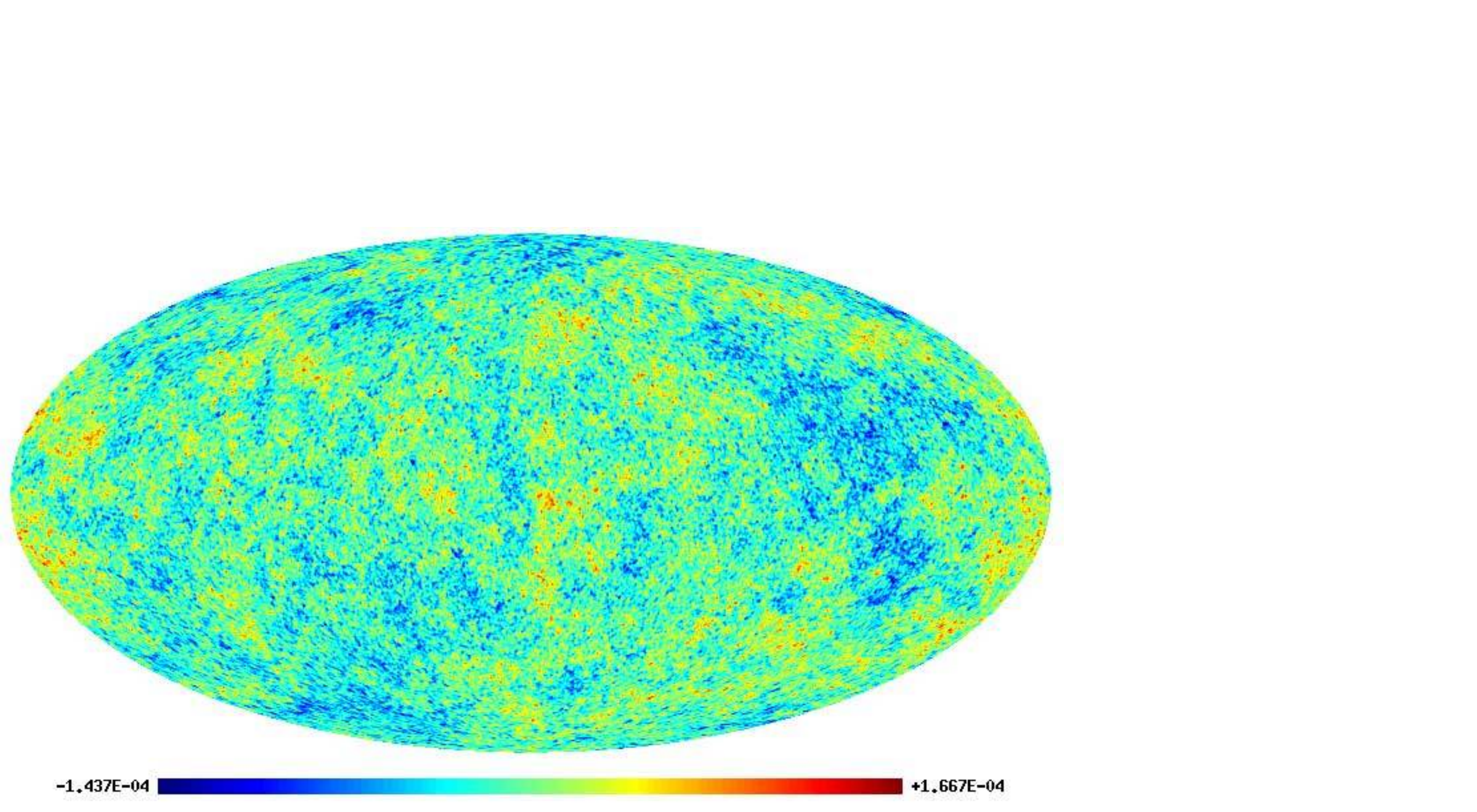}
\includegraphics[width=.75\linewidth]{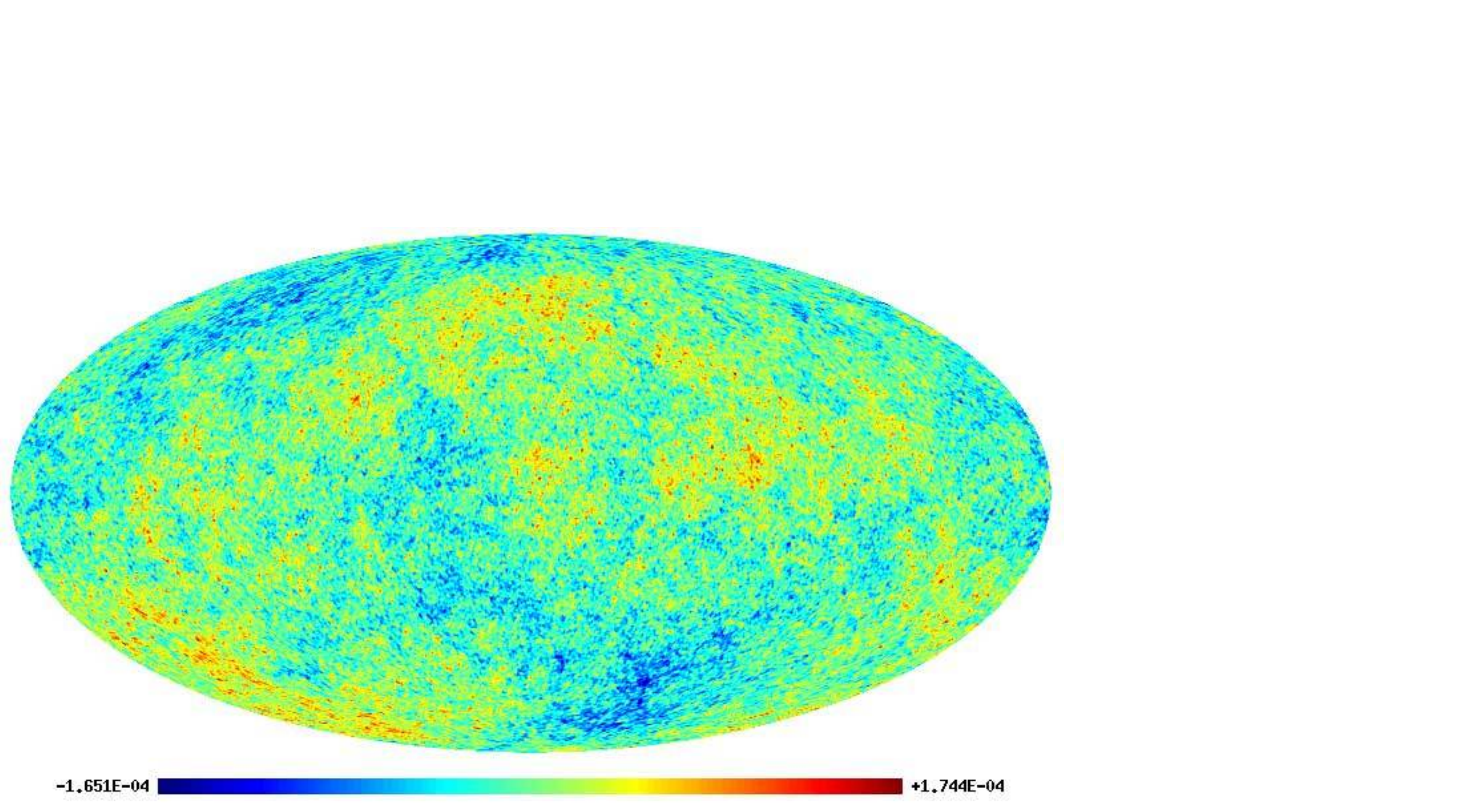}
\includegraphics[width=.75\linewidth]{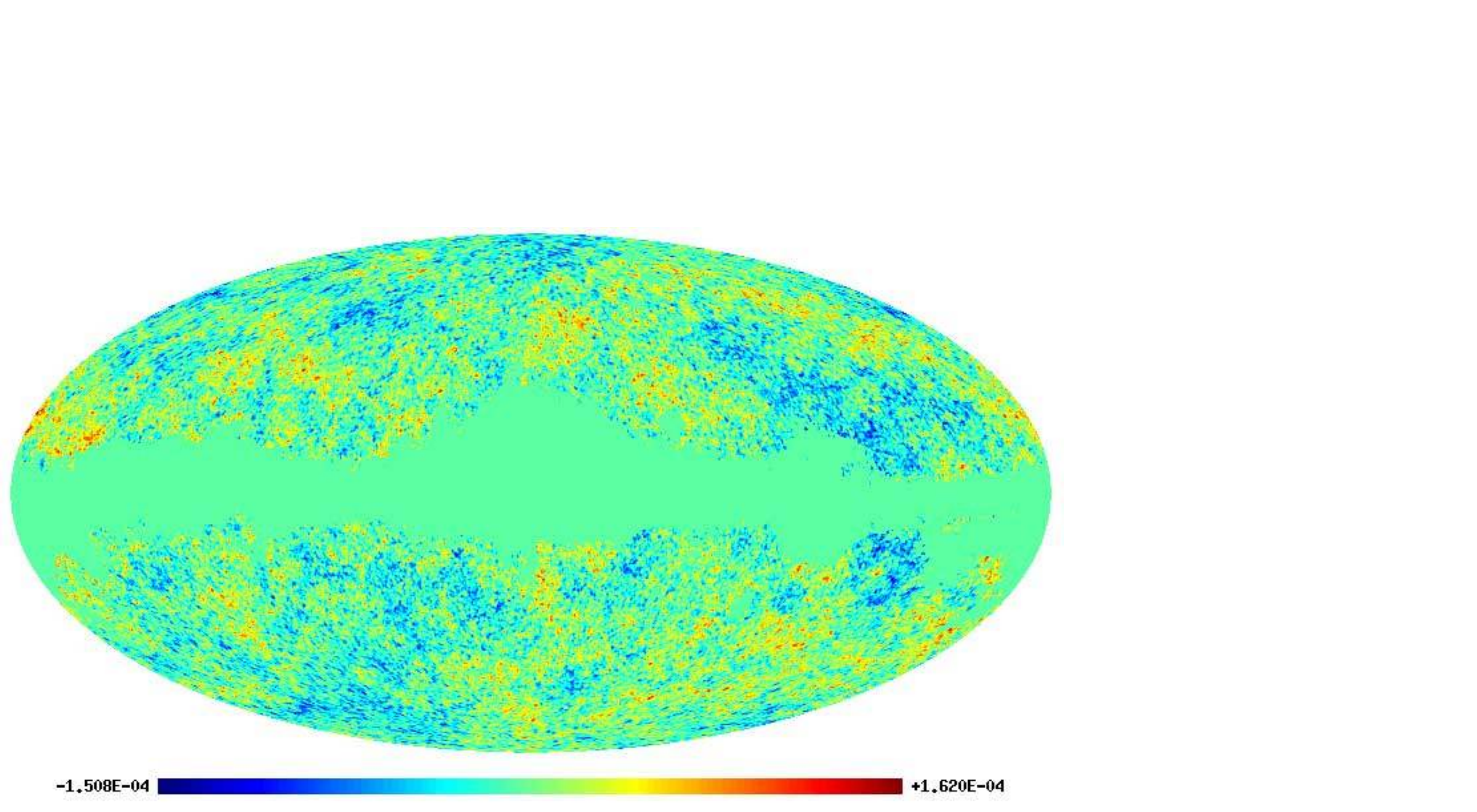}
\caption[Non-Gaussian primordial maps from equilateral model]{\small  Simulated nonGaussian CMB maps
from the equilateral model, created using the primordial map-making method (\ref{eq:almNGsplit}).  The upper panel shows
a map simulation with $\fnl = 400$ (barely discernable from the underlying Gaussian template), the middle panel shows a map
with a large NG signal with $\fnl=4000$, while the lower panel shows the $\fnl =400$ case above in a WMAP-realistic context using 
the KQ75 mask and with inhomogeneous noise added.}
\label{fig:equilmaps}
\end{figure}

In principle, the numerical instabilities which cause problems for the standard separable approximations, could now 
affect the angular integrals over the polynomials $q_p$ given in (\ref{eq:almNGsplit}). However, 
as shown in fig.~\ref{fig:Qconvergence}, this is not the case.
The key point is that all the functions $q_p^l(x)$ now scale as $\frac{1}{l(l+1)}$ (see fig.~\ref{fig:Qconvergence}), that is, in the 
same way as the non-pathological 
 $\bbeta(x)$ term in the standard local and equilateral decompositions. For this reason the spherical harmonic 
projection of a product of two $M^G_p(\un,x)$ maps is expected to have similar scaling properties as the term 
$\int d^2n Y^{m*}_l(\hat{\mathbf{n}}) \Mbeta^2$. This last integral was previously shown 
to be stable at low multipoles, as discussed for the local case in ref.~ \cite{Hanson:2009kg}).   Thus
all the integrals in equation (\ref{eq:almNGsplit}) are going to be well-behaved at low l's. Since the shape-dependent
information is in the coefficients of the expansion $\aQn$ and not in the precomputed $q_l^p(x)$ modes, we are 
able to produce numerically stable results for any possible shape. Numerical tests were carried out for 
the local and equilateral case, confirming the previous statements. We suggest, therefore, that the 
eigenmode expansion provides a 
numerically stable and efficient means by which to generalize the algorithm in ref.~\cite{0612571} to non-separable bispectrum shapes.

\subsection{Simulated maps from general CMB bispectra}

It is useful to recap the discussion above by using separable mode expansions 
to create simulated maps at late times from a given CMB power spectrum $C_l$ and  reduced bispectrum $\blll$. 
As before with the $\fnl$ estimator, removal of the convolution with transfer functions, makes the late-time 
method much simpler and more transparent.  We begin with the same expression
\begin{align}
a_{lm} = a^G_{lm} + f_{NL} a^{NG}_{lm}\,,
\end{align}
where
\begin{align}
a^{NG}_{lm} = \int d\un \sum_{l_1,l_2,m_1,m_2} b_{l_1 l_2 l_3}Y_{l_1 m_1}(\un)\frac{a_{l_2 m_2} Y_{l_2 m_2}(\un)}{C_{l_2}}\frac{a_{l_3 m_3} Y_{l_3 m_3}(\un)}{C_{l_3}}.
\end{align}

Now we expand the CMB bispectra using the eigenmode decomposition using weight functions motivated by the estimator (refer 
to (\ref{eq:cmbestmodes})
\begin{align}
\frac {v_{l_1}v_{l_2}v_{l_3}}{\sqrt{C_{l_1}C_{l_2}C_{l_3}}} b_{l_1 l_2 l_3} = \sum \a_n \bar{\curl{Q}}_n,
\end{align}      
where $v_l$ is a separable weight factor chosen to remove  scaling from the CMB bispectrum, improving decomposition 
convergence.  These weight factors are important for this late-time map-making method because they help remove the scaling 
of the  $\sqrt{C_l}$ term in the $\bar M^G_p(\un)$ filtered maps,  making there power spectrum flatter (the analogue 
of the problem discussed above for primordial map simulations). We can rewrite the non-Gaussian part as
\begin{align}\label{eq:latetimemaps}
a^{NG}_{lm} = \sum_{pqr\leftrightarrow n}\kern-6pt  \aQn \frac{q_{\{p}(l)}{\sqrt{C_l}v_l}  \int d\un Y_{l m}(\un) \bar M^G_q(\un) \bar M^G_{r\}}(\un) \,,
\end{align}
where the $\bar M^G_p(\un)$ are defined in (\ref{eq:barmapfilter}) and summed with Gaussian $\alm^G$'s.

\begin{figure}[th]
\centering
\includegraphics[width=.85\linewidth]{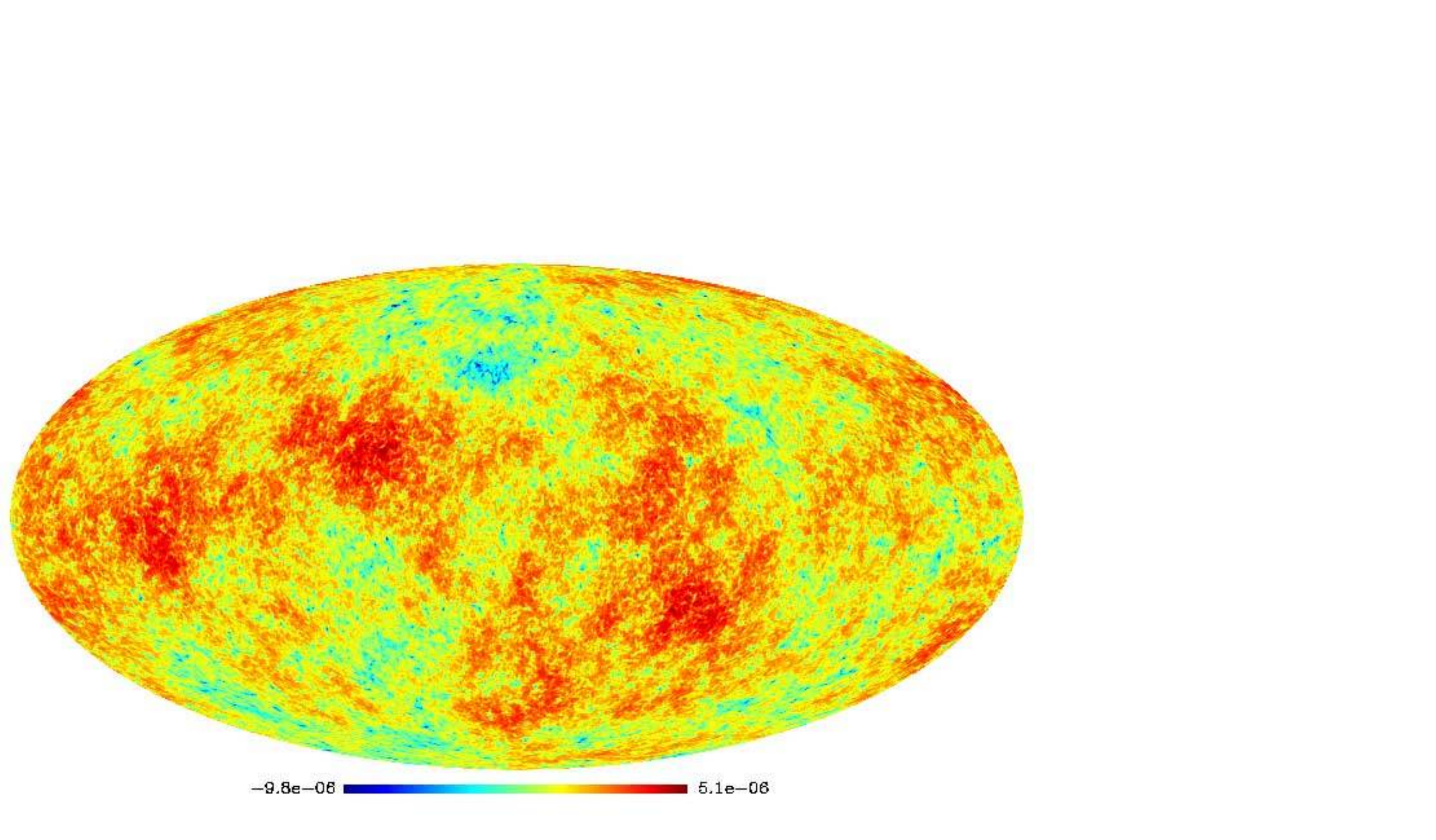}
\includegraphics[width=.85\linewidth]{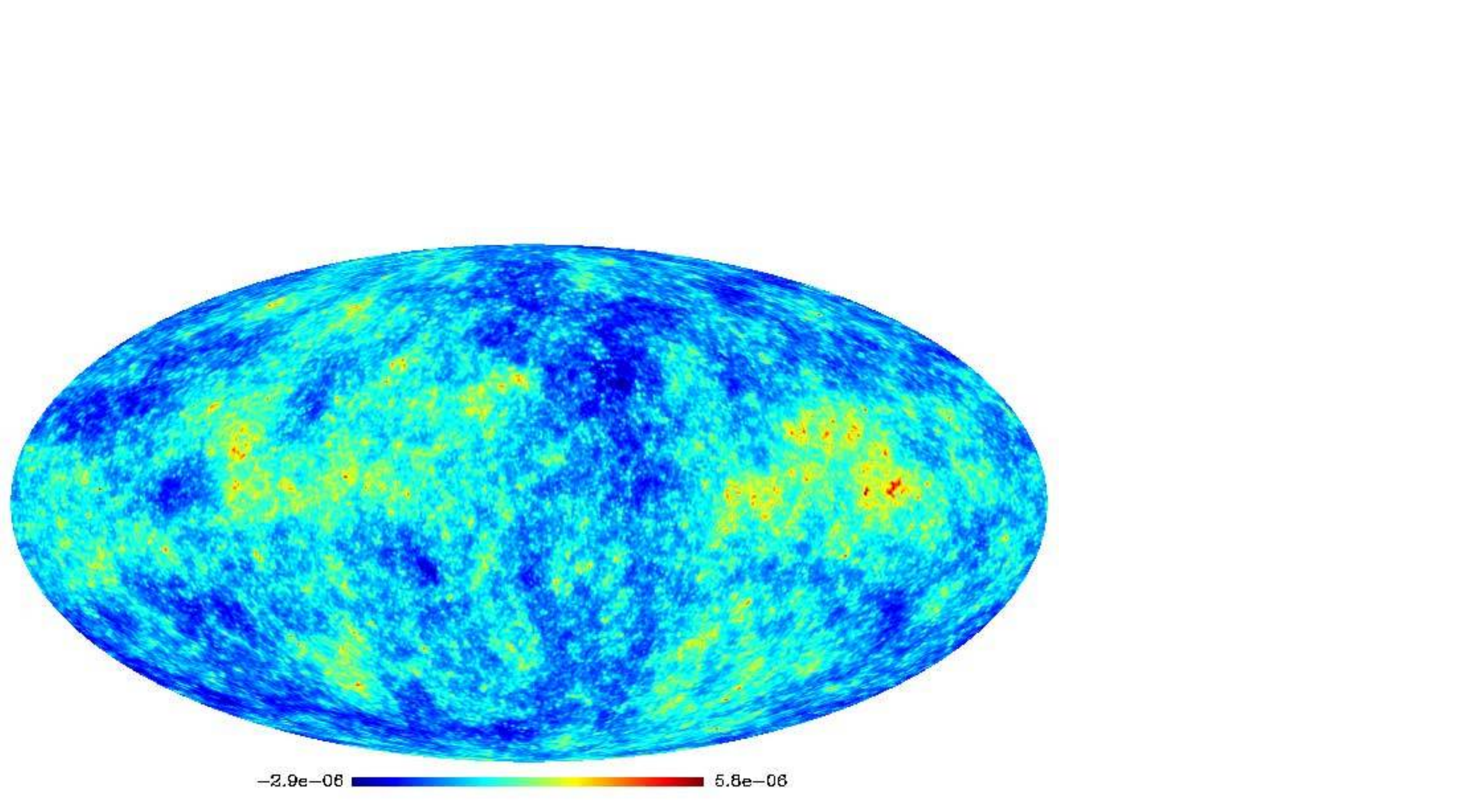}
\caption[Non-Gaussian maps from several different models]{\small  Simulated maps for nonGaussian models using the
late-time map-making method (\ref{eq:latetimemaps}); this only includes the $\alm^{\rm NG}$ contribution.  The upper panel shows a non-Gaussian CMB map 
from cosmic strings obtained using the analytic expression for the string bispectrum (\ref{eq:stringbispect}). The lower panel 
shows a simulated nonGaussian map for an equilateral model.  When added to its Gaussian counterpart map from 
$\alm^{\rm G}$ at an amplitude  $\fnl =600$,
this equilateral map was used for the bispectrum recovery illustrated in  figs.~\ref{fig:reconalpha}, \ref{fig:3dalpha}.   
Note the red colour cast from negative $\fnl$ and blue from positive.}
\label{fig:ngmaps}
\end{figure}

This method is straightforward to implement for a given theoretical $\blll$ and it is highly efficent.  For example, it can 
produce simulated maps in 64 seconds for $l=500$ with 16 eigenmodes (6 polynomials). It has the advantage that, as it depends only on the CMB  bispectra, it can also be used to simulate maps for bispectra produced by late time effects, like cosmic strings,
gravitational lensing and secondary anisotropies. Plots of the non-Gaussian part of simulated maps can be seen in fig.~\ref{fig:ngmaps}
for the non-separable DBI inflation and cosmic string models.

\section{Direct comparison of bispectrum estimators}

We have developed two complete numerical pipelines, implementing 
the eigenmode decomposition methods described in the previous sections. 
For a generic primordial shape $S(\kall)$ or a given CMB bispectrum $\blll$, 
an expansion in monomial symmetric polynomials $\Qn$ is performed 
followed by the generation of nonGaussian map simulations. Bispectrum
estimators are then applied to the map simulations in order to verify that the
input $f_{\rm NL}$ can be properly recovered together with the
expected variance. Both  `early time decomposition' and  `late
time decomposition' $\fnl$ estimators have been fully implemented. 
The former starts from an
expansion of the primordial shape $S(k_1,k_2,k_3)$ in Fourier space
while the latter starts from an expansion of the reduced angular
bispectrum $b_{l_1 l_2 l_3}$ in harmonic space, where in the second
case radiation transfer functions have already been included in the
expression for $b_{l_1 l_2 l_3}$. The redundancy provided by the two
alternative pipelines provide a further check of the reliability of the
final results. 

Since our purpose in this paper is to introduce the
eigenmode expansion method, and to test its implementation, we 
will primarily apply our pipelines to map simulations,
leaving detailed analysis of real datasets over a wider range of models 
for future publication \cite{inprep}.   As this is a proof-of-concept paper, 
we will mainly limit ourselves to the study of the simple equilateral 
family of models.  This is because it is already well-studied in the literature
(see e.g. 
\cite{0509029,0612571,Komatsu:2008hk}), which enables a useful 
comparison between the outcome of our numerical pipelines and
previously published results for the equilateral shape.  Moreover, 
the equilateral case does not require sophisticated noise analysis, unlike 
the local model.  However, we will 
briefly consider other non-separable models outlined earlier in the 
introduction, such as the related DBI 
model and the cosmic string bispectrum.   We note that from the point 
of view of the eigenmode decomposition, the formal separability of the 
equilateral shape is irrelevant; it does not cause early termination of
the expansion series which is nearly identical 
to the non-separable DBI model (see fig.~\ref{fig:Requildecomp}). 
Having established the reliability of the eigenmode expansion method here, 
in a forthcoming publication \cite{inprep} we will apply it to the study 
of families of non-separable shapes 
using WMAP5 data.

\subsection{Simulated observational maps}

\begin{figure}[t]
\centering
\includegraphics[width=.85\linewidth, height = 6.5cm]{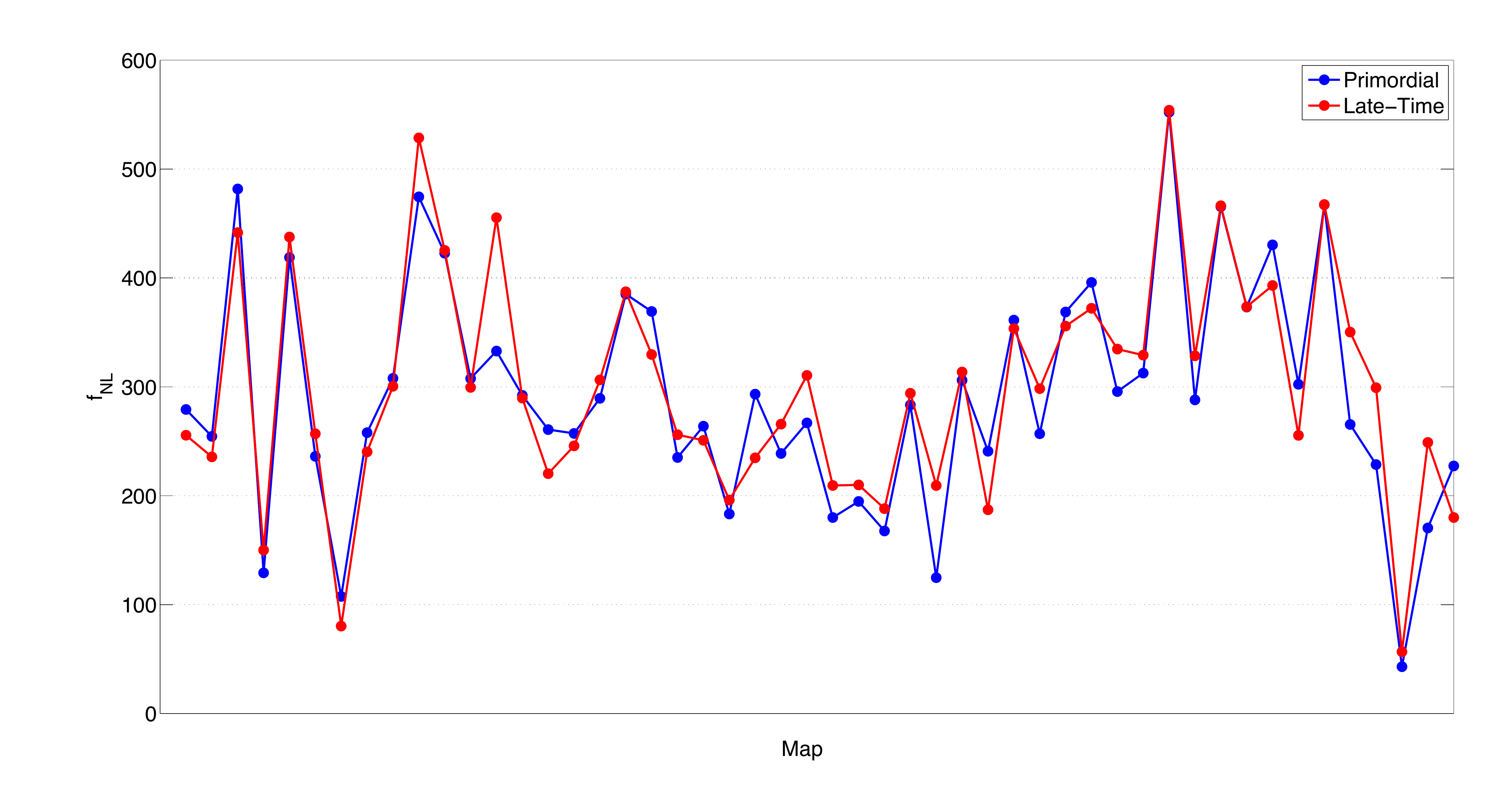}
\includegraphics[width=.85\linewidth, height = 6.5cm]{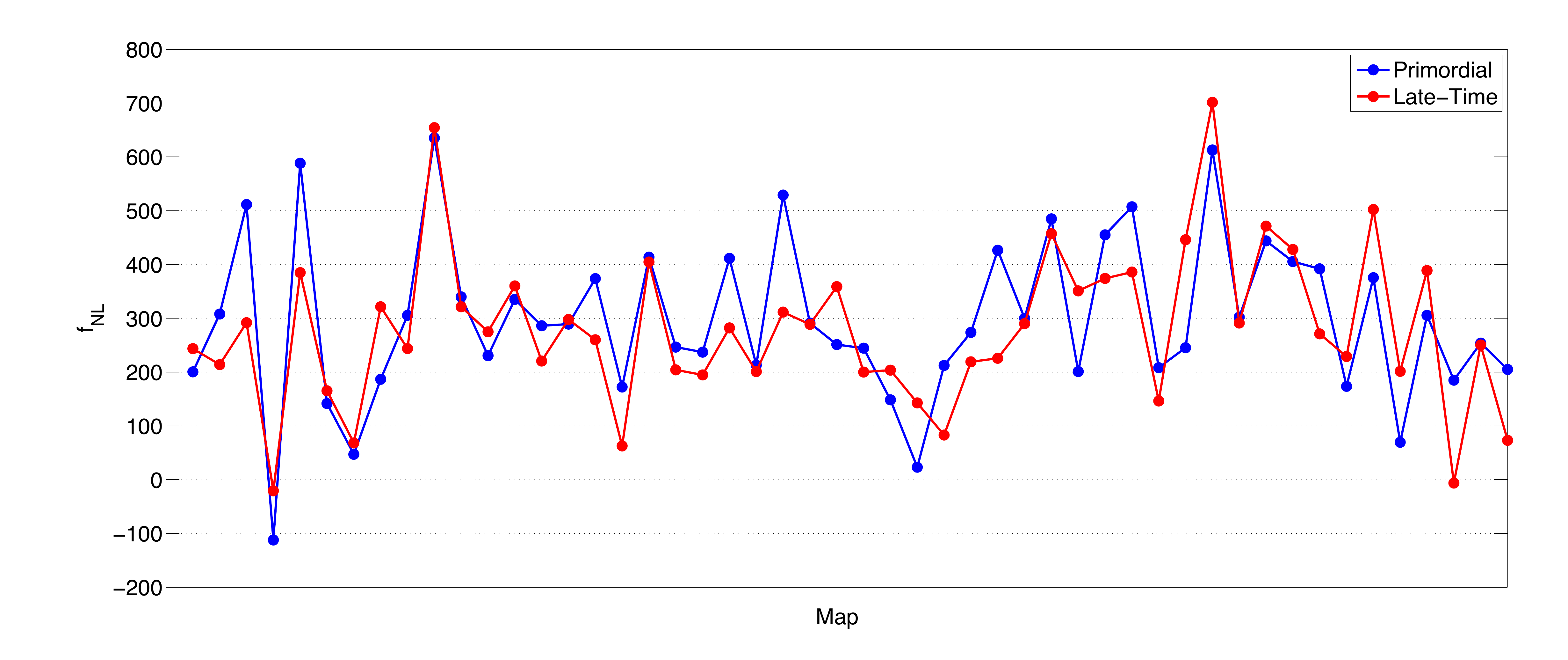}
\caption[Comparison between estimators]{\small  Recovery of $\fnl$ from 50 simulated 
maps of the equilateral model, showing a direct map-by-map comparison between the primordial estimator
(\ref{eq:primestimator}) (blue) and the CMB estimator (\ref{eq:cmbestimator}) (red).  Ideal map 
recovery is shown in the top panel, while recovery for WMAP-realistic maps is shown below 
with beam, inhomogeneous noise and mask included (BNM).   Both methods
recovered the input $\fnl =300$ with a variance of approximately $\Delta \fnl =105$ (clean) and 150 (BNM). Note the 
overall consistency of the two independent estimators with a significantly lower variance evident between 
the methods $\Delta \fnl = 30$ (clean) and $
\Delta \fnl =103$ (BNM). }
\label{fig:estimatorcomparison}
 \end{figure}

Using the algorithm described in section V, we generated a set of $100$ 
equilateral  CMB maps with both the primordial and late-time
decomposition pipelines. We worked at roughly WMAP resolution with 
$l_{max} = 500$ and HEALpix nside $= 512$, corresponding to a pixel number
 $N_{pix} \approx 10^6$. We then applied both our primordial and late-time estimators 
 to both our primordial and late-time sets of simulated maps in all combinations. We found that in all 
 cases the map-making methods gave consistent results, producing 
 simulated maps from which the correct $f_{NL}$ could be reliably recovered 
 with the correct variance. Results for both primordial and late-time estimators 
 on the same set of 50 equilateral maps (with and without the mask and inhomogeneous noise) 
 with $f_{NL} = 300$ can be seen in fig.~\ref{fig:estimatorcomparison}. We observe that the 
two estimators produce consistent results on the same maps. Of course, there is 
some small variation between the results as the two estimators can be regarded 
to be independent but this proved always to be well within the variance.

In addition, we extracted the equilateral configurations 
$B_{lll}$ of the bispectrum from the maps and compared the average over all the
simulations to the semi-analytic expectations obtained 
from the standard decomposition of the equilateral shape in terms 
of $\balpha(x)$, $\bbeta(x)$, $\bgamma(x)$, $\bdelta(x)$ (refer to 
eqns  (\ref{eq:ABterm}) and (\ref{eq:CDterm})). 
The recovered equilateral bispectrum values were in very good agreement
between the semi-analytic prediction from the ``standard''
$\alpha$,$\beta$,$\gamma$,$\delta$ decomposition and the
simulations, based on our eigenmode expansion, thus showing
consistency with previous approaches.

Finally, we reiterate that this general approach to map simulation was
highly efficient, producing Planck resolution maps for the equilateral model on 
short timescales. This made estimator validation through Monte Carlo simulatoins 
easily achievable with only modest resources.   For other well-behaved bispectra, 
such as the cosmic string model, the general method proved robust.  
Examples of non-separable maps already have been 
discussed and shown in fig.~\ref{fig:ngmaps}.


\begin{figure}[th]
\centering
\includegraphics[width=.85\linewidth, height = 6.25cm]{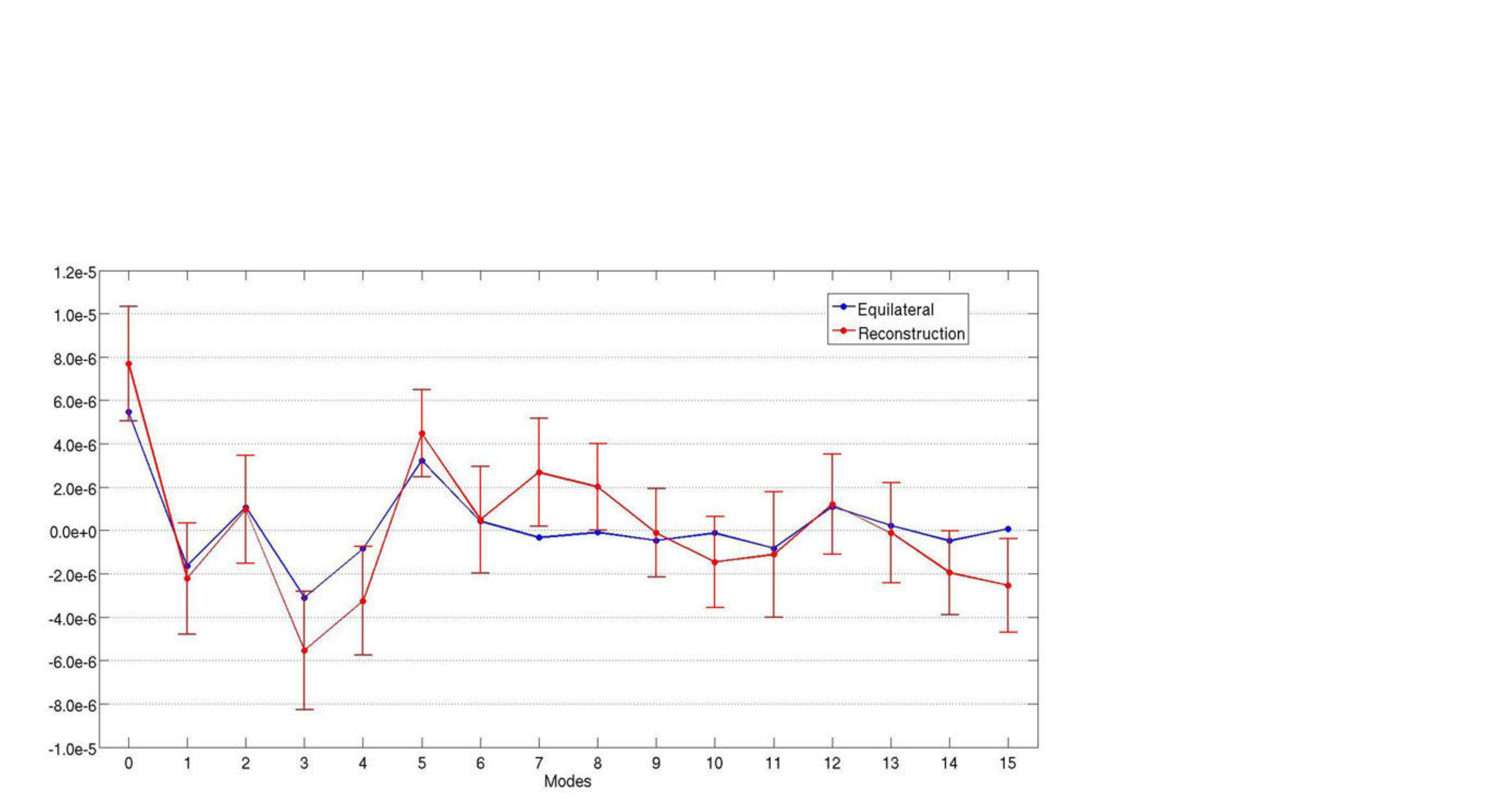}
\includegraphics[width=.85\linewidth, height = 6.25cm]{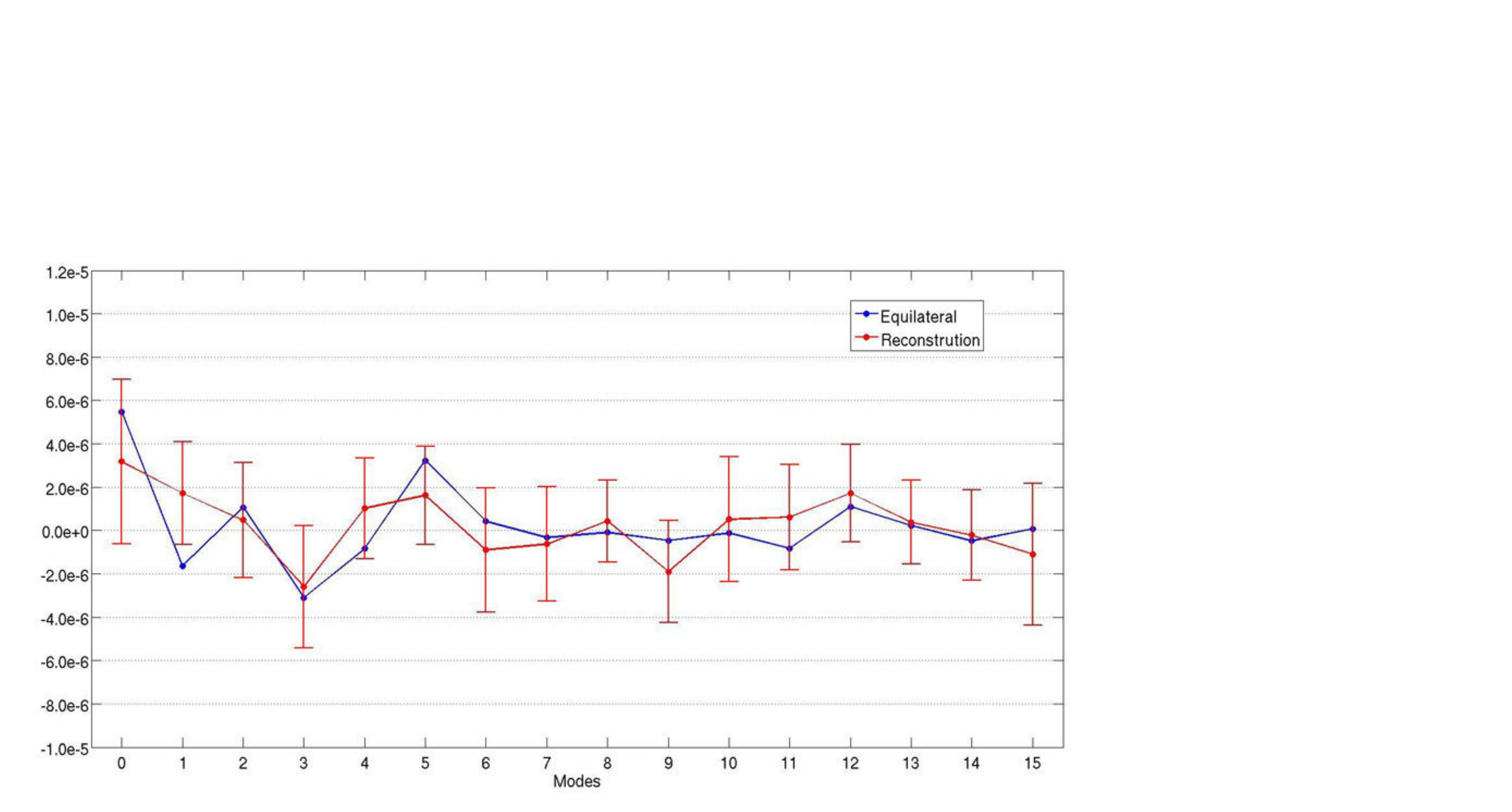}
\caption[]{\small  Recovered spectral coefficients $\bbRn$ from the late time estimator (\ref{eq:cmbestmodes}) from a single
map simulation for an equilateral model with $\fnl = 600$ (or 
normalised relative to the local model $\Fnl \approx 110$);  NG map simulation shown in the lower panel of fig.~\ref{fig:ngmaps}.  In both panels, the original $\baRn$ decomposition coefficients for the 
theoretical model are shown for comparison (blue).  In the upper panel, the $\bbRn$ coefficients recovered from the single realisation
are shown, with error bars (2$\sigma$) estimated from 100 Gaussian maps.   In the lower panel,  the $\bbRn$ are recovered in a WMAP-realistic 
context using the KQ75 mask with inhomogeneous noise added.  Note that the $\bbRn$ provide a remarkably good fit to the$\baRn$ given 
the significance of the non-Gaussian signal.}
\label{fig:reconalpha}
\end{figure}

\subsection{Primordial and late-time $f_{\rm NL}$ estimators}

\begin{table}[b]
\begin{tabular}{c | c | c || c | c |}
       & \multicolumn{2}{|c||}{\bf Ideal simulations} & \multicolumn{2}{|c|}{\bf WMAP5 simulations} \\
       \cline{2-5}\cline{2-5}
                         & $\;$ Average $\;$   & $\;$ St. Dev. $\;$ &
                         $\;$ Average $\;$ & St. Dev. $\;$ \\
       \hline                 
   {\bf Primordial estimator} &  $292.9$  &  $104.8$   & $297.7$  & $152.1$
   \\
  \hline 
   {\bf Late-time estimator}  &  $300.6$  &  $104.9$   & $278.7$  & $160$
   \\
  \hline
   {\bf Internal st.\ dev.} &  \multicolumn{2}{|c||}{$38.5$}  &  \multicolumn{2}{|c|}{$102.6$} \\
   \hline
\end{tabular}
\caption{Results obtained from the application of the primordial and late-time estimators 
as described in the text. In the first two
  columns, labeled by `Ideal simulations', we consider ideal full-sky noiseless
  measurements, while in the last two columns, labeled by `WMAP5
  simulations' we include noise and sky coverage in order to simulate a
  WMAP5-realistic experiment (see text for further explanation). We apply
  both estimators to a single set of maps, in this case created using the late-time
  mode expansion approach.  
  In the last row, we calculate the difference
  between the $f_{\rm NL}$ recovered by the two techniques, map by
  map for 100 maps, and report the final internal standard deviation
  between the methods.}
  \label{tab:fnlcomparisons}
\end{table}

Choosing an input value $f_{\rm NL} = 300$ for the sets of equilateral 
map simulations described above, we compared results from both the primordial and late-time 
bispectrum estimators.  In order
to verify the consistency of the two methods we selected 
the late-time map sets and applied both estimators to it. 
The tests were performed starting 
from a noiseless full-sky map and then more realistic simulations
were used, including partial sky-coverage and an anisotropic 
noise component. 
The rms noise was obtained by coadding WMAP V and W channel using the
same scheme as the one adopted for nonGaussian analysis by the WMAP
team \cite{Komatsu:2008hk}. The sky-coverage was done using the KQ75 mask, 
also adopted by the WMAP5 team for their $\fnl$ analysis. 
Only the approximate form (\ref{eq:approxestimator}) of the
estimator is used, and not the full form (\ref{eq:optimalestimator}) including the
full covariance matrix and a linear term. Note however that this
approximation has been demonstrated in several previous studies to 
work well for equilateral
shapes. Moreover, for our purposes the approximate nearly-optimal
estimator is all we need since it contains all the dependence on the
theoretical ansatz and thus all the dependence on our eigenmode
expansion, which is the primary concern for this initial validation 
process.

We compared the $f_{NL}$ recovered from each map
using the two methods, as well as the
final averages and variances. The variances were compared to
expectations from Fisher matrix forecasts obtained both from our
eigenmode expansion and from the `standard' $\balpha$, $\bbeta$,
$\bgamma$, $\bdelta$ decompostion of the equilateral shape used 
to date in other nonGaussian analysis. In all cases the results were
internally consistent and in agreement with Fisher matrix expectations, as
summarized in table (\ref{tab:fnlcomparisons}).  This led us to
conclude that the eignemode expansion method appears to be a reliable way to
produce non-Gaussian CMB simulations and $f_{\rm NL}$ estimators for
primordial models, whether separable or otherwise.

\begin{figure}[th]
\centering
\includegraphics[width=.325\linewidth]{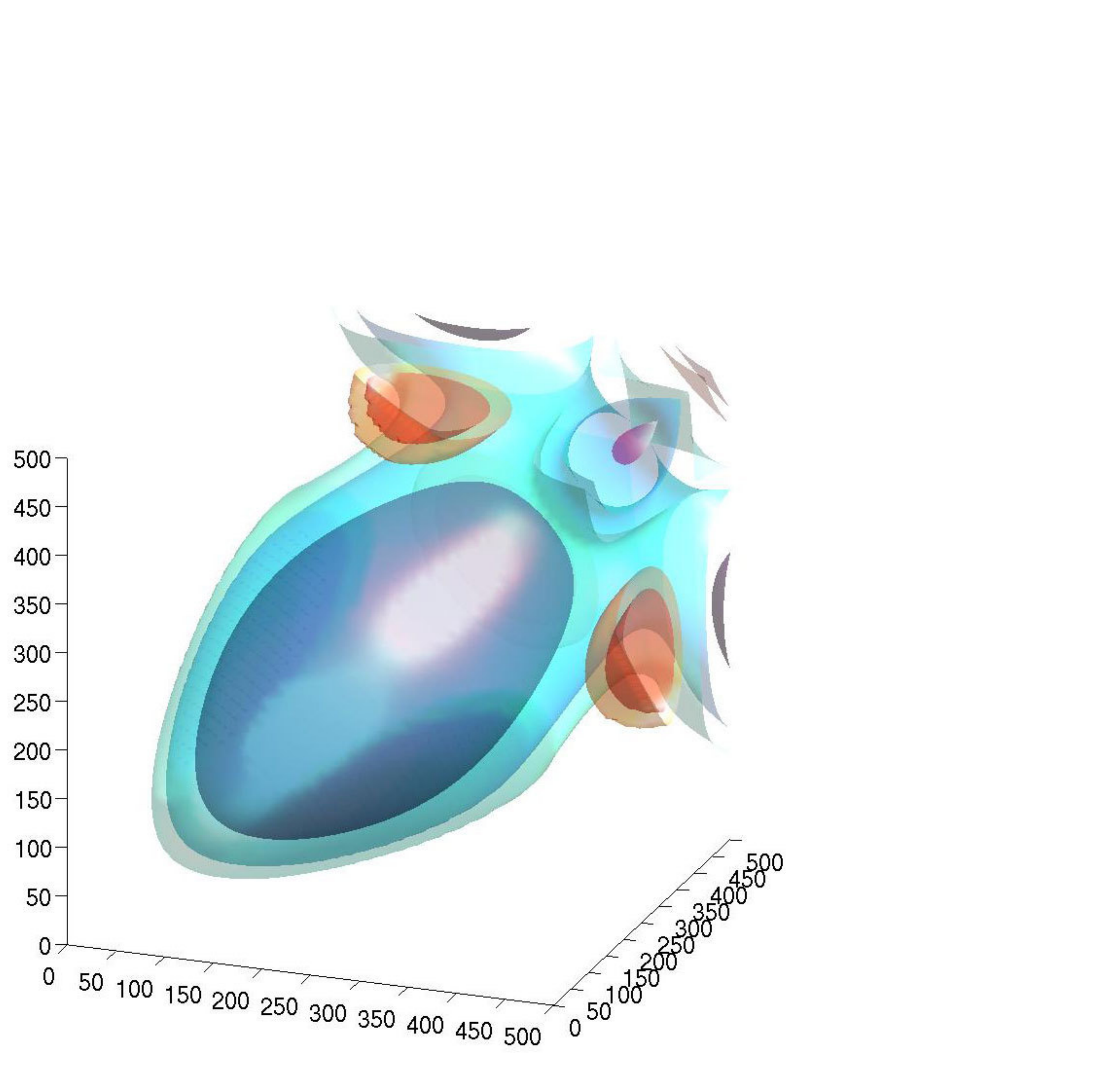}
\includegraphics[width=.325\linewidth]{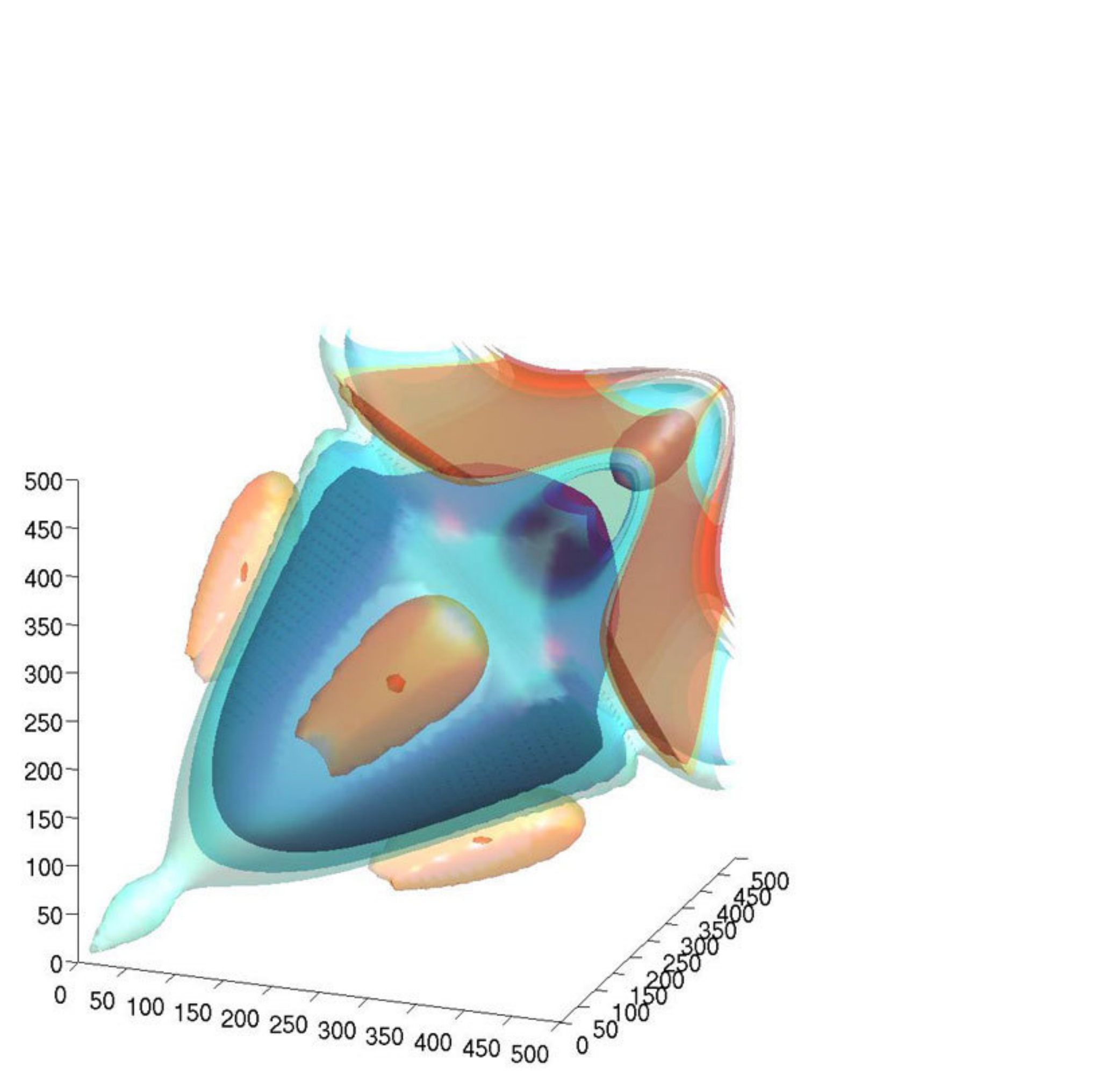}
\includegraphics[width=.325\linewidth]{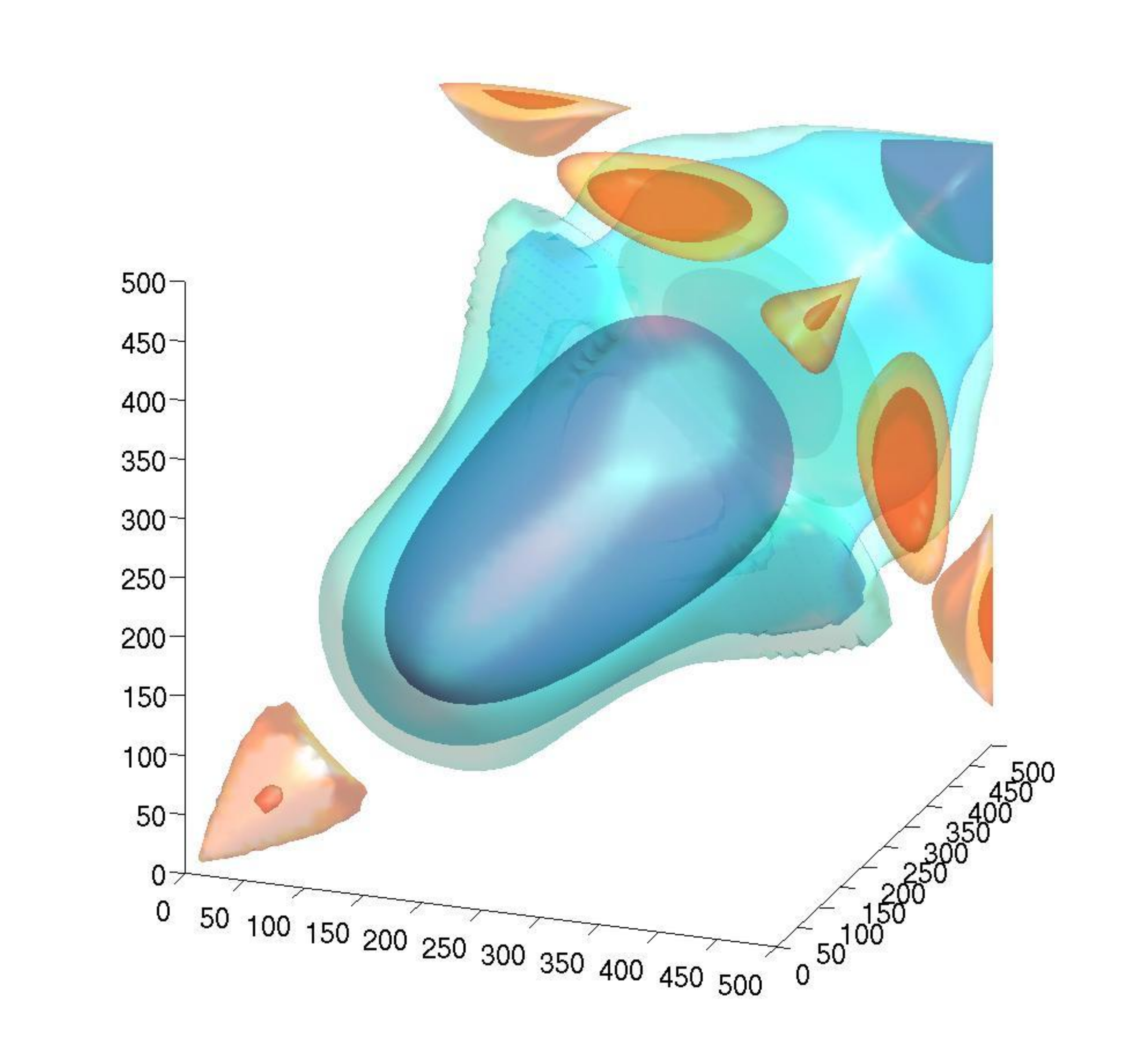}
\caption[Recovered bispectrum]{\small  Recovered 3D bispectrum using the late-time mode decomposition method (\ref{eq:laterecon}) from a single
map simulation for an equilateral model with $\fnl = 600$ (or relative to the local model,  $\Fnl \approx 110$); this figure shows the
reconstruction of the bispectrum from the $\bbRn$ expansion modes illustrated in fig.~\ref{fig:reconalpha}.  The left panel represents the original theoretical 
bispectrum used to construct a single realisation of the map (just like those shown in fig.~\ref{fig:ngmaps}).  The middle
panel represents the recovered bispectrum from the ideal map, while the right panel represents the recovery in a 
WMAP-realistic context using the KQ75 mask with inhomogeneous noise added. The main feature of the bispectrum, 
that is, the primary acoustic peak appears to be 
evident even in the noisy cut-sky case, given the significant nonGaussian signal.  }
\label{fig:3drecon}
\end{figure}

Having verified the two estimator's performance on simulated equilateral maps we then  applied both of them to the  WMAP5 data, coadding the V and W channels as discussed above. 
The primordial estimator obtained the result $-174< \fnl^{\rm equil} < 434$, which is consistent with the existing constraints obtained using standard separable primordial approach (given the caveat that a number of these 
results have now been superseded \cite{Senatore:2009gt}). The constraint using the late-time estimator was  $-90 < \fnl^{\rm equil}  < 550$ which is slightly larger than the primordial result but still well within the $1\s$ range.  As the late-time estimator can be regarded as independent of the primordial 
estimator, some variation is to be expected as we have seen already in fig.~\ref{fig:estimatorcomparison}. 
The difference between the estimators is consistent with the internal variance of 103 noted in table 1 for equilateral map
simulations in a WMAP-realistic context.   We conclude that both the 
primordial and late-time estimators appear to be performing up to expectation.

Using the method described in eqn~(\ref{eq:laterecon}), we can endeavour to recover the full bispectrum from a given map.  To illustrate
this capability, we created a single map realization from an equilateral bispectra with $\fnl = 600$, that is, a map with 
a $4\s$ nonGaussian signal.   We then used the late-time 
estimator to recover the $\bbQn$ and $\bbRn$ mode coefficients described in (\ref{eq:cmbestmodes}). Recall that 
for results of sufficient significance, the $\bbRn$ should approximate the original theoretical 
model coefficients $\baRn$, that is, 
those used to generate the simulated map.   We estimated the variance in each of the eigenmodes 
by applying the same method to 100 Gaussian simulations. The results for the orthornomal coefficients
$\baRn$ are plotted in fig.~\ref{fig:reconalpha} for both ideal maps and for maps with inhomogeneous noise added and a mask applied. We see that we recover the first 7 modes well from the ideal map but the results from the map containing noise and mask are somewhat less encouraging.  Clearly, more work is required to control noise and mask effects
at higher mode numbers. By plotting the 3D bispectra from the reconstructions, see figure \ref{fig:3drecon}, we observe that it is possible to  recover the main accoustic peak and some basic features of the CMB bispectra.  We will address the 
challenging issues associated with bispectrum reconstruction 
in greater detail elsewhere \cite{inprep}.

\section{Conclusions}

We have now implemented two comprehensive and independent pipelines for 
the analysis and estimation of general primordial or CMB bispectra.  Both methods
are based on dual mode expansions, exploiting a complete orthonormal eigenmode basis 
to efficiently decompose arbitrary bispectra into a separable polynomial expansion.  
These separable mode expansions, whether at late or early times, allow a reduction 
of the computational overhead to easily tractable levels, whether calculating the reduced
bispectrum $\blll$, generating Planck resolution nonGaussian map simulations, or 
directly estimating $\fnl$ from simulations or real data sets.   The method exploits 
the smoothness of the pattern of acoustic peaks observed in calculations reviewing
all well-behaved primordial models, implying the rapid convergence of the corresponding 
mode expansions.  While most calculations are performed in the separable basis, a final
rotation of mode coefficients into the orthonormal frame allows for a simple interpretation 
of the contributions to $\fnl$ using Parseval's theorem.  In fact, the completeness of the 
orthonormal eigenmodes means, in principle, that it is straightforward to extract and reconstruct the
full CMB bispectrum from the data, assuming the presence of a sufficiently significant
nonGaussian signal. 

The main purpose of this paper has been to present a detailed theoretical 
framework for $\fnl$ estimation using separable eigenmode expansions, irrespective 
of the specific polynomials or other basis functions employed.  However, we have 
also presented some numerical results from the pipelines we have implemented, 
chiefly for the equilateral model where there are extensive published results for 
direct comparison.   An important milestone for the validation of this approach  
has been the development of a robust and reliable mode expansion method for generating map simulations from
arbitrary bispectra.   While generalising previous methods applied to specific 
 separable cases, we noted that the scale-invariance of the polynomial expansion modes
eliminates numerical instabilities that previously had to be circumvented on a case-by-case basis.  
Given convergent mode expansions for well-behaved bispectra, high resolution
map simulations for a wide variety of models can easily and efficiently be  generated,  with several 
examples illustrated here including late-time cosmic strings.  The many map simulations
created for the equilateral model with both primordial and late-time methods 
showed consistency  in expected variance and $\fnl$ recovery.   

The primordial and late-time $\fnl$ estimators using mode expansions were tested 
successfully on the simulated equilateral maps, matching expectations for semi-analytic 
Fisher matrix forecasts and providing consistent unbiased results for $\fnl$.  This was
achieved for both ideal maps and in a WMAP-realistic context, incorporating beams, 
anisotropic noise and a mask.   Application of the estimators to the WMAP5 data gave
constraints on the equilateral model consistent with each other and previously published
results. These encouraging results suggest that the approach will provide a robust and general framework 
for $\fnl$ estimation for the wide variety of non-separable models which remain to be constrained \cite{Fergusson:2008ra}.  For single equilateral map simulation with $\fnl =600$, we were able to demonstrate 
a reasonable correspondence between the theoretical and recovered mode expansion coefficients, 
while also being able to recover key features of the full CMB bispectrum.   However, a detailed
discussion of such prospects has been left for a future publication \cite{inprep}.   We have also 
left aside for discussion elsewhere a more sophisticated treatment of sky cuts and inhomogeneous 
noise, which is more important for the analysis of the local model, as well as the potential for 
incorporating polarisation data.   Challenges remain for the full 
implementation of the primordial and late-time pipelines at Planck resolution, but the
generality and robustness of this methodology suggests that it should prove to  be a useful tool 
for exploring and constraining  a much wider class of nonGaussian models.

\section{Acknowledgements}

We are very grateful for informative discussions with Xingang Chen and Kendrick Smith
and we have also benefitted from useful conversations  with  Martin Bucher, Anthony Challinor, Olivier 
Forni, Enrique Martinez-Gonzalez and 
Bartjan van Tent.   The ongoing development of these methods has been regularly reported at
 Planck Working Group 4 (NonGaussianity), where we have been grateful for feedback and 
 to learn of related work.  Simulations were performed on the COSMOS supercomputer (an Altix 4700) which is funded by 
STFC, HEFCE and SGI.  JRF, ML and EPS were supported by STFC grant ST/F002998/1 and the 
Centre for Theoretical Cosmology.  

\bibliographystyle{unsrt}
\bibliography{BispectrumI}

\end{document}